\documentclass[trackchanges, twocolumn]{aastex701}
\usepackage{amsmath}
\usepackage{array}
\usepackage{float}
\usepackage{xcolor}
\usepackage{booktabs}
\usepackage{multirow} 

\begin{document}

\title{A Census of Stellar-mass Black Holes in the Milky Way with POPKIN. $\mathbf{I}$. Isolated Black Holes}

\author[orcid=0000-0003-3862-0726]{Jian-Guo He}
\affiliation{Department of Astronomy, Nanjing University, Nanjing 210023, People's Republic of China}
\affiliation{Key Laboratory of Modern Astronomy and Astrophysics, Nanjing University, Ministry of Education, Nanjing 210023, People's Republic of China}
\email{hejg@smail.nju.edu.cn}

\author[orcid=0000-0003-2506-6906]{Yong Shao}
\affiliation{Department of Astronomy, Nanjing University, Nanjing 210023, People's Republic of China}
\affiliation{Key Laboratory of Modern Astronomy and Astrophysics, Nanjing University, Ministry of Education, Nanjing 210023, People's Republic of China}
\email[show]{shaoyong@nju.edu.cn}

\author[orcid=0009-0009-4482-6350]{Yu-Dong Nie}
\affiliation{Department of Astronomy, Nanjing University, Nanjing 210023, People's Republic of China}
\affiliation{Key Laboratory of Modern Astronomy and Astrophysics, Nanjing University, Ministry of Education, Nanjing 210023, People's Republic of China}
\email{1452531593@qq.com}

\author[orcid=0009-0002-3654-8775]{Gui-Yu Wang}
\affiliation{Department of Astronomy, Nanjing University, Nanjing 210023, People's Republic of China}
\affiliation{Key Laboratory of Modern Astronomy and Astrophysics, Nanjing University, Ministry of Education, Nanjing 210023, People's Republic of China}
\email{wangguiyu@smail.nju.edu.cn}

\author[orcid=0000-0001-9565-9462]{Xiao-Tian Xu}
\affiliation{Tsung-Dao Lee Institute, Shanghai Jiao-Tong University, 1 Lisuo Road, Shanghai, 201210, People’s Republic of China}
\email{xxu-tdli@sjtu.edu.cn}

\author[orcid=0000-0002-0716-3801]{Chen Wang}
\affiliation{Department of Astronomy, Nanjing University, Nanjing 210023, People's Republic of China}
\affiliation{Key Laboratory of Modern Astronomy and Astrophysics, Nanjing University, Ministry of Education, Nanjing 210023, People's Republic of China}
\email{chen_wang@nju.edu.cn}

\author[orcid=0000-0002-3614-1070]{Xiao-Jie Xu}
\affiliation{Department of Astronomy, Nanjing University, Nanjing 210023, People's Republic of China}
\affiliation{Key Laboratory of Modern Astronomy and Astrophysics, Nanjing University, Ministry of Education, Nanjing 210023, People's Republic of China}
\email{xuxj@nju.edu.cn}

\author[orcid=0000-0002-0584-8145]{Xiang-Dong Li}
\affiliation{Department of Astronomy, Nanjing University, Nanjing 210023, People's Republic of China}
\affiliation{Key Laboratory of Modern Astronomy and Astrophysics, Nanjing University, Ministry of Education, Nanjing 210023, People's Republic of China}
\email[show]{lixd@nju.edu.cn}

\begin{abstract}

Gravitational-wave observations have revealed hundreds of stellar-mass black holes, yet only about two dozen are known in the Milky Way, almost all in binaries. We present POPKIN, a Python framework that couples single- and binary-star evolution with Galactic orbital dynamics to trace black-hole progenitors from the zero-age main sequence to the present-day isolated black-hole (IBH) population. Across ten models varying the supernova (SN) prescription, mass-transfer efficiency, and common-envelope ejection efficiency, the total IBH abundance is controlled primarily by the SN prescription. Our fiducial model, with a recently proposed metallicity- and stripping-history-dependent SN prescription, predicts \(\sim4\times10^7\) IBHs in the Galaxy, including \(\sim8\times10^4\) within \(1\,\mathrm{kpc}\) of the Sun; alternative SN prescriptions predict \(\sim(1-2)\times10^8\) IBHs. The fiducial model yields a bimodal mass distribution, with peaks near \(9\,M_{\odot}\) and \(20\,M_{\odot}\), and a pronounced deficit at \(13-17\,M_{\odot}\). This feature distinguishes the fiducial model from the alternative prescriptions, some of which substantially populate the \(2-5\,M_{\odot}\) mass-gap region. Non-kicked IBHs follow nearly circular orbits and remain close to the Galactic plane, with typical peculiar velocities of \(20-30\,\mathrm{km\,s^{-1}}\), whereas kicked systems undergo more pronounced radial migration and span a broader velocity range. We estimate \(\sim5\times10^3\) accreting IBHs with \(F_{\rm X}>10^{-14}\,\mathrm{erg\,s^{-1}\,cm^{-2}}\), nearly all non-kicked; this estimate is sensitive to the adopted radiative-efficiency and hot-flow treatments. For a Roman-like bulge survey, our fiducial model predicts \(\sim360\) intrinsic IBH microlensing events over five years in a \(1.70\,\mathrm{deg^2}\) effective area, before survey-selection effects. We propose that long-timescale microlensing events from IBHs can strongly constrain the SN physics governing stellar-mass black hole formation.

\end{abstract}



\section{Introduction} \label{sec:introduction}

Black holes (BHs) are among the most direct astrophysical manifestations of gravitational collapse in general relativity. The theoretical basis for their formation was laid by \citet{Oppenheimer1939}, who derived a relativistic solution for the collapse of massive objects.  The first observational evidence for a BH came from X-ray and optical studies of the X-ray binary Cygnus X-1  \citep{Webster1972,Bolton1972,Miller-Jones2021}. Astrophysical BHs are conventionally divided by mass into three classes: stellar-mass BHs ($\lesssim 10^2\,M_\odot$), intermediate-mass BHs ($\sim 10^2-10^5\,M_\odot$), and supermassive BHs ($\gtrsim10^5\,M_\odot$). Stellar-mass BHs are the evolutionary endpoints of massive stars  \citep{Woosley2002}, and  theoretical estimates suggest that the Milky Way contains on the order of $\sim 10^8$ such objects \citep[e.g.,][]{Brown1994,Timmes1996,Wiktorowicz2019,Olejak2020}.

Hundreds of stellar-mass BHs have now been identified or inferred through electromagnetic observations, gravitational microlensing, and gravitational waves (GWs). Gravitational microlensing, originally proposed as a way to discover dark objects \citep{Paczynski1986,Paczynski1996}, has been widely applied to the search for stellar-mass BHs \citep{Bennett2002,Mao2002,Wyrzykowski2016,Kaczmarek2025}. Recent optical microlensing surveys have probed a broad range of stellar-remnant masses \citep[e.g.,][]{Wyrzykowski2020,Mroz2021}, but have so far led to the robust confirmation of only one well-characterized isolated BH (IBH), OGLE-2011-BLG-0462, with a mass of $\sim 7\,M_\odot$ \citep{Sahu2022,Lam2022,Morz2022,Lam2023,Sahu2025}. 
In X-ray binaries, dynamical mass measurements have firmly identified over 20 BHs \citep{Remillard2006,Casares2014,Corral-Santana2016,Fortin2023,Fortin2024,Avakyan2023,Neumann2023}. Complementary searches for radial-velocity variations of optically bright companions in non-accreting binaries have also been proposed and implemented to uncover quiescent BHs \citep{Trimble1969}. This approach has been pursued through numerous optical surveys \citep[e.g.,][]{Liu2019,Thompson2019,Zheng2019,Giesers2019,Gaia2023,Mahy2022,Shenar2022,El-Badry2023,Nagarajan2025a,An2025,Whitaker2026}, yielding several confirmed systems in the Milky Way \citep[see][for a review]{El-Badry2024}. The advent of GW astronomy, marked by the landmark detection of GW150914 \citep{Abbott2016}, has further opened a new window onto merging BH binaries. To date, near 400 binary merger events involving at least one BH component have been reported in the local universe \citep{Abbott2019,Abbott2021,Abbott2023,LIGO2025a,LIGO2026,LIGO2026b}. Despite this rapidly growing sample, the fundamental properties of stellar-mass BHs, including their masses, spins, and natal kick velocities, remain active topics of investigation and debate.

\subsection{Masses}

One long-standing question is whether a mass gap exists between neutron stars (NSs) and BHs. Observations of X-ray binaries suggest a lower mass limit at $\sim5\,M_\odot$ for BHs \citep{Bailyn1998,ozel2010,Farr2011}. In contrast, constraints from both electromagnetic and GW observations place the maximum NS mass at $\sim 2-2.5\,M_\odot$ \citep{ozel2016,Alsing2018,You2025}. Taken at face value, these measurements imply a possible gap at $\sim 2-5\,M_\odot$ between the heaviest NSs and the lightest BHs  \citep[see][for a review]{Shao2022}.

Because progenitor stars have an approximately continuous initial mass distribution \citep{Kroupa1993}, such a gap would require a non-monotonic mapping between progenitor mass and compact-remnant mass. Several supernova (SN) explosion mechanisms have been proposed to produce this feature \citep[e.g.,][]{Fryer2012,Ugliano2012,Kochanek2014,Fryer2022}. However, 
recent discoveries have begun to populate the putative gap. For example, the accreting BH in the X-ray binary GRO J0422+32 has a mass constrained to $2.7_{-0.5}^{+0.7}\,M_\odot$ \citep{Casares2022}. Using combined spectroscopy and astrometry, \citet{Wang2024} identified a compelling BH candidate with a mass of $3.6_{-0.5}^{+0.8}\,M_\odot$ in a binary system with a red giant companion. In the globular cluster NGC 1851, \citet{Barr2024} reported a compact object of mass $\sim2.1-2.7\,M_\odot$ in a binary with a pulsar companion, which could be either a very massive NS or a low-mass BH. From the GW catalog, events such as GW190814 and GW230529 involve compact objects with masses of $2.59_{-0.09}^{+0.08}\,M_\odot$ \citep{Abbott2020} and $2.5-4.5\,M_\odot$ \citep{Abac2024}, respectively, also placing them in the mass-gap regime. Moreover, the compact-object mass spectrum inferred from GW sources does not appear to contain a completely empty gap between NSs and BHs \citep{LIGO2025b,LIGO2026}.

Post-formation evolution can further complicate the interpretation of the mass gap. Accretion or mergers can increase the masses of NSs, potentially allowing them to enter the $2-5\,M_\odot$ range \citep[e.g.,][]{Gao2022,desa2022,Chen2023,Siegel2023,Zhu2024,Ye2024}. Thus, even if SN explosions naturally produce a dearth of compact remnants in this mass range, later astrophysical processes may partly refill it. It remains uncertain whether SN explosions alone can produce mass-gap compact objects. In addition, an apparent mass gap could arise if binaries containing mass-gap remnants are preferentially disrupted by large SN-driven kicks \citep{Mandel2021,Burrows2025}.

\subsection{Spins}

GW observations of merging BH binaries indicate that their pre-merger components typically have low spins, with an average dimensionless spin parameter of $a_{\rm spin}\sim 0.1-0.2$ \citep{Abbott2023b}. The first-born BHs in these binaries often appear to spin more slowly than the second-born BHs, suggesting that natal BH spins may be intrinsically low and that angular-momentum transport in massive progenitor stars is efficient \citep[e.g.,][]{Qin2018,Fuller2019,Belczynski2020}. By contrast, most published spin measurements for BHs in X-ray binaries report high values, often $a_{\rm spin}\gtrsim 0.7$, especially in systems with high-mass donors \citep{Reynolds2021,Draghis2024}. Because high-mass X-ray binaries are often considered potential progenitors of merging BH binaries, this difference creates an apparent tension between the spin distributions inferred from electromagnetic and GW observations.

Part of this discrepancy may arise from systematic uncertainties in electromagnetic spin measurements. The two primary techniques used for X-ray binaries, the reflection method and the continuum-fitting method, are both subject to significant modeling uncertainties \citep[see][for a review]{Zdziarski2026}. In addition, the observed spin distribution of low-mass X-ray binaries is broad, spanning $a_{\rm spin}\sim 0.1$ to $\gtrsim 0.9$, and is commonly interpreted as the result of spin-up by long-term accretion \citep[e.g.,][]{Podsiadlowski2003,Fragos2015,Shao2020}.

Several astrophysical explanations have also been proposed. Super-Eddington accretion during a temporary Roche-lobe overflow phase may explain the high spins observed in some high-mass X-ray binaries \citep{Qin2022,Xing2025}, and may also account for the minority of GW sources that contain high-spin BHs \citep[e.g.,][]{Shao2022b,Briel2023}. Alternatively, the currently observed high-mass X-ray binaries and the merging BH binaries detected by LIGO and Virgo may not be directly linked by a single evolutionary channel \citep{Gallegos2022,Liotine2023}. If they arise from distinct channels or are observed at different evolutionary stages, the apparent spin tension would be naturally alleviated.

\subsection{Kick velocities}

The natal kick imparted to a stellar-mass BH depends sensitively on its formation channel. Direct collapse is expected to produce little or no kick, whereas an asymmetric SN explosion with substantial mass ejection can impart a significant kick to the remnant \citep{Burrows2025,Popov2025}. If the BH forms in a binary, such a kick can alter both the systemic velocity and the orbital configuration of the post-SN system \citep{Blaauw1961,Nelemans1999}.

Historically, natal kicks in BH X-ray binaries have often been inferred from the current height of the system above the Galactic plane, under the assumption that most progenitors form in the disk \citep{White1996,Jonker2004,Repetto2012,Repetto2017}. This diagnostic is degenerate, however, because a system observed close to the plane may simply be passing through it rather than having formed with a small kick. Using very long baseline interferometry and Gaia astrometry, \citet{Atri2019} reconstructed full three-dimensional Galactocentric orbits for 16 systems and estimated their peculiar velocities at Galactic-plane crossings. They found that most BHs in X-ray binaries likely receive substantial natal kicks, with the population statistically favoring a unimodal Gaussian kick distribution with a mean velocity of $\sim 110\,{\rm km\,s^{-1}}$.

More recently, \citet{Nagarajan2025} analyzed the kinematics of 12 BH binaries relative to their local stellar environments. They found that about half of the systems show at least weak evidence for natal kicks, whereas the other half have kinematics consistent with their local populations and disfavor kicks $\gtrsim 50\,{\rm km\,s^{-1}}$. In particular, Swift J1727.8--162 and GRO J1655--40 show strong evidence for high kicks ($\gtrsim 100\,{\rm km\,s^{-1}}$), while V404 Cyg and VFTS 243 independently indicate very weak kicks ($\lesssim 10\,{\rm km\,s^{-1}}$). This mixed evidence suggests that the BH kick distribution may be bimodal, with some BHs forming through direct collapse and others through energetic, asymmetric SN explosions. At present, however, the sample remains too small to distinguish cleanly between unimodal and bimodal models. Overall, the inferred kicks of BHs appear to be lower than the characteristic velocities observed for isolated pulsars \citep{Hobbs2005,Verbunt2017,Igoshev2020}.

Constraining BH formation physics and SN-explosion mechanisms will require larger samples from electromagnetic, microlensing, and GW observations \citep[see][and references therein]{Shao2022}. In recent years, our team has conducted a series of population-synthesis studies to assess the detectability of different classes of BH binaries, including X-ray-quiet systems with optical companions \citep{Shao2019}, X-ray binaries \citep{Shao2020}, and double compact objects detectable as binary radio pulsars \citep{Shao2018} or GW sources \citep{Shao2021}. These studies generally assumed a constant star-formation rate (SFR) and a fixed stellar metallicity throughout the Milky Way, approximating conditions in the thin disk. For Galactic IBHs, however, the present-day distributions of mass, position, and velocity are shaped jointly by stellar evolution, binary interactions, SN kicks, the Galactic star-formation history (SFH), metallicity evolution, and orbital motion in the Galactic potential. To capture these coupled effects, we have developed POPKIN (see details in Section \ref{sec:popkin}), a new framework that combines population synthesis with stellar kinematics.

\subsection{Galactic IBHs}

Observational constraints on stellar-mass black holes have traditionally been derived from binary systems, including X-ray binaries, non-accreting binaries with luminous optical companions, and merging binaries detected through GWs. These systems represent only a subset of the underlying BH population, and their observed properties reflect both binary interactions and observational selection effects. Population-synthesis studies suggest that the Milky Way contains $\sim 10^8$ stellar-mass BHs, the vast majority of which are single or isolated \citep{Wiktorowicz2019,Olejak2020,Sweeney2022}. IBHs therefore provide a complementary probe of BH-formation physics that is less directly affected by biases associated with binary interactions and the detectability of a companion.

IBHs embedded in the interstellar medium (ISM) can, in principle, emit electromagnetic radiation by accreting surrounding gas, predominantly in the X-ray band. Previous studies have explored the detectability of such accreting IBHs. \citet{Agol2002} showed that observatories such as \textit{Chandra} and \textit{XMM-Newton} may detect tens of IBHs per year, with a strong concentration toward the Galactic plane and Galactic center. \citet{Tsuna2018} found that the proposed all-sky survey \textit{FORCE} could detect approximately $30-100$ IBHs under optimistic assumptions; conversely, non-detections could constrain BH kick velocities and accretion efficiencies. Recently, \citet{Rodriguez2024} reported a non-detection of X-ray emission from Gaia BH2, suggesting that the accretion rate near the event horizon may be much lower than the traditional Bondi rate and thereby challenging searches for IBHs accreting from the ISM. Beyond X-rays, \citet{Kimura2025} proposed that IBHs passing through molecular clouds (MCs) could be sources of PeV cosmic rays and LHAASO dark sources. Using the Low-Luminosity AGN Spectral Energy Distribution code to model the spectrum of an accreting IBH, \citet{Murchikova2025} concluded that about 16 IBHs within 200 pc of the Sun could be detectable with X-ray or radio telescopes. More broadly, \citet{Martinez2025} semi-analytically modeled accretion and explored the emission of IBHs from radio to $\gamma$-ray wavelengths \citep[see also][]{Fender2013,Scarcella2021,Kin2025,Nosirov2026}. Most of these studies, however, assume simplified BH mass and velocity distributions and do not include progenitor evolution or dynamical evolution in the Galactic gravitational potential, both of which can substantially affect the observable properties of IBHs.

Gravitational microlensing provides a complementary, emission-independent channel for detecting IBHs, because the signal is produced by the gravitational deflection of light from a background source star rather than by radiation from the vicinity of the BH. Although the observability of accreting IBHs has been widely discussed, no isolated stellar-mass BH has yet been securely identified through ISM accretion. To date, OGLE-2011-BLG-0462 remains the only unambiguously confirmed IBH, identified through astrometric microlensing \citep{Sahu2025}. Previous studies have therefore explored microlensing as a promising route to detect and characterize the Galactic IBH population. \citet{Wiktorowicz2019} estimated that BH lenses could produce tens of microlensing events per year toward the Galactic bulge in OGLE-like surveys. Using the PopSyCLE framework, \citet{Lam2020} showed that long-timescale events are efficient BH-lens candidates and predicted that the \textit{Roman} microlensing survey could measure the masses of \(\mathcal{O}(10^2-10^3)\) IBHs. A more recent Roman/GBTDS-tailored simulation by \citet{Kaczmarek2026} predicted a corrected yield of \(152\pm 21\) BH-lens events. From the current OGLE sample, \citet{Kaczmarek2025} developed a probabilistic classification method for photometric microlensing events and identified 23 high-probability BH candidates. Independently, \citet{Howil2025} analyzed the long-timescale event Gaia18ajz as a candidate IBH. These studies demonstrate the power of microlensing to probe the abundance, mass distribution, and kinematics of IBHs, but they also rely on Galactic-population assumptions that can strongly influence the inferred lens properties \citep[see also][]{Rybicki2018,Sajadian2023,Koshimoto2024}.

The number of identified IBHs is expected to grow substantially in the coming years. To interpret these detections and use them to constrain SN-explosion mechanisms, it is necessary to connect the intrinsic properties, spatial distributions, and dynamical histories of Galactic IBHs within a unified framework. In this first paper of our series on Galactic BHs, we use our new POPKIN code to conduct a comprehensive study of the IBH population in the Milky Way, focusing on how SN explosions shape the properties and observability of IBHs.

Section~\ref{sec:popkin} describes the POPKIN framework and its main modules. Section~\ref{sec:method} presents our method for generating the Galactic IBH population. Section~\ref{sec:IBH} examines the mass, spatial, and kinematic distributions of IBHs, as well as the relation between natal kicks and peculiar velocities. Section~\ref{sec:accrete ISM} evaluates the detectability of IBHs accreting from different ISM phases. Section~\ref{sec:lens-BH} discusses microlensing prospects and the possible origin of OGLE-2011-BLG-0462. We discuss model uncertainties in Section~\ref{sec:discussion} and summarize our conclusions in Section~\ref{sec:conclusion}.

\section{POPKIN: a Python Code for Population Synthesis and Stellar Kinematics} \label{sec:popkin}

\begin{figure*}
    \centering
    \includegraphics[width=\linewidth]{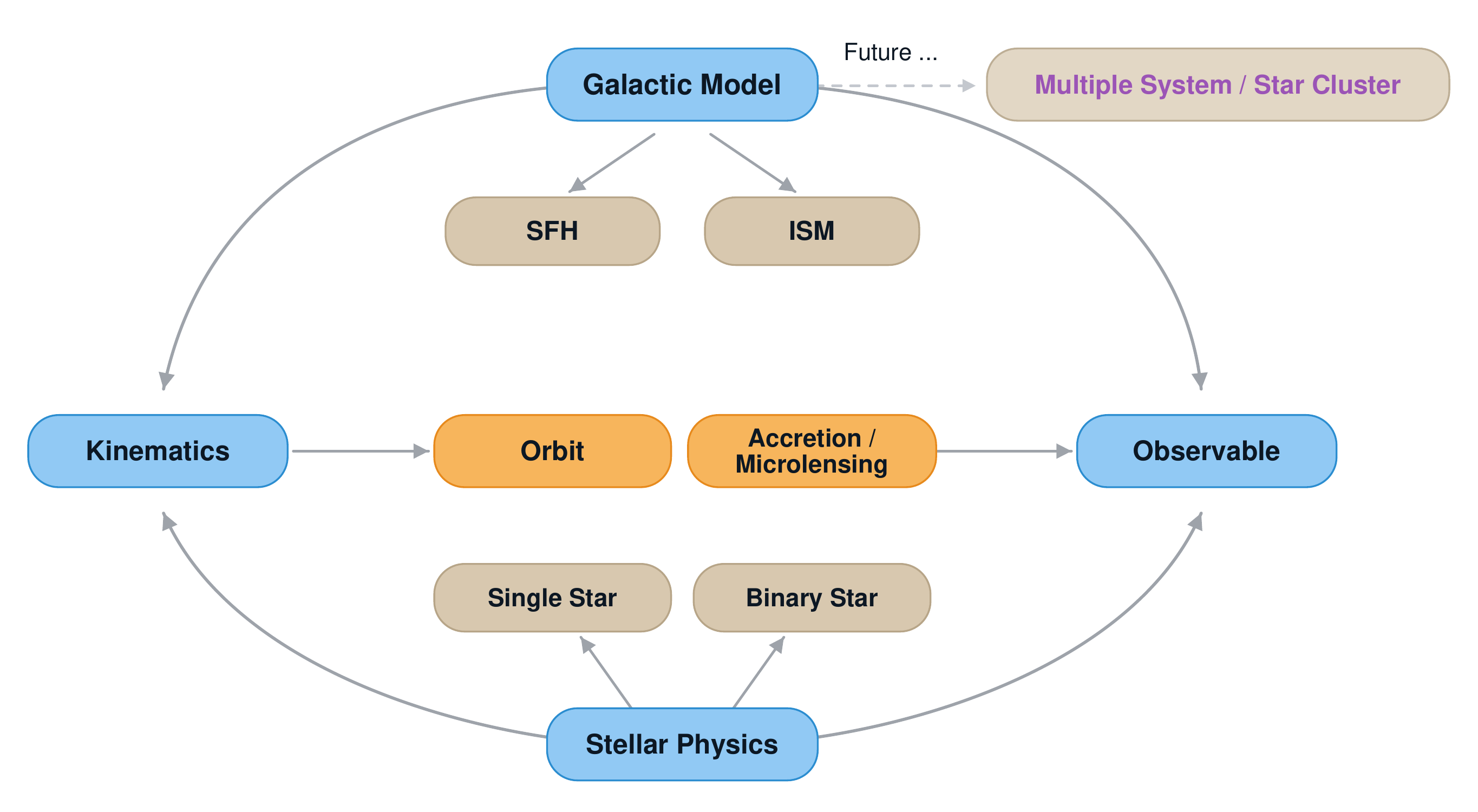}
    \caption{Architecture of POPKIN. The framework consists of four core components: the Galactic model, which specifies the SFH and ISM; stellar physics, which includes single- and binary-star evolution; kinematic evolution, which follows the orbits of stellar systems in the Galactic gravitational potential; and observable post-processing, which maps synthetic populations onto measurable quantities. Future releases will extend POPKIN to include multiple-star systems and star clusters. The direction of each arrow indicates a determining or containment relation, with the preceding component determining or containing the subsequent one.} 
    \label{fig:POPKIN_structure}
\end{figure*}

Stellar and binary population synthesis is a fundamental tool in astrophysics. By connecting assumptions about stellar and binary evolution to the statistical properties of observable populations, it provides a way to constrain uncertain evolutionary processes through comparison with data \citep[see][for a review]{Han2020}.

Over the past decades, rapid population-synthesis codes such as BSE
\citep{Hurley2000, Hurley2002}, StarTrack \citep{Belczynski2008},
MOBSE \citep{Giacobbo2018}, COSMIC \citep{Breivik2020}, COMPAS
\citep{Riley2022, Compas2025}, and SEVN \citep{Iorio2023} have been
widely used in stellar astrophysics. These parametric codes are
computationally efficient, but rely on simplified fitting formulae and
approximate prescriptions for binary interactions, which can introduce
systematic uncertainties. In contrast, BPASS \citep{Eldridge2017} and
POSYDON \citep{Fragos2023} use precomputed grids of detailed
binary-star evolutionary calculations generated with the STARS
\citep{Eggleton1971} and MESA \citep{Paxton2011} stellar evolution
codes, respectively, to address some of these limitations. These
grid-based approaches enable a more self-consistent treatment of stellar
structure and binary interactions than is possible with purely
parametric prescriptions.

Despite the success of traditional population-synthesis codes, modern surveys increasingly require models that connect stellar evolution to the Galactic environment. This connection is especially important for populations whose present-day positions, velocities, and observability depend not only on progenitor evolution but also on their birth locations, orbital evolution in the Galactic potential, and the Galaxy's chemical-enrichment history \citep[see also the \texttt{cogsworth} framework developed by][]{Wagg2025}. These ingredients are often simplified or omitted in frameworks that focus primarily on stellar and binary evolution alone.

Motivated by these needs, we have developed POPKIN (POPulation synthesis and stellar KINematics), a modular Python framework for large-sample population synthesis. POPKIN builds on the BSE code \citep{Hurley2000, Hurley2002}, but reorganizes the calculation in an object-oriented architecture that is easier to inspect, extend, and couple to additional physical modules. The current framework combines single- and binary-star evolution with a Galactic model for the SFH and chemical enrichment, stellar kinematics, accretion from the ISM, gravitational microlensing, and GW signal-to-noise ratio (SNR) calculations. In this way, POPKIN links legacy rapid population synthesis to a modern, extensible astrophysical software ecosystem.

This section is organized as follows. Section~\ref{subsec:overview} gives an overview of POPKIN and its main features. Section~\ref{subsec:galaxy} describes the Galactic model, including the SFH and the ISM. Sections~\ref{subsec:single} and~\ref{subsec:binary} summarize the stellar-physics framework for single and binary stars, respectively. Section~\ref{subsec:kinematic} introduces the treatment of stellar kinematics through orbital integration. Section~\ref{subsec:ps} describes the population-synthesis workflow. Section~\ref{subsec:observables} presents the post-processing modules that connect synthetic populations to observables, and Section~\ref{subsec:use} provides a brief guide to running POPKIN.

\subsection{Overview and Features} \label{subsec:overview}

Figure~\ref{fig:POPKIN_structure} presents the overall architecture of POPKIN. The code is organized into four core components: the Galactic model, the stellar-physics module, the kinematic-evolution module, and the observable post-processing layer. The Galactic model specifies the SFH, chemical enrichment, and ISM, while the stellar-physics module handles both single- and binary-star evolution.

The stellar-physics module is the computational core of POPKIN and can be coupled to the other modules for two main purposes. First, when combined with the Galactic model and kinematic-evolution modules, it traces the orbital evolution of stellar populations, enabling studies of post-SN binaries, isolated NSs and BHs, and runaway stars. Second, when combined with the observable post-processing module, it can estimate the observational signatures of selected stellar populations, such as IBHs studied in this paper. The main features of POPKIN are summarized below.

\paragraph{Core Capabilities}
POPKIN follows single-star evolution from the zero-age main sequence (ZAMS) to remnant formation, as well as binary-star evolution through processes such as Roche-lobe overflow, common-envelope (CE) evolution, and tidal interactions through rapid population synthesis approach. Building on these foundations, the code incorporates stellar kinematics through orbit integration in the Galactic potential and implements a Galactic model that tracks SFH and chemical enrichment in the thin disk, thick disk, and bulge.

\begin{figure}
    \centering
    \includegraphics[width=\linewidth]{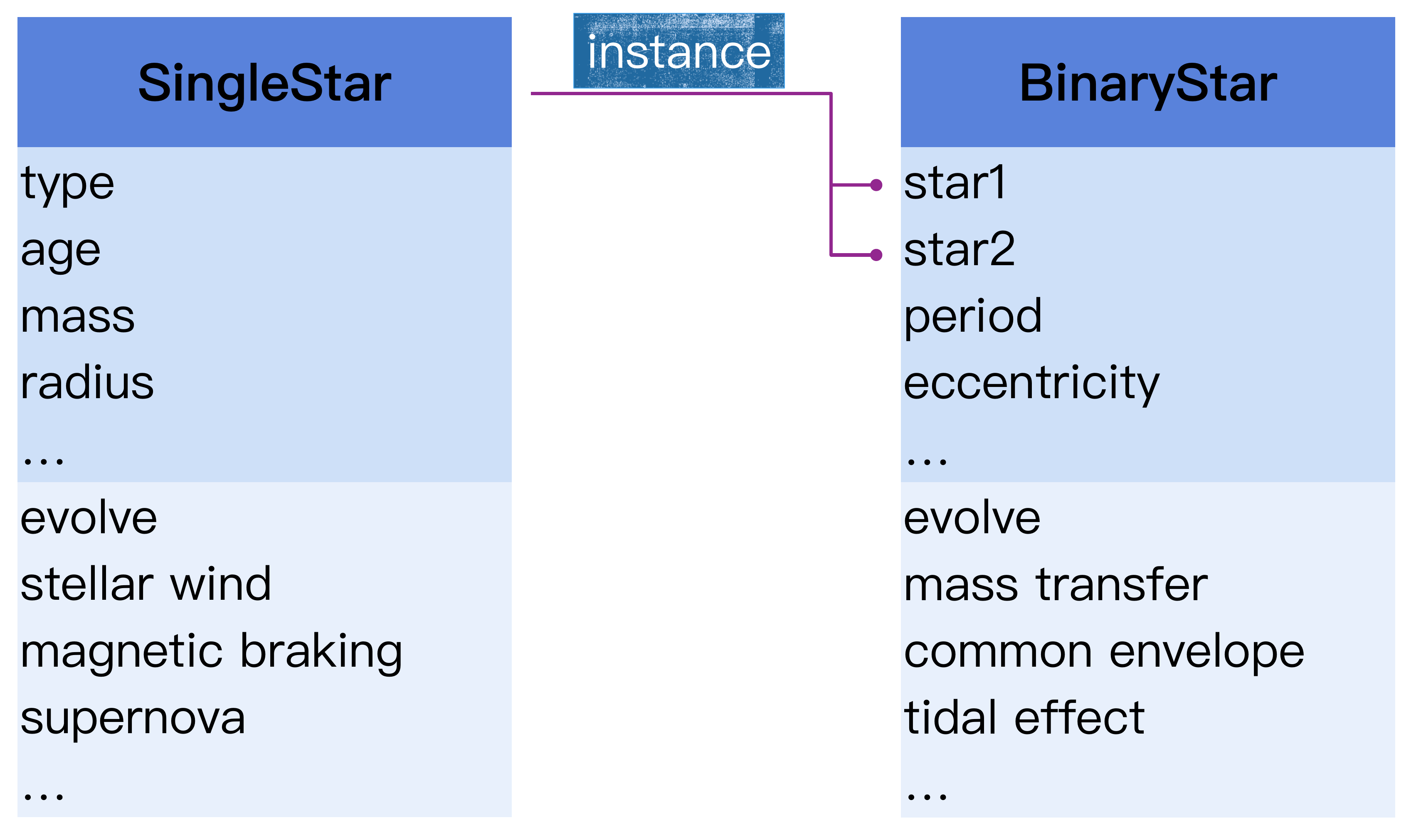}
    \caption{Class structure of the stellar-physics module in POPKIN. The \texttt{SingleStar} class stores the properties and evolutionary methods for a single star, while the \texttt{BinaryStar} class contains two \texttt{SingleStar} instances (\texttt{star1} and \texttt{star2}) together with binary orbital parameters and interaction methods.} 
    \label{fig:stellar_classes}
\end{figure}

\paragraph{Stellar Classes}
The stellar-physics framework in POPKIN is built around two core classes: \texttt{SingleStar} and \texttt{BinaryStar}. As illustrated in Figure~\ref{fig:stellar_classes}, the \texttt{SingleStar} class stores the fundamental properties and evolutionary processes of a single star, including its type, age, mass, radius, and methods for stellar winds, magnetic braking, and SN explosions. The \texttt{BinaryStar} class contains two \texttt{SingleStar} instances (\texttt{star1} and \texttt{star2}), together with binary-specific attributes such as orbital period and eccentricity, and implements binary-interaction processes including tidal effects, mass transfer, and CE evolution. All physical prescriptions are implemented as modular options. Detailed descriptions of these prescriptions are given in Sections~\ref{subsec:single} and~\ref{subsec:binary}. This object-oriented design provides an extensible foundation for coupling stellar evolution to Galactic orbit integration and population synthesis.

\begin{table*}
\caption{Fiducial physical assumptions and runtime configuration settings adopted for POPKIN calculations.}
\label{tab:popkin_settings}
\centering
\footnotesize
\begin{tabular}{lll}
\hline  \hline
\multicolumn{1}{c}{Name} & 
\multicolumn{1}{c}{Option} & 
\multicolumn{1}{c}{Description} \\
\hline
\multicolumn{3}{c}{Physical Assumptions} \\
\hline
$\alpha_{\rm CE}$    & 1.0     &    CE ejection efficiency, set to unity by default    \\
$\lambda_{\rm bind}$   & \citet{XuXJ2010}       &  Binding-energy parameter from \citet{XuXJ2010}    \\
HG survive {CE} &  True    &  Whether Hertzsprung-gap (HG) stars are allowed to survive CE evolution  \\ 
CCSN remnant prescription & \citet{Maltsev2025} & Metallicity- and stripping-history-dependent prescription,  \\
 	&		& 		based on \citet{Maltsev2025}\\
CCSN natal kick model & single Maxwellian & Single-Maxwellian kick distribution with 1D rms velocity \(\sigma_{\rm CCSN}=217~{\rm km~s}^{-1}\), \\
 	&		& 	following the revised single-peak model of \citet{Disberg2025} \\
ECSN \& AIC natal kick & $30 \mathrm{~km~s}^{-1}$ &  Maxwellian natal kick with 1D rms velocity from \citet{Pfahl2002} \\
mass transfer efficiency & rotation dependent &  Rotation-dependent prescription from \citet{Shao2014} \\
WD supercritical accretion & CE-wind &  Prescription adopted when a WD undergoes supercritical accretion, \\
 	&		& 		based on \citet{Cui2022}\\
stellar wind & \citet{Merritt2026} & Phase-dependent stellar-wind prescription following \citet{Merritt2026} \\
magnetic braking  & \citet{Hurley2002} &  Magnetic-braking prescription from \citet{Hurley2002} \\
maximum NS mass & 2.5 $M_{\odot}$ &  Following \citet{Fryer2012} \\ [2pt]
\hline
\multicolumn{3}{c}{Runtime Configuration} \\
\hline
program & sse & Single-star evolution driver \\
max step & 20000 & Maximum number of evolution iterations, following \citet{Hurley2002}  \\
max time & 12000 $\rm Myr$ & Maximum evolution time, set to the Galactic lifetime \\
parallel processes & 10 & Number of parallel processes for population synthesis \\
jit decorator & disable & Whether to enable the Numba \texttt{@jit} decorator for performance acceleration \\
orbital integration & disable & Whether to enable orbital integration for stellar kinematics \\
GW SNR & disable & Whether to calculate compact-binary SNRs \\ [2pt]
\hline
\multicolumn{3}{c}{Binary Population Synthesis} \\
\hline
Z & $0.0001 - 0.03$ & Stellar metallicity \\
chemical evolution & constant & Galactic metallicity-evolution model used in population synthesis   \\
initial mass function (IMF) & \citet{Kroupa2002} & IMF from \citet{Kroupa2002} \\
binary fraction & variable & Mass-dependent binary fraction from \citet{Haaften2013} \\
$P_{\rm orb}$ & $\log_{10}P_{\rm orb}=(0.15, 5.5)$ & Initial orbital-period range [days], based on \citet{deMink2015}\\
eccentricity & 0 & Initially circular orbit \\
spin of stars & fitting & Derived from the main-sequence fitting formulae of \citet{Hurley2000} \\ [2pt]
\hline  \hline

\end{tabular}

\tablecomments{For a full list of parameters and detailed descriptions, see \texttt{src/popkin/config/controls\_default.py}.}
\end{table*}

\paragraph{Technical Highlights}
POPKIN builds on a substantially updated version \citep[][and the current paper]{Shao2014,Shao2021} of the rapid population-synthesis code BSE \citep{Hurley2000,Hurley2002}. Its core single- and binary-evolution algorithms trace back to BSE, while the implementation has been rewritten from Fortran to Python and reorganized into a more readable and extensible architecture. Optional acceleration through the Numba just-in-time compiler\footnote{\url{https://numba.pydata.org/}} helps maintain high performance for the most computationally intensive stellar-evolution routines. When enabled, Numba's \texttt{@njit} decorator is applied to the stellar-physics module, substantially accelerating repeated single-star and binary-evolution calculations after the initial compilation overhead\footnote{
The JIT acceleration is applied only to the stellar-physics module. The kinematic-evolution module, including Galactic orbit integration, is not JIT-accelerated. Since the compilation overhead is incurred at runtime, JIT acceleration is most beneficial for large binary population-synthesis calculations, where the cost is amortized over many evolved systems.
}.

Designed for high-performance computing environments, POPKIN supports multi-process parallel execution to efficiently simulate large stellar populations\footnote{
On Unix-like systems, POPKIN uses the \texttt{fork} multiprocessing start method. When JIT compilation is enabled, the main process warms up the compiled routines before creating the worker pool, allowing child processes to inherit the compiled state through copy-on-write memory sharing. This optimization is specific to \texttt{fork}-based multiprocessing and to signatures compiled before the worker processes are created.}. The code adopts a modular architecture that separates stellar physics, the Galactic model, kinematic evolution through orbit integration, observable post-processing, and the infrastructure for configuration, logging, and output management. Physical assumptions and runtime options are controlled through human-readable configuration files, and runtime logs record information useful for debugging, reproducibility, and performance analysis. Table~\ref{tab:popkin_settings} summarizes the fiducial model assumptions and runtime configuration settings.


\subsection{Galactic Model} \label{subsec:galaxy}

The Galactic model in POPKIN specifies both the SFH and the ISM. This model allows us to follow the birth sites, ages, and chemical evolution of stellar populations in the Milky Way. Extensions to extragalactic stellar populations are planned for future releases.

\subsubsection{SFH of the Milky Way} \label{subsec:sfh}

\begin{figure}
	\includegraphics[width=\columnwidth]{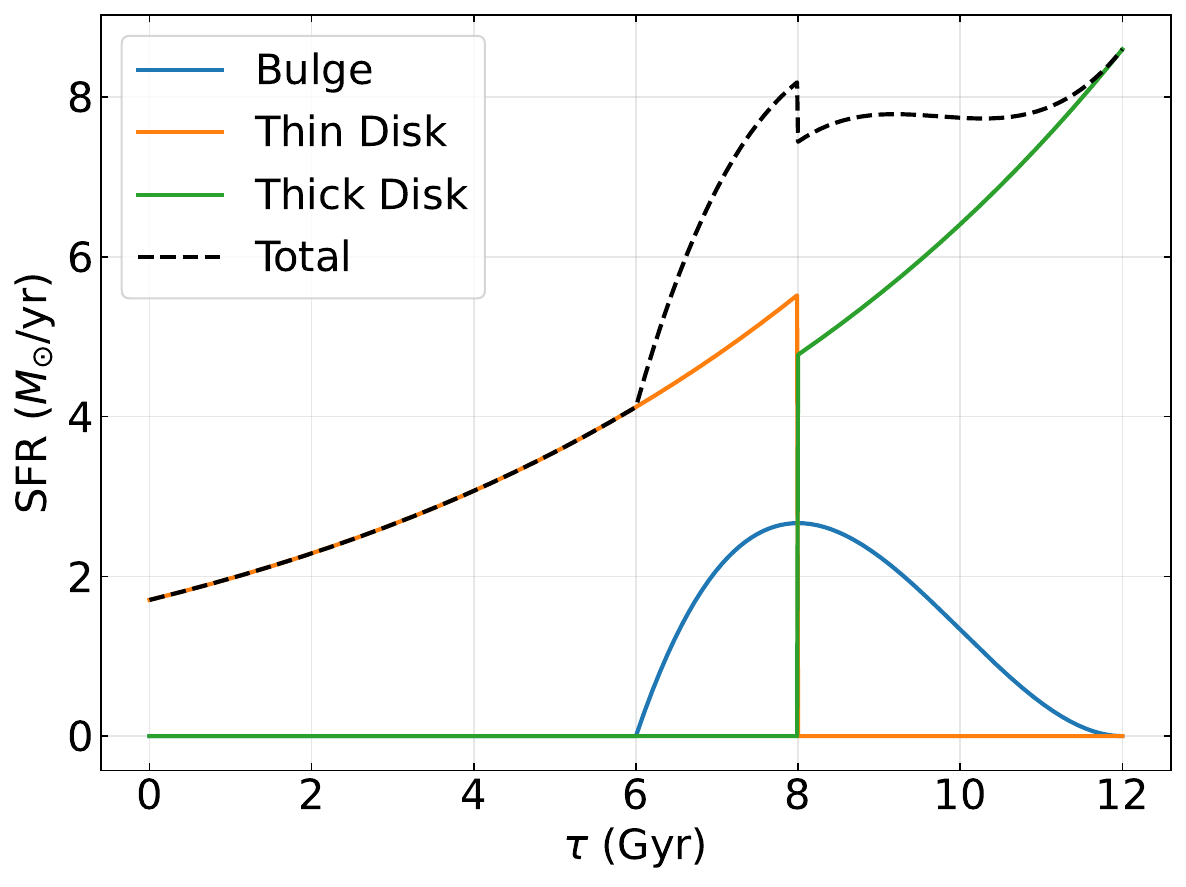}
    \caption{SFRs of the Galactic bulge, thin disk, thick disk, and total Milky Way as functions of lookback time \(\tau\).} 
    \label{fig:sfr}
\end{figure}

To model the SFH and metallicity evolution of the Milky Way, we adopt the framework of \citet{Wagg2022}.

We include three Galactic components: a relatively metal-poor thick disk, a relatively metal-rich thin disk, and a central bulge. The thick disk and bulge dominate star formation at early times, whereas the thin disk dominates during the past $\sim 6-8$ Gyr. The SFRs of the thin and thick disks are written as
\begin{equation}
\mathrm{SFR}(\tau) =\mathrm{SFR}_{\mathrm{thin}/\mathrm{thick}}  \exp \left[-\frac{\left(\tau_{\mathrm{MW}}-\tau\right)}{\tau_{\mathrm{SFR}}}\right],
\end{equation}
where $\tau$ is the lookback time, with $\tau = 0$ at the present day, $\tau_{\mathrm{MW}}=12\, \mathrm{Gyr}$ is the adopted age of the Milky Way, and $\tau_{\mathrm{SFR}}=6.8\, \mathrm{Gyr}$ is the star formation timescale \citep{Frankel2018}. The adopted formation times are $0\,\rm Gyr <\tau<8\,\rm Gyr$ for the thin disk and $8\,\rm Gyr <\tau<12\,\rm Gyr$ for the thick disk. After normalization, $\mathrm{SFR}_{\mathrm{thin}}=9.96\, M_{\odot}\,\mathrm{yr}^{-1}$ and $\mathrm{SFR}_{\mathrm{thick}}=8.6\, M_{\odot}\,\mathrm{yr}^{-1}$, assuming stellar masses of $2.6 \times 10^{10}\, M_{\odot}$ for both disks \citep{Licquia2015}. The stellar mass of the bulge is taken to be $0.9 \times 10^{10}\, M_{\odot}$, and its SFR follows a $\beta(2,3)$ distribution over the lookback-time interval $6\,\rm Gyr <\tau<12\,\rm Gyr$. Figure~\ref{fig:sfr} shows the SFRs of the bulge, thin disk, and thick disk.

The radial probability density functions (PDFs) of the three components follow a gamma distribution $\Gamma(2, R_d)$,
\begin{equation}
p(R)= \frac{R}{R_d^2} \exp \left(-\frac{R}{R_d}\right),
\end{equation}
where $R$ is the Galactocentric radius and $R_d$ is the scale length. For the thin disk, $R_d$ is time dependent and is given by \(R_{d}(\tau)=4\, \mathrm{kpc}\left[1-0.3\left(\tau/8\, \mathrm{Gyr}\right)\right]\). For the thick disk and bulge, we adopt \(R_{d}=(1 / 0.43)\,\mathrm{kpc}\) \citep{Bovy2016} and \(R_{d}=1.5\,\mathrm{kpc}\) \citep{Bovy2019}, respectively.

\begin{figure}
	\includegraphics[width=\columnwidth]{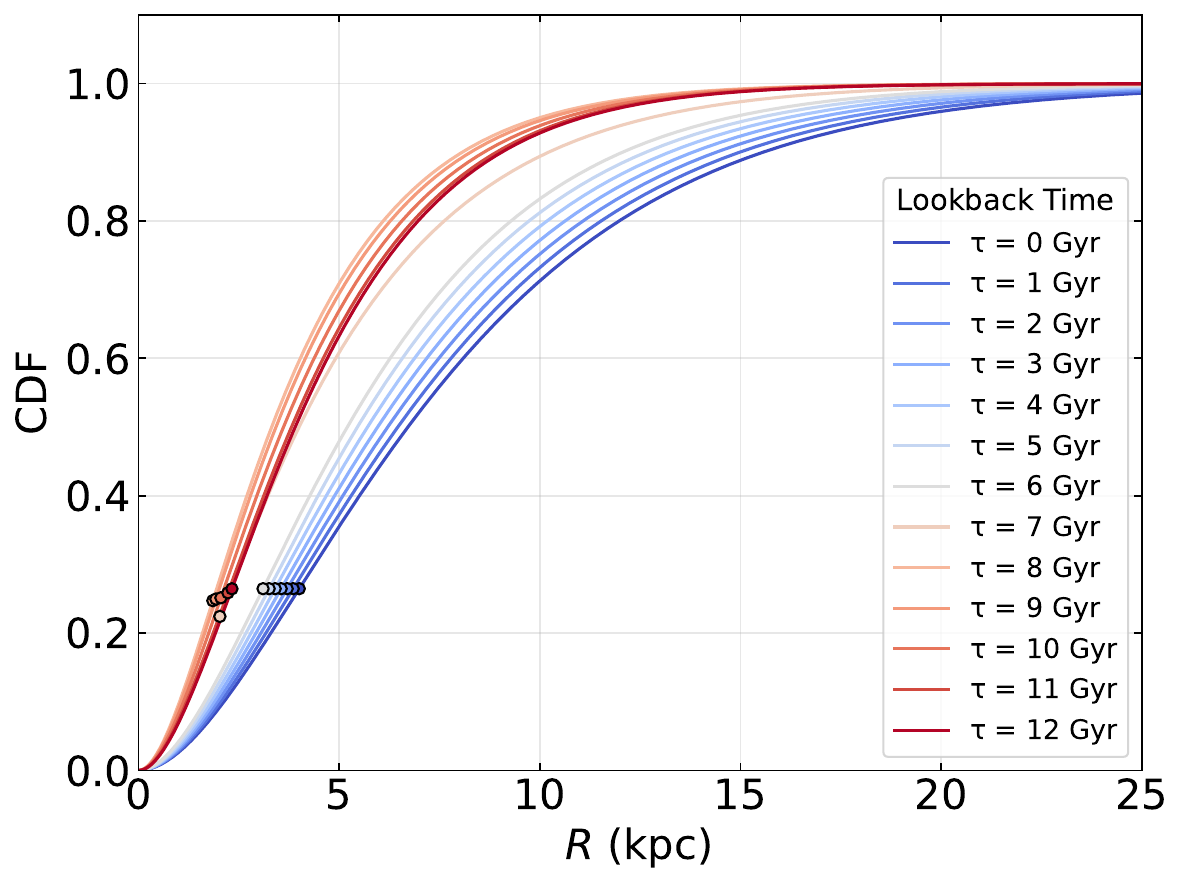}
    \caption{Radial CDFs of SFR at selected lookback times \(\tau\). Filled circles mark the radius at which the corresponding radial PDF peaks.} 
    \label{fig:radial CDF}
\end{figure}

Figure~\ref{fig:radial CDF} shows the radial cumulative distribution functions (CDFs) of SFR for different lookback times $\tau$, with filled circles marking the radii of peak probability. The present-day radial distribution peaks at approximately \(4\,\mathrm{kpc}\), consistent with the radial distribution adopted by \citet{Faucher2006} to reproduce the observed pulsar population. This radius is also close to the peak surface density of molecular hydrogen, \(\mathrm{H}_2\) \citep{Nakanishi2016}, the radius of maximum far-infrared luminosity from massive stars, \(4.7\,\mathrm{kpc}\) \citep{Bronfman2000}, and the peak in the distribution of shell-type SN remnants, \(4.8\,\mathrm{kpc}\) \citep{Case1998}. The radial extent of star formation increases toward more recent epochs, reflecting the inside-out growth of the Galactic disk.

The vertical distribution of each component is described by
\begin{equation}
p(|z|) = \frac{1}{z_d} \exp \left(-\frac{|z|}{z_d}\right),
\end{equation}
where \(z\) is the height above the Galactic mid-plane and \(z_d\) is the scale height. Following \citet{Wagg2022}, we set \(z_d=0.3\, \mathrm{kpc}\) for the thin disk, \(z_d=0.95\, \mathrm{kpc}\) for the thick disk, and \(z_d=0.2\, \mathrm{kpc}\) for the bulge.

\begin{figure*}
	\centering
	\includegraphics[width=16cm]{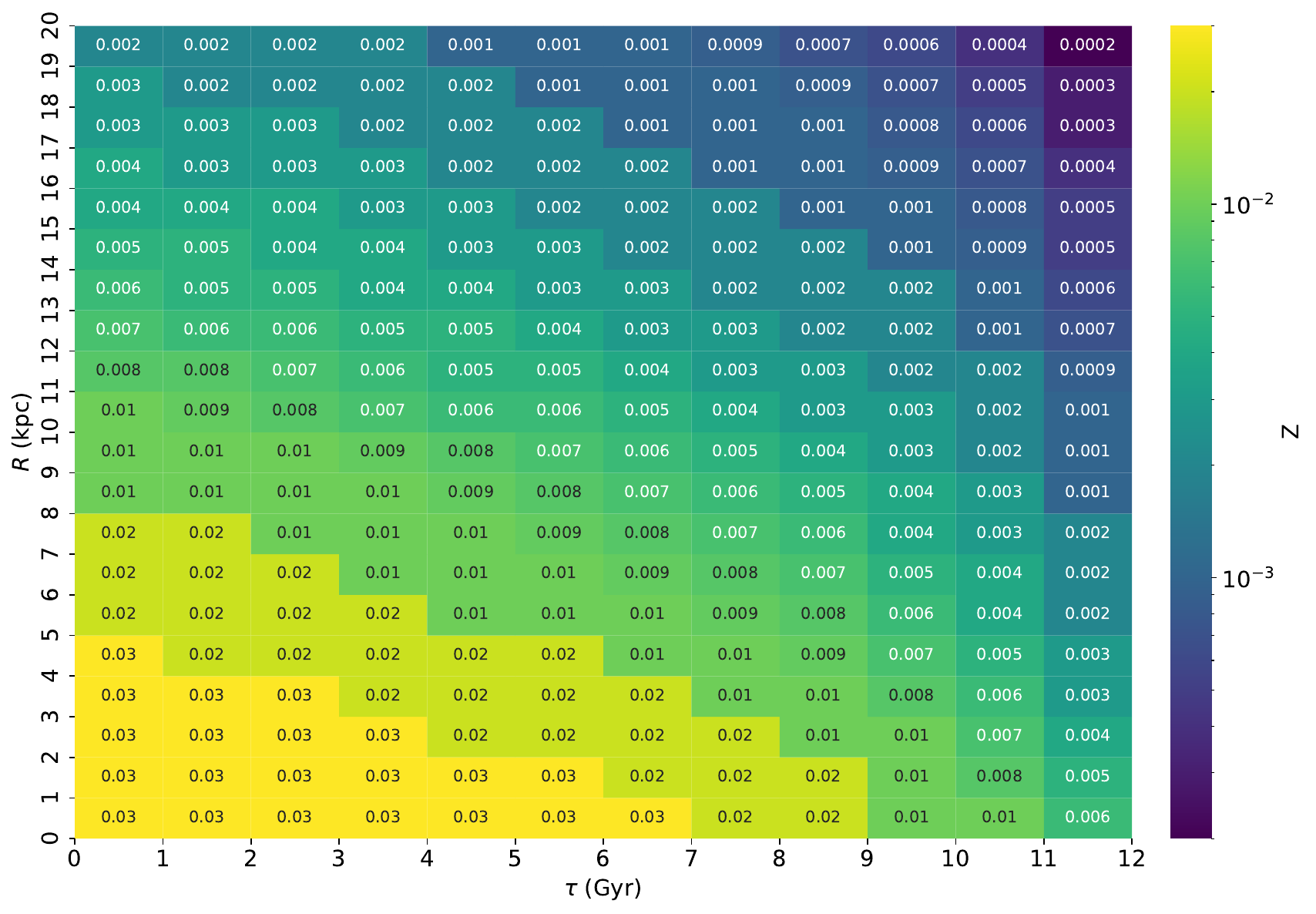}
    \caption{Mean stellar metallicity \(Z\) assigned to each spatiotemporal grid cell as a function of lookback time \(\tau\) and Galactocentric radius \(R\). Each cell is labeled with the corresponding adopted metallicity. 
    }
    \label{fig:Metallicity}
\end{figure*}

For Galactic metallicity evolution, we use a relation that depends on both lookback time and Galactocentric radius \citep{Frankel2018}. To facilitate population synthesis, we discretize the metallicity history on a spatiotemporal grid, as shown in Figure~\ref{fig:Metallicity}. The lookback-time grid contains 12 bins of width \(1\,\mathrm{Gyr}\), and the radial grid contains 20 bins of width \(1\,\mathrm{kpc}\), covering the region within \(20\,\mathrm{kpc}\) where most stars form (Figure~\ref{fig:radial CDF}). In each spatiotemporal cell, the mean metallicity is assigned as the initial metallicity of stars born in that cell.

\subsubsection{The ISM} \label{subsec:ism}

\begin{deluxetable*}{cccccccc}
\caption{Adopted properties of the five ISM phases: scale height \(H_{\mathrm{ISM}}\), mean molecular weight \(\mu\), hydrogen number-density range \(n_{\rm H}\) \citep{Agol2002}, effective sound speed \(c_s\) \citep{Tsuna2018}, and solar-neighborhood mid-plane volume filling fraction \(f_v\).}
\label{tab:ISM}
\tablehead{
\colhead{Phase} & \colhead{$H_{\mathrm{ISM}}$} & \colhead{$\mu~(m_p)$} & \colhead{$n_1\left(\mathrm{~cm}^{-3}\right)$} & \colhead{$n_2\left(\mathrm{~cm}^{-3}\right)$} & \colhead{$\beta$} & \colhead{$f_v(R=\mathrm{8\,kpc}, z=0)$} & \colhead{$c_{\mathrm{s}}\left(\mathrm{~km} \mathrm{~s}^{-1}\right)$} 
}
\startdata
    MCs & $75 \mathrm{~pc}$ & $2.72$ & $10^2$ & $10^5$ & 2.8 & 0.0006 & $3.7\left(n_{\rm H} / 100 \mathrm{~cm}^{-3}\right)^{-0.35}$ \\
    Cold $\mathrm{H}_{\text {I}}$ & $150 \mathrm{~pc}$ & $1.36$ & $10^1$ & $10^2$ & 3.8 & 0.01 &  10 \\
    Warm $\mathrm{H}_{\text {I}}$ & $500 \mathrm{~pc}$ & $1.36$ & \multicolumn{2}{c}{0.3} & $-$ & 0.59 & 10 \\
    Warm $\mathrm{H}_{\text {II}}$ & $1 \mathrm{~kpc}$ & $1.36$ & \multicolumn{2}{c}{0.15} & $-$ & 0.2 & 10 \\
    Hot $\mathrm{H}_{\text {II}}$ & $3 \mathrm{~kpc}$ & $1.36$ & \multicolumn{2}{c}{0.002} & $-$ & 0.4 & 150 \\
\enddata
\end{deluxetable*}

Following \citet{Agol2002}, we divide the ISM into five phases: molecular clouds (MCs) composed of \(\mathrm{H}_2\), cold neutral medium (cold \(\mathrm{H}_{\text{I}}\)), warm neutral medium (warm \(\mathrm{H}_{\text{I}}\)), warm ionized medium (warm \(\mathrm{H}_{\text{II}}\)), and hot ionized medium (hot \(\mathrm{H}_{\text{II}}\)). The adopted properties of these phases are summarized in Table~\ref{tab:ISM}. For simplicity, we assume that the ISM is axisymmetric and follows an exponential vertical profile, although this idealization does not capture all observed ISM structure \citep{Nakanishi2016}.

Following \citet{Agol2002}, we adopt scale heights of \(75 \, \mathrm{pc}\), \(150 \, \mathrm{pc}\), \(500 \, \mathrm{pc}\), \(1 \, \mathrm{kpc}\), and \(3 \, \mathrm{kpc}\) for MCs, cold \(\mathrm{H}_{\text{I}}\), warm \(\mathrm{H}_{\text{I}}\), warm \(\mathrm{H}_{\text{II}}\), and hot \(\mathrm{H}_{\text{II}}\), respectively. The hydrogen number-density distributions of MCs and cold \(\mathrm{H}_{\text{I}}\) are modeled as power laws with index \(\beta\) over the range \(n_1 < n_{\rm H} < n_2\). For MCs, we take \(n_1 = 10^2 \, \mathrm{cm}^{-3}\), \(n_2 = 10^5 \, \mathrm{cm}^{-3}\), and \(\beta = 2.8\). For cold \(\mathrm{H}_{\text{I}}\), we take \(n_1 = 10 \, \mathrm{cm}^{-3}\), \(n_2 = 10^2 \, \mathrm{cm}^{-3}\), and \(\beta = 3.8\). We assume constant hydrogen number densities of \(0.3 \, \mathrm{cm}^{-3}\), \(0.15 \, \mathrm{cm}^{-3}\), and \(0.002 \, \mathrm{cm}^{-3}\) for warm \(\mathrm{H}_{\text{I}}\), warm \(\mathrm{H}_{\text{II}}\), and hot \(\mathrm{H}_{\text{II}}\), respectively. The mean molecular weight, in units of the proton mass \(m_{\rm p}\), is set to \(\mu = 2.72\) for MCs and \(\mu = 1.36\) for cold and warm \(\mathrm{H}_{\text{I}}\), reflecting their low ionization fractions (\(\sim 1 \times 10^{-4} - 3 \times 10^{-3}\)). For warm and hot \(\mathrm{H}_{\text{II}}\), we follow \citet{Tsuna2018} and adopt the same \(\mu\) values as for the neutral phases.

We use the results of \citet{Nakanishi2016} to specify the surface density \(\Sigma\) of MCs and \(\mathrm{H}_{\text{I}}\) as a function of Galactocentric radius. \citet{Smith2023} found that the ratio of cold \(\mathrm{H}_{\text{I}}\) to total \(\mathrm{H}_{\text{I}}\) has little radial dependence except near the Galactic center. Based on their results, we assume that the cold-\(\mathrm{H}_{\text{I}}\) fraction is \(0.5\) for \(R \leq 0.5 \, \mathrm{kpc}\), \(0.2\) for \(1 \, \mathrm{kpc} < R \leq 11 \, \mathrm{kpc}\), and \(0\) for \(R > 13 \, \mathrm{kpc}\), with smooth transitions over \(0.5 \, \mathrm{kpc} < R \leq 1 \, \mathrm{kpc}\) and \(11 \, \mathrm{kpc} < R \leq 13 \, \mathrm{kpc}\).

The volume filling fraction of each ISM phase at a given location is expressed as \citep{Agol2002}
\begin{equation}
f_v(R, z) = f_v(R, z=0) \cdot \exp\left(-\frac{|z|}{H_{\mathrm{ISM}}}\right),
\end{equation}
where \(R\) is the Galactocentric radius, \(z\) is the height above the disk, \(H_{\mathrm{ISM}}\) is the scale height, and \(f_v(R, z=0)\) is the mid-plane volume filling fraction, given by
\begin{equation}
f_v(R, z=0) = 
\begin{cases}
    0.2, & \text{for warm } \mathrm{H}_{\text{II}} \\
    0.4, & \text{for hot } \mathrm{H}_{\text{II}} \\
    \frac{\Sigma(R)}{2 H_{\mathrm{ISM}} \langle n_{\rm H} \rangle \mu m_{\rm p}}, & \text{for other phases}
\end{cases}
\end{equation}
where \(\langle n_{\rm H} \rangle\) is the average hydrogen number density of each ISM phase. For MCs and cold \(\mathrm{H}_{\text{I}}\), which have broad density distributions, the average hydrogen number density is
\begin{equation}
\langle n_{\rm H} \rangle = \frac{\beta - 1}{\beta - 2} \frac{n_1^{2 - \beta} - n_2^{2 - \beta}}{n_1^{1 - \beta} - n_2^{1 - \beta}}.
\end{equation}
The filling fraction of MCs and cold \(\mathrm{H}_{\text{I}}\) also depends on the number density \(n_{\rm H}\). The differential filling fraction is
\begin{equation}
{\rm d}f_v(R,z,n_{\rm H}) =
f_v(R,z)\,
\frac{\beta-1}{n_1^{1-\beta}-n_2^{1-\beta}}\,
n_{\rm H}^{-\beta}\,{\rm d}n_{\rm H},
\end{equation}
where \(f_v(R,z)\) is the total filling fraction integrated over \(n_1\leq n_{\rm H}\leq n_2\).

\begin{figure}
	\includegraphics[width=\columnwidth]{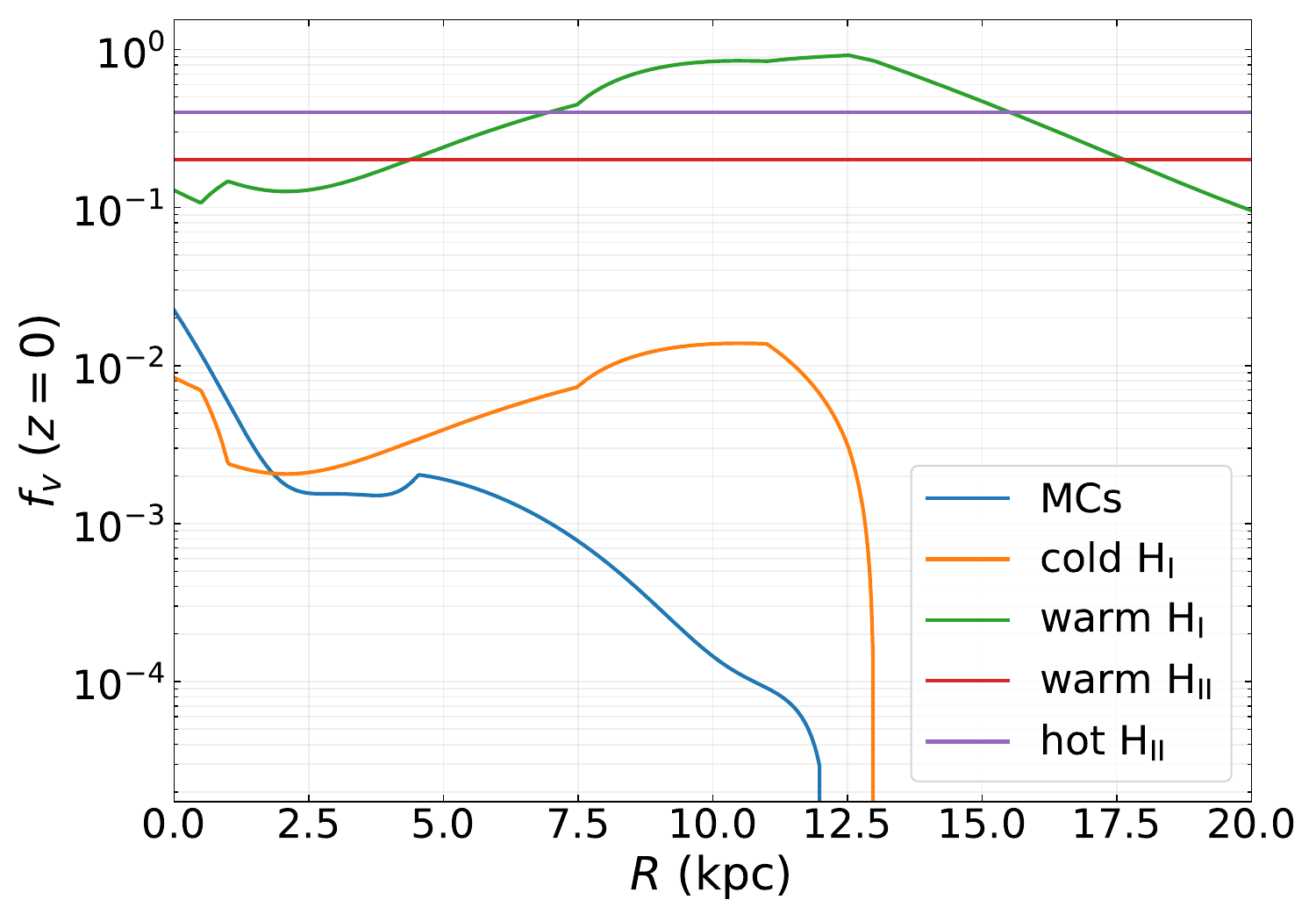}
    \caption{Mid-plane volume filling fractions of the five ISM phases as functions of Galactocentric radius \(R\).} 
    \label{fig:filling fraction}
\end{figure}

Figure~\ref{fig:filling fraction} shows the mid-plane filling fractions of the five ISM phases. MCs have the smallest filling fraction because of their high number densities; they are concentrated in the inner disk and peak near the Galactic center. The cold \(\mathrm{H}_{\text{I}}\) phase is truncated beyond \(R\sim 12 \, \mathrm{kpc}\), whereas the warm \(\mathrm{H}_{\text{I}}\) phase extends farther into the outer disk \citep{Nakanishi2016}. Cold \(\mathrm{H}_{\text{I}}\) also tends to cluster with MCs \citep{Smith2023}. Although MCs and cold \(\mathrm{H}_{\text{I}}\) have low filling fractions, their high densities and low sound speeds make them favorable environments for producing detectable emission from accreting compact objects such as IBHs and NSs \citep{Agol2002, Rodriguez2024}. In some regions of the mid-plane disk, the sum of the adopted filling fractions exceeds unity. This mainly reflects the approximate filling fractions assumed for warm and hot \(\mathrm{H}_{\text{II}}\), together with azimuthal variations in the \(\mathrm{H}_{\text{I}}\) surface density.

\subsection{Single-Star Evolution} \label{subsec:single}

\begin{figure}
    \centering
    \includegraphics[width=\linewidth]{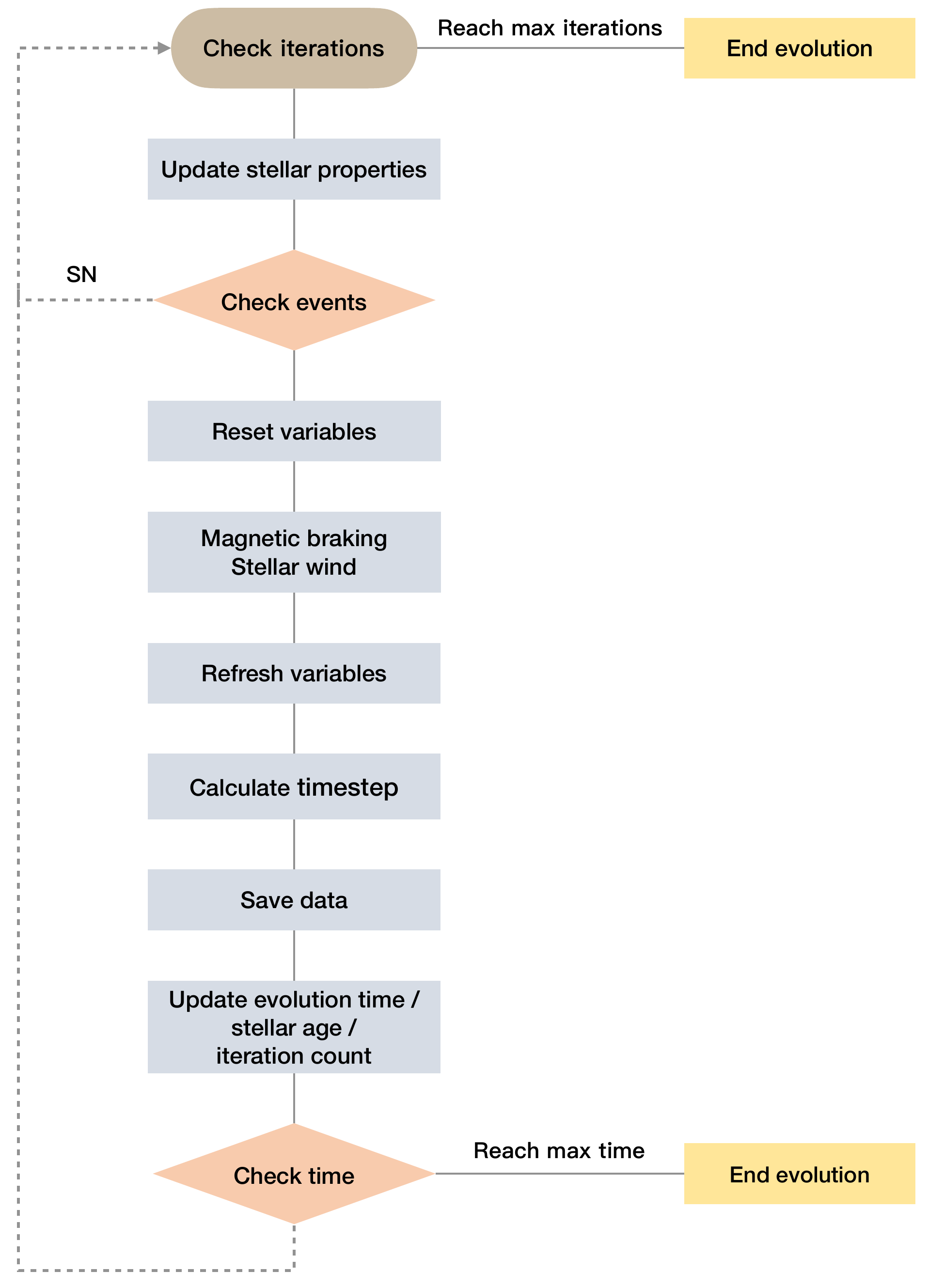}
    \caption{Flowchart of the single-star evolution algorithm implemented in POPKIN.} 
    \label{fig:POPKIN_single_flow}
\end{figure}

Single-star evolution is the foundation of population synthesis. POPKIN follows the evolution of a single star from the ZAMS to its final remnant, either a white dwarf (WD), NS, or BH, including stellar winds, magnetic braking, and SN explosions. The single-star prescriptions follow the fitting formulae of \citet{Hurley2000}. The overall single-star evolutionary algorithm is illustrated in Figure~\ref{fig:POPKIN_single_flow}.

\subsubsection{Stellar Wind} \label{subsec:wind}

Stellar winds are a major source of uncertainty in massive-star evolution. By removing mass and angular momentum before core collapse, they modify the pre-SN mass, the core structure, and consequently the remnant mass and degree of envelope stripping. POPKIN implements three alternative stellar-wind prescriptions, based on \citet{Hurley2000}, \citet{Belczynski2010a}, and \citet{Merritt2026}.

The prescription of \citet{Hurley2000} combines several phase-dependent components. For luminous stars, it adopts the empirical mass-loss calibration of \citet{NJ1990}; for post-main-sequence stars, a Reimers-type rate \citep{KR1978}; and for asymptotic giant branch stars, the prescription of \citet{VW1993}. Hydrogen-stripped Wolf--Rayet (WR) stars are assigned a Hamann-type wind \citep{Hamann1998}. An additional luminous-blue-variable (LBV)-like contribution is applied when the Humphreys--Davidson criterion is satisfied \citep{Humphreys1994}.

The prescription of \citet{Belczynski2010a} provides an updated treatment of the most massive stars. Hydrogen-rich OB stars follow the line-driven wind rates of \citet{Vink2001}, WR winds include an explicit metallicity dependence, and LBV mass loss is described by a calibrated rate that accounts for both steady outflows and possible eruptive shell ejections. For evolutionary phases not covered by these updates, the calculation reverts to the corresponding rates of \citet{Hurley2000}.

The phase-dependent implementation of \citet{Merritt2026} assigns dedicated prescriptions to different stellar classes. OB-star winds follow \citet{Vink2021}, which revises the temperature boundaries and metallicity scalings near the bi-stability jumps introduced by \citet{Vink2001}. Red-supergiant winds are described using the empirical calibration of \citet{Decin2024}, whereas very massive stars follow \citet{Sabhahit2023}. For stripped-helium stars, including WR stars, the adopted mass-loss rate is the larger of the low-luminosity prescription of \citet{Vink2017} and the optically thick WR prescription of \citet{Sander2020}, supplemented by the high-temperature correction of \citet{Sander2023}. Stars outside these classes retain the legacy Hurley-type terms, while the LBV-like contribution is included whenever the Humphreys--Davidson criterion is met.

Unless stated otherwise, POPKIN adopts the prescription of \citet{Merritt2026} as its default stellar-wind treatment because it provides the most explicit phase- and stellar-type-dependent description of mass loss across the massive-star evolutionary sequence.

\subsubsection{Magnetic Braking} \label{subsec:mb}

Magnetic braking is the process by which stars with appreciable convective envelopes lose rotational angular momentum through magnetized stellar winds. \citet{Rappaport1983} proposed an empirical magnetic-braking formula based on observations of cataclysmic variables, which has since been widely adopted in stellar-evolution codes. \citet{Hurley2002} further incorporated the mass of the convective envelope into the magnetic-braking formula. More recently, \citet{Van2019} proposed the Convection And Rotation Boosted magnetic-braking prescription for persistent NS low-mass X-ray binaries.

Magnetic braking has its strongest impact in close binaries. When coupled with tidal interactions that exchange angular momentum between stellar spin and the orbit, magnetic braking usually drives orbital contraction and can strongly affect the subsequent binary evolution. This mechanism is often invoked to explain the observed properties of close binaries such as cataclysmic variables and low-mass X-ray binaries \citep{DengZL2021, FanYN2024}.

\subsubsection{SN Explosion} \label{subsec:sn}

\begin{figure*}
    \centering
    \includegraphics[width=0.8\textwidth]{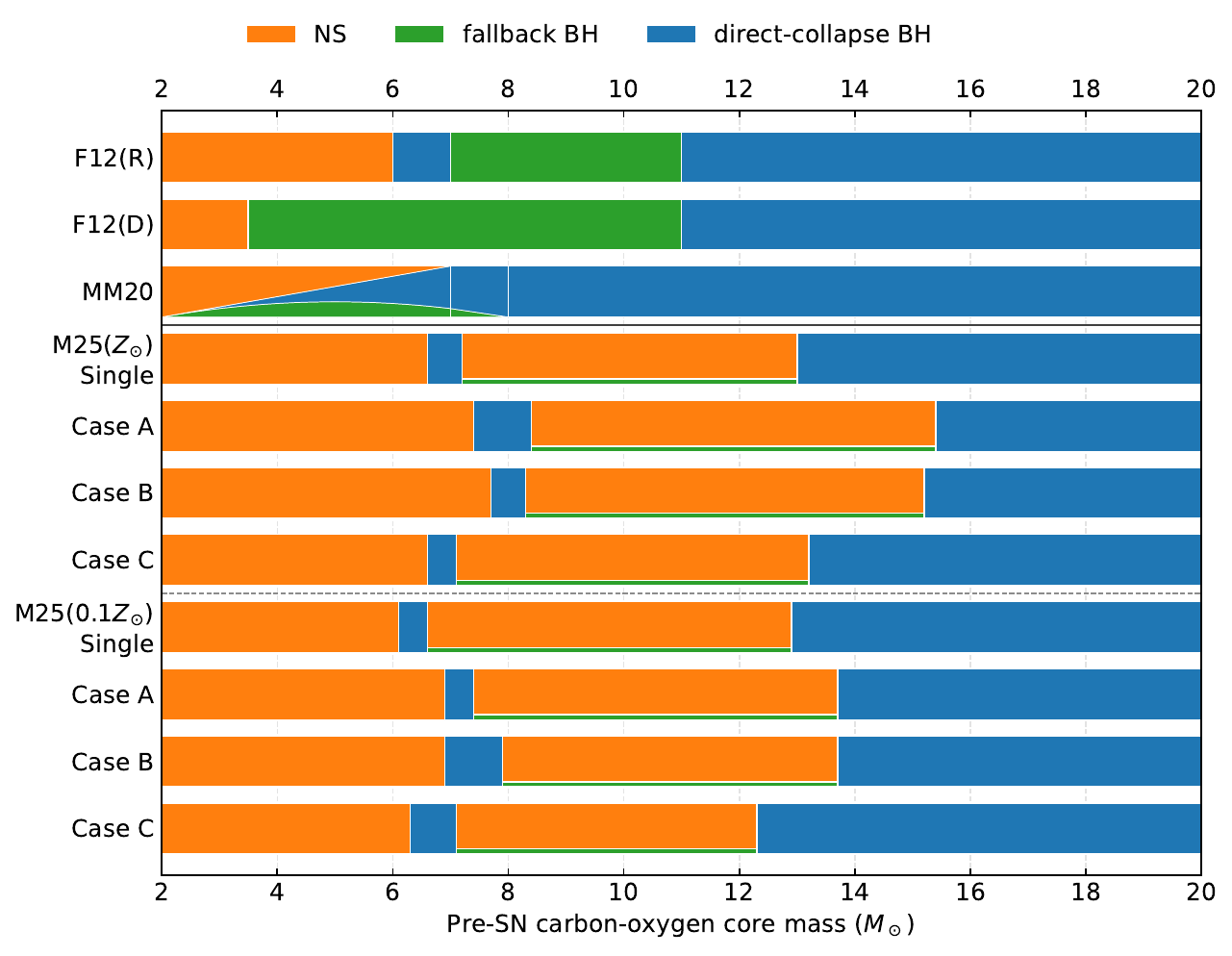}
    \caption{Compact-remnant outcomes as a function of pre-SN carbon-oxygen core mass for the rapid and delayed prescriptions of \citet[][denoted as F12(R) and F12(D), respectively]{Fryer2012}, the stochastic prescription of \citet[][MM20]{Mandel2020}, and the prescription of \citet[][M25]{Maltsev2025}. For M25, outcomes are shown separately for single stars and for binary stars stripped through Case A, B, or C mass transfer, at metallicities \(Z=Z_\odot\) and \(Z=0.1Z_\odot\). Where MM20 and M25 permit more than one outcome, the vertical subdivisions in each mass bin indicate the relative probabilities of the corresponding remnant types.}
    \label{fig:remnant_type_vs_co_core}
\end{figure*}

When a massive star undergoes a core-collapse supernova (CCSN), it may leave behind an NS or BH. Asymmetric mass ejection or anisotropic neutrino emission can impart a natal kick to the remnant, whose mass and velocity depend sensitively on the physics of the explosion.

Early population-synthesis calculations generally relied on deterministic prescriptions for compact remnants. Among the most widely used are the rapid and delayed prescriptions of \citet{Fryer2012}, which determine the remnant mass primarily from the pre-SN carbon-oxygen core mass. The rapid prescription naturally produces a dearth of remnants at \(\sim 2-5\,M_{\odot}\), between the NS and BH populations \citep{Shao2022}, whereas the delayed prescription tends to populate this interval. By contrast, the stochastic treatment introduced by \citet{Mandel2020} assigns the remnant mass and natal kick probabilistically at a given carbon-oxygen core mass rather than mapping each progenitor to a unique outcome.

Recent work indicates that the explodability of stripped stars depends not only on the carbon-oxygen core mass but also on metallicity and the history of binary stripping \citep[e.g.,][]{Schneider2021}. In the calculations of \citet{Maltsev2025}, the collapse outcome is determined by the progenitor metallicity and the evolutionary stage at the onset of mass transfer. This dependence produces distinct regions of parameter space for NS formation, fallback BH formation, and direct-collapse BH formation, allowing binary stripping to reshape the low-mass BH population and the overall remnant-mass spectrum. We assign remnant masses to these three outcomes as follows. A failed SN followed by direct collapse produces a BH with \(M_{\rm BH}=M_{\rm He}\), where \(M_{\rm He}\) is the progenitor He-core mass at core collapse. A successful SN that forms an NS yields \(M_{\rm NS}=1.4\,M_{\odot}\). For a successful SN that forms a BH through fallback, we follow \citet{Willcox2025} and set
\begin{equation}
    M_{\rm BH}=M_{\rm NS}+f_{\rm fb}(M_{\rm He}-M_{\rm NS}),
\end{equation}
where \(f_{\rm fb}\) is the fallback fraction, set to \(0.5\) by default in POPKIN. Following \citet{Mandel2020}, the direct-collapse branch limits the BH mass to the He-core mass. This treatment differs from that of \citet{Fryer2012}, which assumes that the hydrogen envelope also collapses onto the remnant. The fraction of the hydrogen envelope ultimately accreted during direct collapse remains uncertain \citep{Mapelli2020}.

Figure~\ref{fig:remnant_type_vs_co_core} compares the compact-remnant outcomes as a function of the pre-SN carbon–oxygen core mass for the rapid and delayed prescriptions of \citet[][denoted as F12(R) and F12(D), respectively]{Fryer2012}, the stochastic prescription of \citet[][MM20]{Mandel2020}, and the prescription of \citet[][M25]{Maltsev2025}. The F12(D) results are schematic because the exact remnant type and mass also depend on the total pre-SN mass, especially near the NS-BH transition at a carbon-oxygen core mass of about \(3.5\,M_{\odot}\). Overall, F12(D) produces a smoother remnant-mass distribution and populates the $2-5\,M_\odot$ mass-gap region,  whereas direct collapse is disfavored at lower core masses. The M25 prescription yields qualitatively different behavior: remnant classification depends strongly on both metallicity and stripping history; the second direct-collapse regime shifts to larger carbon–oxygen core masses; and a small probability of fallback BH formation emerges in the intermediate-mass regime.

Natal kicks are similarly uncertain. The kick velocities of newly formed NSs are commonly drawn from either a single Maxwellian distribution \citep{Hobbs2005} or a bimodal Maxwellian distribution \citep{Verbunt2017,Igoshev2020}. Based on a reanalysis of the proper motions and distances of young isolated pulsars, \citet{Disberg2025} showed that the data can be described by a single Maxwellian distribution with a one-dimensional velocity dispersion of \(\sigma_{\rm CCSN}=217~{\rm km~s^{-1}}\), lower than the widely used value of \citet{Hobbs2005}. Kick directions are assumed to be isotropic, although anisotropic alternatives, including spin-aligned kicks \citep{Biryukov2025} and polar kicks \citep{Valli2025}, have also been proposed. POPKIN adopts this single-Maxwellian distribution by default and provides several alternative kick models. These include the lognormal distribution proposed by \citet{Disberg2025}, parameterized by \(\mu_{\rm k}=5.60\) and \(\sigma_{\rm k}=0.68\) for \(\ln(v_{\rm k}/{\rm km~s^{-1}})\).

In contrast to NS natal kicks, BH natal kicks are generally expected to be suppressed by fallback \citep{Fryer2012}. For a BH formed through fallback, POPKIN first draws \(v_{\rm k}\) from the same single-Maxwellian distribution adopted for CCSNe and, by default, reduces it by a factor of \((1-f_{\rm fb})\). A direct-collapse BH receives no natal kick. Because the strength of fallback-induced suppression remains uncertain, POPKIN also allows BH kicks to be retained in full or set to zero. No additional reduction is applied when the kick prescription of \citet{Mandel2020} is used, because it specifies the BH kick internally. POPKIN also permits the remnant-mass prescriptions of \citet{Mandel2020} and \citet{Maltsev2025} to be paired with alternative kick treatments.

For NSs formed through electron-capture supernovae (ECSNe) or accretion-induced collapse (AIC), we adopt a fixed remnant mass of \(1.3\,M_{\odot}\) and draw the natal kick from a Maxwellian distribution with \(\sigma_{\rm ECSN/AIC}=30~{\rm km~s^{-1}}\).

Overall, the physics of SNe remains uncertain, and existing prescriptions may not fully capture the connection between remnant mass and natal kick. Binary interactions can further modify the pre-SN stellar structure and, consequently, the properties of the resulting compact objects \citep{Schneider2021,Maltsev2025}. The statistical properties of compact-remnant populations, including the IBHs studied here, therefore provide valuable constraints on the physics of SN explosions.

\subsubsection{Pre-SN and Remnant Mass Relation} \label{subsec:mass_relation}

\begin{figure*}
    \centering
    \includegraphics[width=0.8\linewidth]{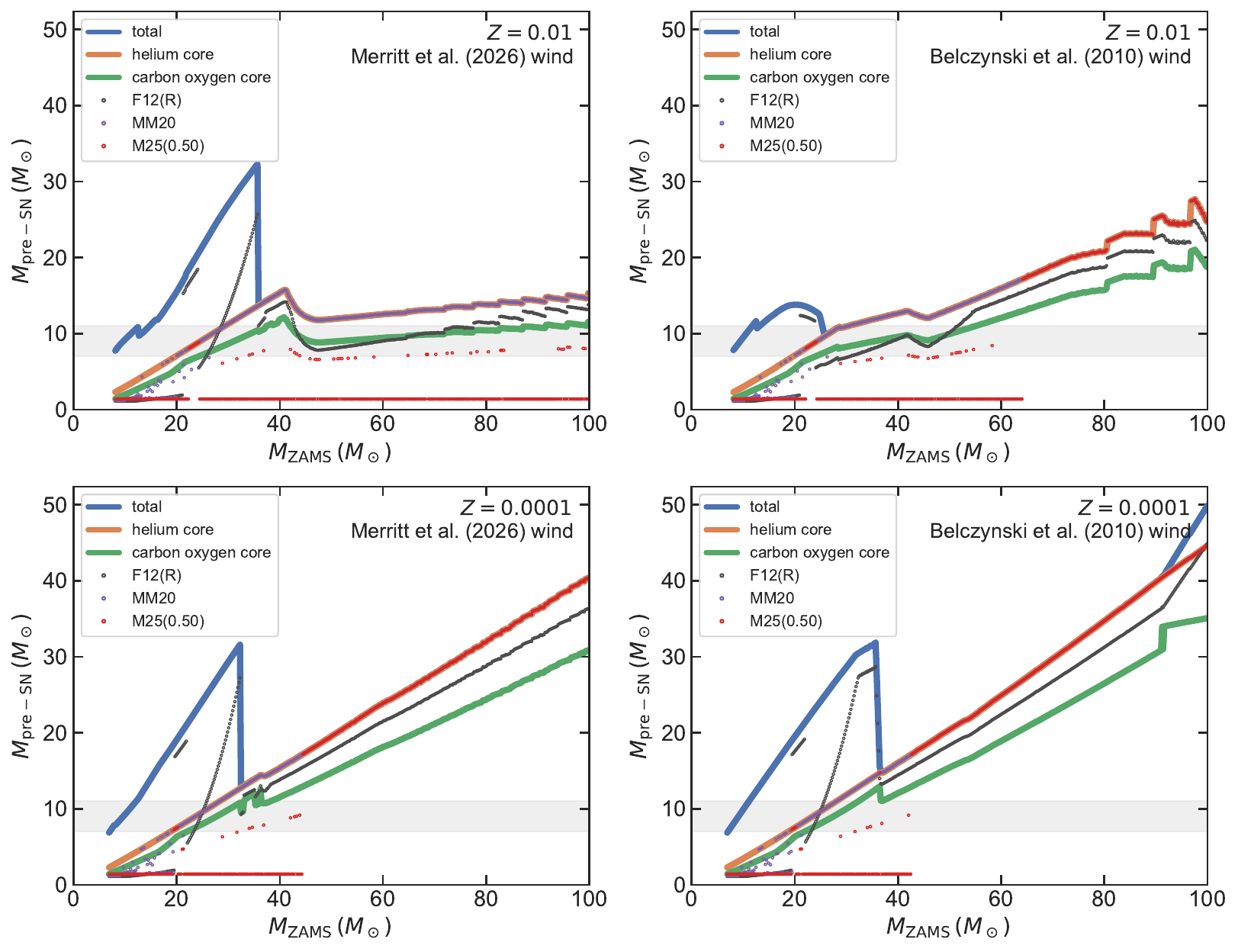}
    \caption{Relations between the ZAMS mass and the pre-SN mass for single-star evolution. The upper and lower panels correspond to $Z=0.01$ and $Z=0.0001$, respectively, while the left and right panels adopt the stellar-wind prescriptions of \citet{Merritt2026} and \citet{Belczynski2010a}, respectively. Solid curves show the total mass immediately before core collapse, the helium-core mass, and the carbon-oxygen-core mass. Open circles denote the compact-remnant masses predicted by the rapid prescription of \citet[][F12(R)]{Fryer2012}, the stochastic prescription of \citet[][MM20]{Mandel2020}, and the prescription of \citet[][M25(0.50)]{Maltsev2025}; for the latter, $f_{\rm fb}=0.5$. The gray shaded band marks the carbon-oxygen-core-mass interval from $7$ to $11~{\rm M_\odot}$, within which fallback BHs may form in the rapid prescription of \citet{Fryer2012}.}
    \label{fig:stellar_mass_relation}
\end{figure*}

For single-star evolution, the mapping from ZAMS mass to compact-remnant mass is governed by the interplay between wind-driven mass loss and the adopted core-collapse prescription. Figure~\ref{fig:stellar_mass_relation} illustrates this relation for initial masses from $0.1$ to $100~{\rm M_\odot}$ by comparing the total mass immediately before core collapse, the helium-core mass, the carbon-oxygen-core mass, and the compact-remnant mass. The left and right panels use the wind prescriptions of \citet{Merritt2026} and \citet{Belczynski2010a}, respectively, while the upper and lower panels correspond to $Z=0.01$ and $Z=0.0001$.

Lower metallicity generally weakens stellar winds, allowing massive stars to retain more mass until core collapse. The contrast between the two wind prescriptions is particularly pronounced at $Z=0.01$. In the calculations using the Merritt wind prescription, the pre-SN mass increases up to $M_{\rm ZAMS}\simeq35~{\rm M_\odot}$ and then drops sharply at $M_{\rm ZAMS}\simeq36~{\rm M_\odot}$. This transition occurs when the star satisfies the Humphreys--Davidson criterion and the additional LBV-like mass-loss term from \citet{Hurley2000} is activated. The resulting removal of the hydrogen-rich envelope reduces the pre-SN mass from approximately $32~{\rm M_\odot}$ to $14~{\rm M_\odot}$, after which it closely follows the helium-core mass. The star then continues to lose mass through WR winds. The additional decline at \(M_{\rm ZAMS}\gtrsim40\,M_{\odot}\) arises because more massive stars
enter the WR phase earlier and therefore experience strong WR
winds for a longer period before core collapse. Their cumulative wind
mass loss consequently outweighs the increase in initial stellar mass.

At the same metallicity, the \citet{Belczynski2010a} prescription produces two declines in the pre-SN mass below $M_{\rm ZAMS}=50~{\rm M_\odot}$. The first, at $M_{\rm ZAMS}\simeq25-26~{\rm M_\odot}$, is driven by the \citet{Hurley2000} wind treatment, whose mass loss is dominated by the \citet{NJ1990} term and removes most of the hydrogen-rich envelope. The second begins near $M_{\rm ZAMS}\simeq42~{\rm M_\odot}$ and results from LBV-driven stripping while the progenitor is still hydrogen-rich; WR winds subsequently dominate the evolution of the stripped helium star.

These wind-driven differences in the pre-SN structure strongly affect the remnant-mass distribution. Under the rapid prescription of \citet{Fryer2012}, BHs with $M_{\rm BH}\gtrsim20~{\rm M_\odot}$ in the $Z=0.01$ calculation with Merritt winds originate mainly from progenitors with $M_{\rm ZAMS}\simeq32-35~{\rm M_\odot}$, whose carbon-oxygen core masses fall in the partial-fallback regime. More massive progenitors undergo more efficient envelope stripping, leading to lower pre-SN masses and, consequently, lighter remnants. By contrast, the \citet{Belczynski2010a} prescription produces more massive carbon-oxygen cores at the high-mass end. Once the carbon-oxygen core mass exceeds the complete-fallback threshold, the rapid prescription yields a direct-collapse BH.

The prescriptions of \citet{Mandel2020} and \citet{Maltsev2025} additionally allow stochastic remnant outcomes. In our implementation of the \citet{Maltsev2025} prescription, systems in the intermediate carbon-oxygen core mass regime are assigned a fallback-BH outcome with a uniform probability of $10\%$. In the $Z=0.01$ calculation with Merritt winds, progenitors with $M_{\rm ZAMS}\simeq30-100~{\rm M_\odot}$ predominantly occupy this regime, so fallback BHs with $f_{\rm fb}=0.5$ can arise across this broad initial-mass interval. This behaviour is qualitatively different from the deterministic rapid prescription. Because these are single-star calculations, the \citet{Maltsev2025} thresholds depend only on metallicity and are not affected by any prior binary mass-stripping history.

Overall, the compact-remnant mass of a massive star is set jointly by its metallicity, wind-driven mass loss, and the adopted SN prescription. In binaries, mass transfer, envelope stripping, common-envelope evolution, and stellar mergers add further pathways that can substantially reshape the compact-remnant mass distribution.

\subsection{Binary-Star Evolution} \label{subsec:binary}

\begin{figure}
    \centering
    \includegraphics[width=\linewidth]{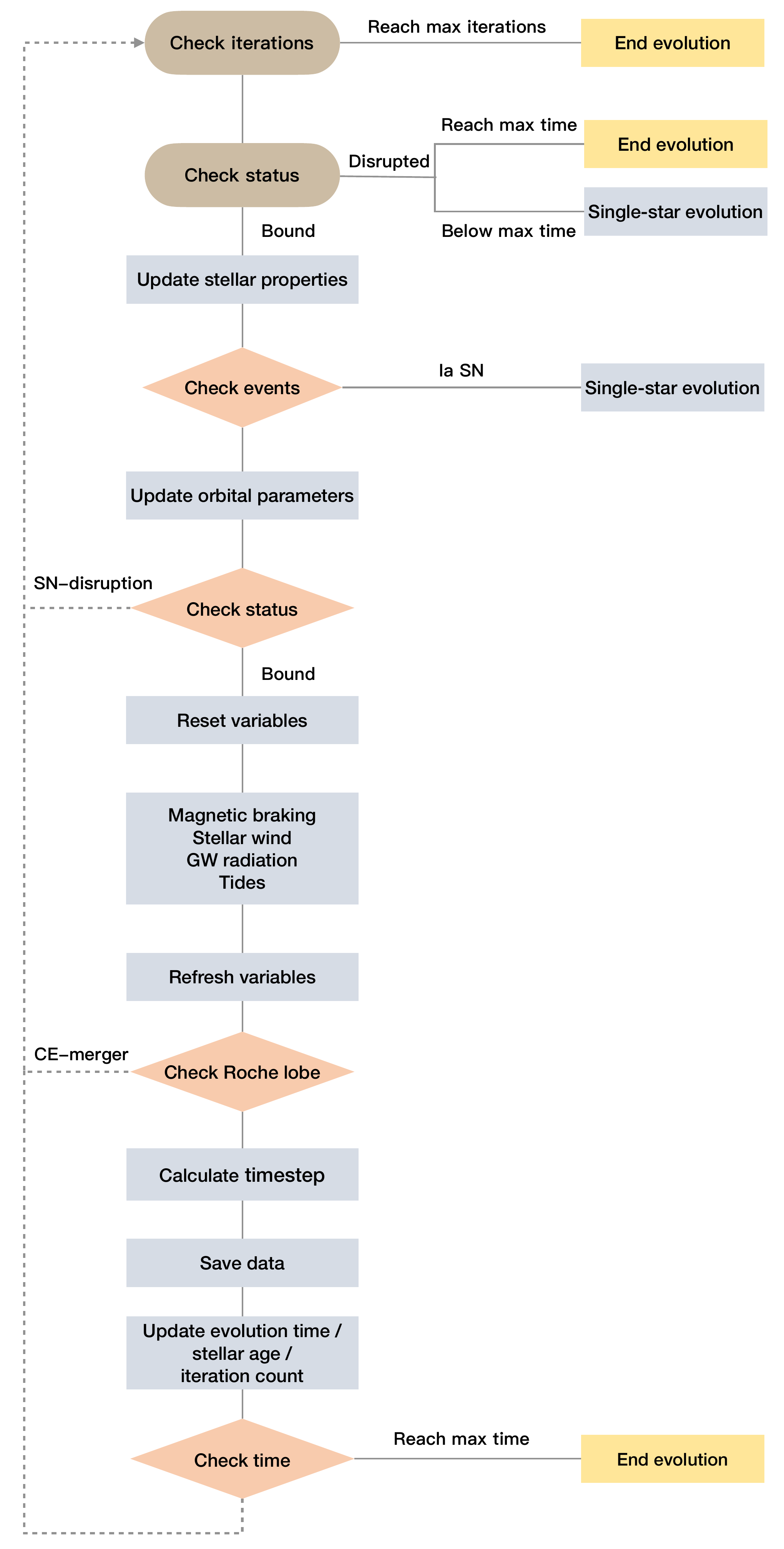}
    \caption{Flowchart of the binary-star evolution algorithm implemented in POPKIN.} 
    \label{fig:POPKIN_binary_flow}
\end{figure}

In binary evolution, interactions between the two components often determine the final outcome of the system and give rise to phenomena such as kilonovae, short gamma-ray bursts, GW sources, and Thorne-{\.Z}ytkow objects. The overall binary-evolution algorithm is illustrated in Figure~\ref{fig:POPKIN_binary_flow}.

\subsubsection{Tidal Effect} \label{subsec:tide}

The treatment of tidal interactions in POPKIN largely follows the BSE framework of \citet{Hurley2002}, including orbital circularization and spin synchronization. For stars with convective envelopes, we follow the equilibrium-tide prescription with convective damping of \citet{Hut1981}, while for compact degenerate stars we retain the weak-damping prescription of \citet{Campbell1984}. For stars with radiative envelopes, we replace the original \citet{Hurley2002} treatment with the dynamical-tide prescription and radiative damping developed by \citet{Zahn1975,Zahn1977}, adopting the revised secular equations of \citet{Sciarini2024}. These equations yield a more reliable tidal torque away from exact spin--orbit synchronization than the simplified synchronization formula commonly used in BSE implementations.

By redistributing angular momentum between stellar rotation and the orbit, tides can alter the orbital separation and eccentricity and thereby influence both the onset and stability of mass transfer. More recent tidal prescriptions have been developed (e.g., \citealt{Kapil2026}), but their implementation in POPKIN is deferred to future work. The present calculation also assumes solid-body rotation. This approximation may be inadequate when the core and envelope are weakly coupled, particularly during giant-branch evolution.

\subsubsection{The Stability of Mass Transfer} \label{subsec:mt}

\begin{table*}
\caption{Mass-transfer stability prescriptions adopted in POPKIN.}
\label{tab:MT_stability}
\centering
\footnotesize
\setlength{\tabcolsep}{5pt}
\begin{tabular}{llll}
\toprule
Donor & Accretor & Adopted criterion & Reference \\
\midrule
WD & WD & \(q_{\rm crit}=0.628\) & \citet{Hurley2002} \\
WD & NS or BH & \(M_{\rm d}\leq 1.25\,M_\odot\) & \citet{He2024} \\
NS or BH & NS or BH & Direct merger & This work \\
\midrule
Hydrogen-rich star & NS or BH &
Stable if \(q<2\); Unstable if \(q>2.1+0.8(M_{\rm acc}/M_\odot)\); &
\citet{Shao2021} \\ 
 & & Otherwise radius-dependent & \\
Helium star & NS or BH &
Stable except \(M_{\rm d}>2.7\,M_\odot\) and \(P_{\rm orb}<0.06\,{\rm days}\) &
\citet{Tauris2015} \\
\midrule
MS or HG star& MS or HG star&
prescription-dependent \(q_{\rm crit}\) &
\citet{Shao2014} \\
HG star & Other accretor&
\(q_{\rm crit}=4\) &
\citet{Hurley2002} \\
(Asymptotic) Giant-branch star & Other accretor&
\(q_{\rm crit}=0.362+\left[3\left(1-M_{\rm d,core}/M_{\rm d}\right)\right]^{-1}\) &
\citet{Hurley2002} \\
Other donor types & Other accretor&
\(q_{\rm crit}=3\) &
This work \\
\bottomrule
\end{tabular}

\tablecomments{
Here \(q\equiv M_{\rm d}/M_{\rm acc}\). Unless otherwise specified, mass
transfer is stable for \(q\leq q_{\rm crit}\).
}
\end{table*}

In a binary system, once a star overfills its Roche lobe, mass transfer can proceed on a nuclear, thermal, or dynamical timescale, depending on the donor's evolutionary state and the binary parameters. The stability of this process is commonly assessed by comparing the mass–radius response of the donor with that of its Roche lobe \citep[e.g.,][]{Soberman1997}. \citet{Hurley2002} provided critical mass ratios under the assumption of conservative mass transfer, although this assumption is not always appropriate. 

For post-main-sequence donors, \citet{Temmink2023} found that the outer layers of giants can thermally readjust during mass loss, permitting stable transfer over a wider parameter space than fully adiabatic estimates suggest. Conversely, in binaries with non-degenerate accretors, spin-up and thermal swelling of the accretor may drive the system into contact even when the donor alone might otherwise be stripped stably \citep{Shao2014,Schurmann2024}. Consequently, POPKIN does not apply a universal stability threshold but instead adopts distinct prescriptions for different donor-accretor combinations (see Table~\ref{tab:MT_stability}).

For mass transfer between two hydrogen-rich, non-degenerate stars, POPKIN adopts the stability regions calculated by \citet{Shao2014} for three mass-transfer prescriptions. In the rotation-dependent prescription, accretion halts once the accretor reaches critical rotation, resulting in highly non-conservative mass transfer. In the half-accretion prescription, the accretor retains half of the transferred mass. In the thermal-equilibrium-limited prescription, the accretion efficiency is determined by the thermal response of the accretor, making mass transfer nearly conservative across much of the parameter space. In the calculations of \citet{Shao2014}, mass transfer was classified as unstable if the accretor filled its Roche lobe or if the transfer rate rose rapidly to approximately \(10^{-3}\,M_{\odot}\,{\rm yr}^{-1}\), with either condition indicating the onset of contact or CE evolution \citep[see also][]{demink2007}. The corresponding stability regions depend on the donor mass, orbital period, and mass ratio. For reference, the rotation-dependent prescription permits stable transfer at donor-to-accretor mass ratios as high as approximately 6 in some regions of parameter space, whereas the half-accretion and thermal-equilibrium-limited prescriptions typically allow stable transfer only up to mass ratios of approximately 2.5 and 2.2, respectively. The choice of mass-transfer prescription affects the mass growth of the accretor and can consequently alter the final compact-remnant mass \citep[see also][]{Xu2025,Wang2026,Xing2026}. At present, POPKIN approximates the post-accretion structure and subsequent evolution of the accretor using single-star models following \citet{Hurley2002}; a self-consistent treatment of the accretor's structural response to mass gain is deferred to future work \citep{Xu2025b,Wang2026}.

For mass transfer from a hydrogen-rich donor to a BH or NS, POPKIN adopts the stability criterion of \citet{Shao2021}, which is based on an extensive MESA grid for BH binaries with donors initially on the ZAMS. In these calculations, accretion onto the compact object is Eddington-limited, and any excess material is ejected with the specific orbital angular momentum of the accretor. A CE is assumed to begin if the photon-trapping radius exceeds the Roche-lobe radius of the accretor \citep{King1999}, or if the mass-transfer rate exceeds $2\%$ of the donor mass per orbital period while the donor expands to its volume-equivalent outer Lagrange lobe \citep{Pavlovskii2015,Ge2020}. In practice, the stability criterion can be summarized as follows. Systems with donor-to-accretor mass ratios below $\sim 2$ are always stable, whereas systems above the mass-dependent upper boundary fitted by \citet{Shao2021} are always unstable. In the intermediate regime, stability is determined by whether the donor radius lies between the lower and upper boundaries defined in that work; donors outside this interval are treated as unstable. The lower boundary corresponds mainly to expansion instability in close systems, while the upper boundary is associated with the development of deep convective envelopes in wider systems \citep[see also][]{Pavlovskii2017}.  We apply the same formalism to NS accretors, for which the maximum stable mass ratio is $\sim 3-3.5$ \citep[see also][]{Tauris2000,Shao2012,Misra2020}. 

For helium-star donors transferring material to a BH or NS, we follow \citet{Tauris2015} and assume that Case BB/BC mass transfer is stable unless the helium-giant mass exceeds \(2.7\,M_{\odot}\) and the orbital period is shorter than \(0.06\) days. For mass transfer between WDs, we adopt a critical mass ratio of \(0.628\) \citep{Hurley2002}. For WDs transferring mass to an NS or BH, the stability condition is more uncertain, with critical WD donor masses in the literature ranging from \(0.2\) to \(1.25\,M_{\odot}\) \citep{He2024}.

Motivated by the weak metallicity dependence of the stable mass-transfer parameter space reported by \citet{Shao2021}, we provisionally neglect any explicit metallicity dependence in all of the above stability criteria and apply them uniformly across the stellar metallicities considered in this work. This assumption may be revisited in future work as improved constraints on the metallicity dependence of mass-transfer stability become available.

\subsubsection{Roche-lobe Overflow} \label{subsec:rlof}

Overall, POPKIN follows the Roche-lobe overflow treatment of \citet{Hurley2002}, while updating the accretion prescriptions for different types of accretors. For stable mass transfer between two non-degenerate stars, we adopt the prescriptions described above. For NS or BH accretors, we assume that half of the transferred material is accreted \citep{ChenWC2020}. In non-conservative transfer, the remaining material is expelled through isotropic re-emission, carrying away the specific orbital angular momentum of the accretor. The accretion rate onto an NS or BH is further limited by the corresponding Eddington rate.

For Roche-lobe overflow in binaries containing an accreting WD, we adopt the stable-burning-band prescription of \citet{WangB2018}. If the mass-accretion rate exceeds the stable-burning limit, we follow the standard treatment of \citet{Hurley2002}, in which the inflated material may form an envelope around the WD and trigger a CE phase. Because the stable-accretion regime is narrow, the classical CE model predicts a low Type Ia SN birth rate. \citet{Kato1994} proposed the optically thick wind model, but its required wind speed is difficult to reconcile with observations and it struggles to explain Type Ia SNe at high redshift and low metallicity. To address these issues, \citet{Cui2022} proposed the CE-wind model, in which the envelope remains small (\(<0.03\,M_{\odot}\)) and excess material is ejected from the CE surface by dynamical instabilities and stellar pulsations. The resulting spiral-in timescale can exceed \(\sim 10^5\) yr, allowing the WD to accrete enough mass to approach the Chandrasekhar limit. Mass loss in the CE-wind model is primarily driven by the self-excited opacity mechanism operating in the hydrogen and helium partial-ionization zones, and is independent of metallicity. POPKIN adopts the CE-wind model by default.

\subsubsection{CE Evolution} \label{subsec:ce}

When mass transfer becomes dynamically unstable, a CE phase may be triggered. In this phase, the companion is engulfed by the expanding envelope of the giant donor and spirals inward under frictional drag. The orbital energy released during the inspiral heats the envelope. If the envelope is successfully ejected, the binary survives as a post-CE system; otherwise, the two stars merge. Because CE evolution is short-lived, typically lasting from years to a few hundred years, systems undergoing this phase are difficult to observe directly \citep[e.g.,][]{Ivanova2013}, and only a limited number of post-CE binaries have been identified \citep{Kruckow2021}.

POPKIN uses the standard energy-balance \(\alpha_{\rm CE}-\lambda_{\rm bind}\) prescription \citep{Webbink1984} to determine whether the binary survives CE evolution. Here, \(\alpha_{\rm CE}\) is the CE ejection efficiency, which specifies the fraction of released orbital energy used to unbind the donor envelope. Although its value remains uncertain because of the complex physics of CE evolution, the default value in POPKIN is \(\alpha_{\rm CE}=1.0\). The parameter \(\lambda_{\rm bind}\) describes the envelope binding energy and depends on the donor mass, radius, and evolutionary state. For hydrogen-rich giants, POPKIN includes the prescriptions of \citet{XuXJ2010} and \citet{WangC2016}, with the former adopted by default. For helium giants, we assume \(\lambda_{\rm bind}=0.5\) for simplicity. 

Recent numerical simulations suggest that, once most of the donor's hydrogen envelope has been stripped and only a small residual envelope remains, the binary may exit the CE phase and transition to stable Roche-lobe overflow. This evolutionary stage is referred to as CE decoupling \citep{Nie2025}. It is not captured by the standard energy-balance formalism, which assumes that CE evolution ends in either complete envelope ejection or a merger.

\subsubsection{Orbital Changes Induced by SN} \label{subsec:sn_orbit}

When one component of a binary undergoes an SN explosion, the binary orbit may be disrupted, producing high-velocity objects such as runaway stars, hypervelocity stars, and isolated pulsars with large proper motions. If the binary remains bound after the explosion, it may instead emerge on an eccentric post-SN orbit, as observed in systems such as high-mass X-ray binaries \citep{WangXY2025}. It is therefore essential to account for SN-induced orbital changes.

For CCSNe and ECSNe, the natal kick imparted to the compact remnant, particularly an NS, can strongly perturb the binary orbit. Sudden mass loss \citep{Blaauw1961} can also disrupt the binary even if the explosion is spherically symmetric; this channel is common in the stochastic SN prescription of \citet{Mandel2020}. A Type Ia SN can likewise disrupt a binary. Compared with stars ejected by natal kicks or core-collapse mass loss, the companion in the Type Ia SN channel can reach a higher runaway velocity because the pre-SN orbit is often extremely compact \citep{Yisikandeer2016}. Following \citet{Hurley2002}, we assume that a WD is destroyed in a Type Ia SN when (i) the mass of a helium WD or carbon-oxygen WD exceeds the Chandrasekhar limit, (ii) a helium WD accretes more than \(0.7\,M_{\odot}\) of helium-rich material, or (iii) a carbon-oxygen WD accretes more than \(0.15\,M_{\odot}\) of helium-rich material.

In POPKIN, we calculate the center-of-mass velocity of a bound post-SN system and the runaway velocities of unbound components following the prescription in  Appendix~\ref{appendix:postSNorbit}. These velocities are first evaluated in the rest frame of the pre-SN center of mass and are then transformed into the Galactocentric frame for subsequent orbit integration.

\subsection{Kinematic Evolution} \label{subsec:kinematic}

POPKIN follows the kinematic evolution of stellar systems by integrating their orbits in a realistic Galactic gravitational potential. Two Python packages, \texttt{galpy} \citep{Bovy2015} and \texttt{gala} \citep{gala}, are widely used for Galactic dynamics calculations, each with different advantages. For orbit integrations with uniform time steps, \texttt{gala}, for example with its default leapfrog integrator, is faster than \texttt{galpy}. Stellar-evolution calculations, however, generally require non-uniform time grids because the time step depends on evolutionary phase, mass-loss rate, binary angular-momentum loss, and other processes. For such grids, \texttt{galpy} gives more stable results, although at higher computational cost. We therefore adopt \texttt{galpy}, with the \texttt{dop853} integrator by default, as the primary orbit-integration tool. For flexibility, POPKIN also retains an implementation based on \texttt{gala}, which requires the \texttt{dop853} integrator with tighter accuracy settings on non-uniform time grids to avoid numerical divergence.

We integrate stellar trajectories from birth until the evolutionary time reaches 12 Gyr, approximately the age of the Milky Way, using the \texttt{MWPotential2014} model implemented in \texttt{galpy}. Users can optionally add the Galactic-center supermassive BH, with mass \(4 \times 10^6 M_{\odot}\) \citep{Gillessen2009}, to \texttt{MWPotential2014}. This option improves orbit integration within the supermassive BH influence radius, typically \(\lesssim 0.1 \,\mathrm{pc}\) \citep{Dexter2014}.

The initial positions of newly born stars are generated according to the adopted Galactic model. Their initial velocities in the Galactocentric cylindrical coordinate system \((v_R, v_T,v_z)\) are assigned as
\begin{equation}
\begin{aligned}
& v_{R}=\mathcal{N}(0, \sigma_{\rm init}) \\
& v_{T}=\mathcal{N}(0, \sigma_{\rm init})+v_{\mathrm {circ}} \\
& v_{z}=\mathcal{N}(0, \sigma_{\rm init})
\end{aligned},
\end{equation}
where \(\mathcal{N}\) denotes a normal distribution with velocity dispersion \(\sigma_{\rm init}\), and \(v_{\mathrm {circ}}\) is the circular velocity given by \texttt{MWPotential2014}. Based on the observed velocity dispersion of OB stars \citep{Ramirez-Tannus2021,Bobylev2022}, we adopt \(\sigma_{\rm init}=10 \mathrm{~km} \mathrm{~s}^{-1}\). We also use \texttt{MWPotential2014} to calculate the Galactic escape velocity, which is approximately \(500 \mathrm{~km} \mathrm{~s}^{-1}\) at the Solar position. POPKIN implements two orbit-integration schemes, described below.

\paragraph{Piecewise orbital integration}
Low-mass stars experience negligible kinematic perturbations and therefore require only a single continuous orbit integration. For massive stars, however, SN explosions can change the kinematic state one or more times. After an SN, the binary may be disrupted by the natal kick imparted to the remnant, or it may remain bound but acquire a new systemic velocity \citep{Hurley2002}. In disrupted systems, both components can receive substantial runaway velocities \citep{Tauris1998}. Therefore, whenever the kinematic state of a star changes, POPKIN starts a new orbit-integration segment. This piecewise approach handles a broad range of evolutionary channels and records the orbital properties of each system throughout its evolution. The velocity transformation between the pre-SN frame and the Galactocentric cylindrical frame is described in Appendix~\ref{appendix:frame}.

\paragraph{Multiple orbital integrations}
The radial scale length of the Galactic thin disk evolves with time (Figure~\ref{fig:radial CDF}). Consequently, the initial positions of primordial systems depend on their formation epoch. To reduce computational cost while retaining the metallicity-enrichment history (Figure~\ref{fig:Metallicity}), POPKIN groups formation times into 1 Gyr intervals and performs a full set of orbit integrations for each interval.

\subsection{Population Synthesis} \label{subsec:ps}

\begin{figure*}
    \centering
    \includegraphics[width=\linewidth]{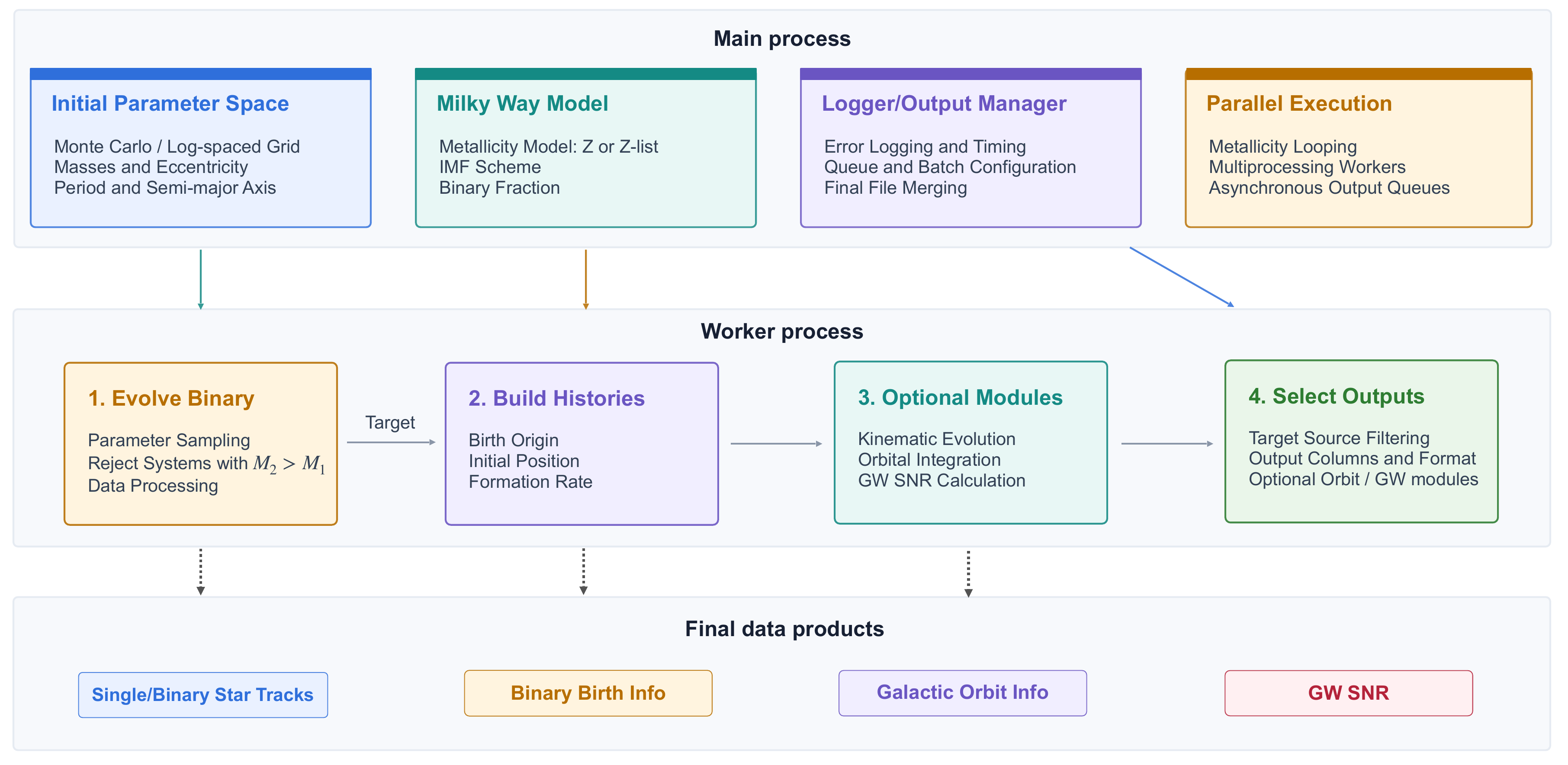}
    \caption{Binary population-synthesis workflow in POPKIN. The workflow connects the primordial binary parameter space to the Milky Way model, stellar and binary evolution, optional kinematic and observable post-processing modules, and user-selected synthetic catalogs.} 
    \label{fig:POPKIN_BPS_workflow}
\end{figure*}

\subsubsection{Initial Stellar Multiplicity} \label{subsec:initialSM}

Stars form as single, binary, and higher-order multiple systems. Observational surveys show that stellar multiplicity depends strongly on primary mass, with binary and multiple systems especially common among massive stars \citep{Sana2012, DucheneKraus2013,Offner2023,Chen2024}. In the current version of POPKIN, all non-single systems are treated as binaries; triples and higher-order multiples will be incorporated in future developments \citep[see e.g.,][]{Li2026}.

By default, POPKIN adopts a mass-dependent binary fraction following \citet{Haaften2013},
\begin{equation}
\mathcal{B}(M_1)=\frac{1}{2}+\frac{1}{4} \log _{10}(M_1) \quad\left(0.08 \leq M_1 / M_{\odot} \leq 100\right).
\end{equation}
Users may alternatively specify a constant binary fraction in the range \(0 \leq \mathcal{B} \leq 1\).

For a stellar population containing both single and binary systems, the mean initial mass of stellar systems is
\begin{equation}
M_* = \int_{M_{\rm min}}^{M_{\rm max}} \xi(M_1) \left[\mathcal{B}(M_1) \cdot \frac{3}{2} M_1 + (1-\mathcal{B}(M_1)) M_1 \right] dM_1
\end{equation}
where \(\xi(M_1)\) is the IMF \citep{Kroupa1993,Kroupa2002}, and \((M_{\rm min}, M_{\rm max})\) defines its mass range. The term \(3M_1/2\) is the mean mass of a binary with primary mass \(M_1\), assuming a uniform mass-ratio distribution on \([0,1]\). Adopting the default IMF of \citet{Kroupa2002} and the binary fraction of \citet{Haaften2013}, we obtain a mean system mass of \(M_*\simeq0.76\,M_{\odot}\).

The formation rate of a sampled stellar system \(i\) is then scaled as
\begin{equation}
\mathcal{R}_i = \frac{\mathrm{SFR}(\tau)}{M_*} W_i,
\end{equation}
where \(\mathrm{SFR}(\tau)\) is the star formation rate at lookback time \(\tau\), corresponding to the evolutionary age of system \(i\) from the ZAMS to the present day. The factor \(W_i\) denotes the dimensionless weight of the progenitor system in the sampled initial-parameter space.

\subsubsection{Initial Parameter Space} \label{subsec:initial}

For binary population synthesis, POPKIN adopts a three-dimensional initial parameter space consisting of the primary mass \(M_1\), the secondary mass \(M_2\), and one orbital parameter. The distribution of \(M_1\) is determined by the adopted IMF \citep{Kroupa1993,Kroupa2002,Weisz2015}. By default, the orbital parameter is the orbital period \(P\). Following the period distribution inferred for massive binaries \citep{Sana2012} and commonly adopted in compact-object population synthesis \citep[e.g.,][]{deMink2015}, we use
\begin{equation}
\psi_P(x) = C_P x^{\alpha_P},
x \equiv \log_{10}\left(P/{\rm day}\right),
\alpha_P=-0.55,
\end{equation}
over the range \(x \in [0.15,5.5]\), where \(C_P=0.2605\) is the normalization constant.

To improve the sampling of low-mass systems, initial masses are sampled on logarithmic grids. The default number of grid points for each initial parameter \(\chi\) is \(N_\chi=50\). For a primordial binary, the corresponding statistical weight is
\begin{equation}
\begin{aligned}
W_{\rm b} =
&\ \,\mathcal{B}(M_1)\,\Phi(\ln M_1)\,\Delta \ln M_1 \\
&\times
\varphi(\ln M_2)\,\Delta \ln M_2\,
\psi_P(x)\,\Delta x ,
\end{aligned}
\end{equation}
where
\begin{equation}
\Phi(\ln M_1)=M_1\xi(M_1),
\quad
\varphi(\ln M_2)=\frac{M_2}{M_1}.
\end{equation}
Here, the quantities \(\Delta \ln M_1\), \(\Delta \ln M_2\), and \(\Delta x\) are the grid intervals in the corresponding logarithmic variables.

Users may instead choose the orbital separation \(a\) as the initial orbital parameter. In this case, POPKIN adopts a log-flat separation distribution over
\(a \in [3,10^4]\,R_{\odot}\), following the commonly used prescription in rapid binary population synthesis \citep{Hurley2002}. The binary weight then becomes
\begin{equation}
\begin{aligned}
W_{\rm b} =
&\ \mathcal{B}(M_1)\,\Phi(\ln M_1)\,\Delta \ln M_1 \\
&\times
\varphi(\ln M_2)\,\Delta \ln M_2\,
\psi_a(\ln a)\,\Delta \ln a ,
\end{aligned}
\end{equation}
where
\begin{equation}
\psi_a(\ln a)=C_a,
\quad
C_a=\left[\ln\left(\frac{10^4}{3}\right)\right]^{-1}=0.12328 .
\end{equation}
The default model assumes initially circular binaries. Users may instead adopt a uniform eccentricity distribution or a thermal distribution, \(p(e)=2e\), over \(0<e<1\).

For single-star population synthesis, the initial parameter space reduces to the stellar mass \(M\). The statistical weight of a primordial single star is therefore
\begin{equation}
W_{\rm s}
=
\Phi(\ln M)\,\Delta \ln M\,[1-\mathcal{B}(M)] .
\end{equation}

\subsubsection{Metallicity Models} \label{subsec:metallicity}

In addition to initial masses and orbital parameters, progenitor metallicity is a key input for stellar and binary evolution. POPKIN currently supports two metallicity prescriptions for population synthesis: a constant-metallicity model and a Galactic chemical-enrichment model. The latter is coupled to the Milky Way SFH described in Section~\ref{subsec:sfh}.

In the constant-metallicity model, all stellar systems form with the same metallicity, independent of birth time and birth location. This prescription is useful for controlled experiments, for isolating the effects of metallicity on stellar evolution, and for modeling populations formed in a relatively narrow environment, such as the Solar neighborhood or a young stellar population. In contrast, the enrichment model is designed for synthetic surveys of the Galactic stellar population, in which systems formed at different lookback times and Galactocentric radii may have different metallicities. When this model is adopted, POPKIN evolves the stellar population over a user-specified list of metallicities.

For each system sampled from the initial parameter space, POPKIN assigns a physically allowed birth region for its progenitor. This requires specifying both an allowed birth time, expressed as lookback time, and an allowed spatial region, expressed here by Galactocentric radius. In the constant-metallicity model, the birth region is unconstrained by metallicity: systems can form over the full adopted lookback-time range, \(0-12~{\rm Gyr}\), and over the full radial range, \(0-20~{\rm kpc}\). In the enrichment model, the allowed birth region depends on metallicity, as illustrated in Figure~\ref{fig:Metallicity}. For example, systems with \(Z=0.02\) can form only within an effective lookback-time range of approximately \(0-9~{\rm Gyr}\), and the allowed Galactocentric radius range varies with lookback time.

For numerical implementation, POPKIN divides the Galactic history into 12 time bins of width \(1~{\rm Gyr}\). For each time bin, POPKIN evaluates the SFRs of the thin disk, thick disk, and bulge across the Galactocentric radius range allowed by the adopted metallicity. These component-wise SFRs are then used to assign the birth component, birth position, and statistical contribution of each sampled stellar system. The same birth information is used as the initial condition for Galactic orbit integration when kinematic evolution is enabled.

\subsubsection{Population-Synthesis Workflow} \label{subsec:workflow}

The binary population-synthesis workflow in POPKIN is summarized in Figure~\ref{fig:POPKIN_BPS_workflow}. For a given metallicity prescription, POPKIN first constructs the primordial binary population from the prescribed distributions of primary mass, secondary mass, and orbital parameters. Each primordial binary is assigned a statistical weight according to the adopted IMF, binary fraction, and initial distributions of mass ratio and orbital parameters.

Stellar and binary evolution form the core of the workflow. Each sampled binary is evolved from the ZAMS, yielding the evolutionary histories of both stellar components and the binary orbital properties. These evolutionary histories are coupled to the Milky Way model, which uses the adopted SFH and metallicity-dependent birth regions to assign each sampled system a Galactic component and birth position. Each evolutionary snapshot therefore carries the stellar and binary properties of the system, its Galactic birth-environment information, and its statistical contribution to the synthetic population.

The same evolutionary histories can then be passed to optional modules. When kinematic evolution is enabled, POPKIN integrates the orbits of the systems in the adopted Galactic potential. Bound binaries are followed as single systems, whereas binaries disrupted by an SN are followed as two independent components. The code then appends the resulting positions and velocities, together with observer-frame quantities such as distance, Galactic longitude, Galactic latitude, proper motion, and radial velocity. When GW calculations are enabled, POPKIN estimates the SNRs of compact binaries for the adopted detector configuration. The final synthetic catalogs are produced by applying user-defined source-selection criteria and retaining the requested output columns. Together, these steps couple stellar and binary evolution to the Galactic birth environment, follow the resulting orbital evolution in the Galactic potential, and map the synthetic systems to observable quantities within a unified population-synthesis workflow.

\subsection{Post-processing and Connections to Observables}
\label{subsec:observables}

In addition to stellar and binary evolution, POPKIN provides a post-processing layer that maps synthetic populations onto observable quantities. This layer operates on catalogs produced by the evolution and population-synthesis calculations and is designed to be extensible to different scientific applications. At present, it includes calculations of compact-binary GW SNRs, IBH accretion from the ISM, and microlensing by compact lenses. Additional capabilities for other electromagnetic observables and survey-selection effects are planned for future versions.

\subsubsection{GW SNRs}
\label{subsec:gw_snr}

In the coming decades, space-based GW detectors such as LISA \citep{LISA2023} and TianQin \citep{TianQin2025} are expected to reveal a large population of Galactic GW sources. Such detections will provide important constraints on compact-binary formation channels and the physical processes governing binary evolution.

POPKIN includes a post-processing capability for estimating the SNRs of potential Galactic GW sources using the Python package \texttt{LEGWORK} \citep{LEGWORK2022}\footnote{We used \texttt{LEGWORK} v0.5.2 with \texttt{NumPy}$<$2.4; users should ensure that the two packages are mutually compatible.}. In the current implementation, we adopt LISA as the detector, assume a mission duration of 4 yr, and include the \citet{Robson2019} model for Galactic foreground confusion noise. These detector and noise-model settings can be modified by the user.

When Galactic orbit integration is disabled, the SNR is computed using the birth positions assigned by the Galactic model. When orbit integration is enabled, POPKIN also computes the SNR using the present-day positions of the sources and outputs both estimates.

\subsubsection{Accreting IBHs in the ISM}
\label{subsec:bh_ism}

IBHs may produce detectable electromagnetic emission by accreting material from the ISM \citep{Agol2002}. A natural first estimate of the gas-capture rate is provided by the Bondi--Hoyle prescription \citep{Bondi1944}. However, this rate does not necessarily correspond to the gas reaching the vicinity of the BH event horizon. In the low-accretion regime relevant for most IBHs, the captured gas is expected to form an advection-dominated accretion flow (ADAF), in which strong outflows can substantially reduce the accretion rate reaching the BH \citep{Yuan2012}. In addition, \citet{Xie2012} showed that the radiative efficiency can decrease to \(\eta \sim 0.1\%-1\%\) at low accretion rates, well below the standard thin-disk value of \(\eta_{\rm std}\simeq 0.1\). These effects imply that accretion-powered emission from IBHs is likely to be very faint, as illustrated by recent constraints on systems such as Gaia BH2 \citep{Rodriguez2024}.

POPKIN therefore provides a post-processing capability for estimating the accretion properties of IBHs moving through different ISM phases. It computes the Bondi--Hoyle gas-capture rate, corrects for mass loss in radiatively inefficient flows, and estimates the resulting bolometric luminosity and flux. These outputs can be used both to calculate cumulative IBH counts above user-defined accretion-rate or flux thresholds and to select candidate systems whose predicted bolometric flux exceeds a specified detection limit.

\paragraph{Accretion rate}

We first estimate the gas-capture rate of an IBH moving through the surrounding ISM using the Bondi--Hoyle prescription \citep{Bondi1944},
\begin{equation}
\label{eq:BondiHoyle}
\dot{M}_{\rm Bondi} =
\frac{4 \pi G^2 M_{\rm BH}^2 \mu m_p n_{\rm H}}
{\left(v_{\rm rel}^2+c_s^2\right)^{3/2}},
\end{equation}
where \(M_{\rm BH}\) is the BH mass, \(n_{\rm H}\) is the hydrogen number density of the ISM, \(\mu\) is the mean molecular weight, \(m_p\) is the proton mass, 
\(v_{\rm rel}\) is the relative velocity between the BH and the ISM, and \(c_s\) is the effective sound speed of the gas. The BH masses and kinematic properties are obtained from the population-synthesis calculation and Galactic orbit integration. In evaluating \(v_{\rm rel}\), we assume that the ISM follows the Galactic rotation curve. Equation~(\ref{eq:BondiHoyle}) can be written in convenient units as

\begin{equation}
\label{eq:BondiHoyle_simple}
\begin{aligned}
\dot{M}_{\rm Bondi} =
&\ 3.7 \times 10^{14}\,{\rm g\,s^{-1}}\left(\frac{M_{\rm BH}}{M_{\odot}}\right)^2 \\
&\times
 \mu \left(\frac{n_{\rm H}}{{\rm cm}^{-3}}\right)
\left(\frac{\sqrt{v_{\rm rel}^2+c_s^2}}{{\rm km\,s^{-1}}}\right)^{-3}.
\end{aligned}
\end{equation}

The Bondi--Hoyle rate describes the gas supply at large radii, but it does not necessarily equal the accretion rate reaching the BH. The captured gas may circularize and form an accretion disk. Following \citet{Agol2002}, the characteristic disk radius is estimated as

\begin{equation}
\label{eq:r_disk}
\frac{r_{\rm disk}}{r_s}
= 10^4
\left(\frac{M_{\rm BH}}{9 M_{\odot}}\right)^{2/3}
\left(\frac{\sqrt{v_{\rm rel}^2+c_s^2}}{40~{\rm km\,s^{-1}}}\right)^{-10/3},
\end{equation}
where \(r_s\) is the Schwarzschild radius.

The accretion mode depends on both the accretion rate and the radial scale of the accretion flow. For accretion rates below a critical value, \(\dot{M}_{\rm crit}\), both a standard thin-disk solution and an ADAF solution may exist \citep{Narayan1995}. A standard thin disk is optically thick, geometrically thin, and radiatively efficient, with dissipated energy radiated locally. In contrast, an ADAF is optically thin and radiatively inefficient, with a large fraction of the dissipated energy advected inward with the gas. This class of low-luminosity accretion solutions is commonly described as a hot accretion flow or, more generally, as a radiatively inefficient accretion flow (RIAF). Thin-disk solutions in this regime may become unstable and transition to the advection-dominated state. Therefore, in our default treatment, the accretion flow below \(\dot{M}_{\rm crit}\) is treated as a hot accretion flow. We adopt a viscosity parameter \(\alpha_{\rm v}=0.3\) and a ratio of gas pressure to total pressure \(\beta_{\rm g}=0.5\), for which the critical accretion rate is approximated by
\begin{equation}
\label{eq:M_crit}
\dot{M}_{\rm crit}
= 0.15 \dot{M}_{\rm Edd}
\left(\frac{r}{1000 r_s}\right)^{\alpha_1},
\end{equation}
where \(\alpha_1=0.33\) for \(10r_s<r<1000r_s\) and \(\alpha_1=-0.47\) for \(r>1000r_s\). The Eddington accretion rate is
\begin{equation}
\dot{M}_{\rm Edd}
= 1.4 \times 10^{18}\,{\rm g\,s^{-1}}
\left(\frac{M_{\rm BH}}{M_{\odot}}\right)
\left(\frac{0.1}{\eta_{\rm std}}\right),
\end{equation}
where \(\eta_{\rm std}\simeq 0.1\) is the characteristic radiative efficiency of a standard thin disk.

Hot accretion flows are expected to lose mass through outflows, causing the accretion rate to decrease inward \citep{Yuan2003, Yuan2012}. Observational evidence for this behavior has been found in systems such as \(\mathrm{Sgr~A}^{\ast}\) \citep{Yuan2003} and NGC 3115 \citep{Wong2011}. Following \citet{Yuan2012}, we describe the radial profile of the accretion rate as \(\dot{M}(r)\propto r^s\), with \(s=0\) for \(r<10r_s\) and \(s\simeq 0.5\) for \(10r_s<r<10^4r_s\). After accounting for mass loss in the hot accretion flow, the accretion rate reaching the inner accretion flow near the BH is estimated as
\begin{equation}
\label{eq:Mdot_BH}
\dot{M}_{\rm BH}
= \dot{M}_{\rm Bondi}
\left(\frac{10r_s}{r_{\rm out}}\right)^{0.5},
\end{equation}
where \(r_{\rm out}\) is the outer radius of the hot accretion flow. If the Bondi rate exceeds \(0.15\dot{M}_{\rm Edd}\), we treat the flow as a standard thin disk and do not apply the outflow correction. Otherwise, \(r_{\rm out}\) is determined from the circularization radius \(r_{\rm disk}\) and the critical radius implied by Equation~(\ref{eq:M_crit}).

For each IBH in the synthetic catalogue, POPKIN evaluates the accretion rate separately for different ISM phases. For phases with a single characteristic density, such as warm \(\mathrm{H}_{\text{I}}\), warm \(\mathrm{H}_{\text{II}}\), and hot \(\mathrm{H}_{\text{II}}\), the statistical contribution of each BH is weighted by the local volume filling fraction \(f_v(R,z)\). The cumulative number of BHs above a given accretion-rate threshold (\(\dot{M}_{\rm th}\)) is therefore written schematically as
\begin{equation}
\begin{aligned}
N_X(>\dot{M}_{\rm th}) =
&\ \sum_j N_{{\rm BH},j}\,f_{v,X}(R_j,z_j)\, 
\\
&\times
 \Theta\!\left(\dot{M}_{{\rm BH},j,X}-\dot{M}_{\rm th}\right),
\end{aligned}
\end{equation}
where \(X\) denotes the ISM phase, \(N_{{\rm BH},j}\) is the population weight of the \(j\)-th BH, and \(\Theta\) is the Heaviside step function.

For MCs and cold \(\mathrm{H}_{\text{I}}\), the gas density spans a broad range and is therefore discretized into logarithmic density bins. Observations show that Galactic density fluctuations approximately follow \(\delta\rho/\rho \sim (L/10^{18}\,{\rm cm})^{1/3}\) \citep{Armstrong1995}. For a \(10M_\odot\) BH moving relative to the ISM at \(v_{\rm rel}=50~{\rm km\,s^{-1}}\), the Bondi radius is only \(\sim 5\times10^{13}\,{\rm cm}\), much smaller than the scale on which large density fluctuations become important. We therefore treat the gas density as locally constant over the accretion region. The contribution from each density bin is weighted by its logarithmic density probability, giving
\begin{equation}
\begin{aligned}
N_X(>\dot{M}_{\rm th}) =
&\ \sum_{j,i} N_{{\rm BH},j}\, f_{v,X}(R_j,z_j)\, w_X(n_i)\,
\\
&\times
\Theta\!\left(\dot{M}_{{\rm BH},j,X}(n_i)-\dot{M}_{\rm th}\right),
\end{aligned}
\end{equation}
where \(w_X(n_i)\) is the weight of the \(i\)-th logarithmic density bin for phase \(X\). In practice, we use ten logarithmic density bins for MCs and cold \(\mathrm{H}_{\text{I}}\). The different ISM phases are treated statistically through their local filling fractions, and overlap between phases is neglected.

\paragraph{Radiative efficiency}

The accretion-rate model above estimates the rate reaching the inner accretion flow near the BH, \(\dot{M}_{\rm BH}\), after accounting for mass loss in hot accretion flows. To convert this accretion rate into luminosity, we specify the radiative efficiency \(\eta\). For a standard thin disk, a typical value is \(\eta_{\rm std}\simeq 0.1\). In hot accretion flows, however, the radiative efficiency can be much lower because a significant fraction of the dissipated energy is advected inward rather than radiated locally \citep{Narayan1995, Yuan2014}.

In POPKIN, we follow \citet{Xie2012} and parameterize the radiative efficiency as a function of the dimensionless accretion rate,
\begin{equation}
f_{\dot{M}} = \frac{\dot{M}_{\rm BH}}{\dot{M}_{\rm Edd}} .
\end{equation}
The adopted relation is
\begin{equation}
\label{eq:eta}
\eta(f_{\dot{M}}) = 
\begin{cases}
0.035 \left( \dfrac{f_{\dot{M}}}{f_1} \right)^{0.65}, & f_{\dot{M}} < f_1, \\
0.035 \left( \dfrac{f_{\dot{M}}}{f_1} \right)^{0.076}, & f_1 \leq f_{\dot{M}} < f_2, \\
0.05 \left( \dfrac{f_{\dot{M}}}{f_2} \right)^{1.12}, & f_2 \leq f_{\dot{M}} < f_3, \\
0.085, & f_{\dot{M}} \geq f_3,
\end{cases}
\end{equation}
where \(f_1 = 2.9 \times 10^{-5}\), \(f_2 = 3.3 \times 10^{-3}\), and \(f_3 = 5.3 \times 10^{-3}\).

The bolometric luminosity is then given by
\begin{equation}
L_{\rm bol} = \eta \dot{M}_{\rm BH} c^2 ,
\end{equation}
and the corresponding bolometric flux is
\begin{equation}
\label{eq:F_bol}
F_{\rm bol}
= \frac{L_{\rm bol}}{4\pi D^2}
= \frac{\eta \dot{M}_{\rm BH} c^2}{4\pi D^2},
\end{equation}
where \(D\) is the distance from the accreting BH to the Sun. POPKIN outputs the cumulative number of IBHs above user-defined bolometric-flux thresholds and can also generate a candidate table for systems whose predicted \(F_{\rm bol}\) exceeds a specified threshold.

\subsubsection{Microlensing of Compact Objects}
\label{subsec:microlensing}

Microlensing offers a powerful way to detect faint or dark compact objects, including WDs, NSs, and BHs, through the transient magnification and astrometric shifts they induce in background source stars. Because the signal depends on the lens mass, lens-source geometry, and relative motion, microlensing can also constrain the physical properties of compact lenses when photometric and astrometric data are available. In this subsection, we summarize the basic microlensing quantities, discuss how observations can constrain compact-object masses, and describe the simplified microlensing post-processing calculation implemented in POPKIN.

\paragraph{Microlensing observables}
For a lens of mass \(M_{\rm L}\) at distance \(D_{\rm L}\) and a background source at distance \(D_{\rm S}\), the angular Einstein radius is
\begin{equation}
\theta_{\rm E}
=
\left[
\frac{4GM_{\rm L}}{c^2}
\left(
\frac{1}{D_{\rm L}}
-
\frac{1}{D_{\rm S}}
\right)
\right]^{1/2}.
\end{equation}
This quantity sets the characteristic angular scale of the event. Introducing the lens-source relative parallax,
\begin{equation}
\pi_{\rm rel}
=
{\rm au}
\left(
\frac{1}{D_{\rm L}}
-
\frac{1}{D_{\rm S}}
\right),
\end{equation}
the angular Einstein radius can also be written as
\begin{equation}
\theta_{\rm E}
=
\left(
\kappa M_{\rm L}\pi_{\rm rel}
\right)^{1/2},
\end{equation}
where
\begin{equation}
\kappa
=
\frac{4G}{c^2{\rm au}}
\simeq
8.14~{\rm mas}~M_\odot^{-1}.
\end{equation}
If \(D_{\rm L}\) and \(D_{\rm S}\) are expressed in kpc, \(\pi_{\rm rel}\) is expressed in mas as
\begin{equation}
\frac{\pi_{\rm rel}}{{\rm mas}}
=
\frac{1}{D_{\rm L}/{\rm kpc}}
-
\frac{1}{D_{\rm S}/{\rm kpc}} .
\end{equation}

The Einstein crossing time, or event timescale, is
\begin{equation}
t_{\rm E}
=
\frac{\theta_{\rm E}}{\mu_{\rm rel}},
\end{equation}
where \(\mu_{\rm rel}\) is the relative proper motion between the lens and the source. For a point lens with uniform rectilinear motion, and neglecting higher-order effects such as annual microlensing parallax, the lens-source angular separation in units of the Einstein radius is
\begin{equation}
u(t)
=
\left[
u_0^2+
\left(
\frac{t-t_0}{t_{\rm E}}
\right)^2
\right]^{1/2},
\end{equation}
where \(u_0\) is the minimum lens-source angular separation in units of the Einstein radius, and \(t_0\) is the time of maximum magnification. The corresponding photometric magnification is
\begin{equation}
A(u)
=
\frac{u^2+2}{u\sqrt{u^2+4}} .
\end{equation}
Microlensing can also produce an astrometric shift of the image centroid,
\begin{equation}
\delta_{\rm c}(u)
=
\frac{u}{u^2+2}\theta_{\rm E},
\end{equation}
which provides a direct way to constrain \(\theta_{\rm E}\) when sufficiently precise astrometric data are available.

\paragraph{Lens-mass constraints}
Photometric microlensing alone mainly constrains \(t_{\rm E}\), which is degenerate with the lens mass, geometry, and relative proper motion. A robust mass measurement therefore requires two complementary observables beyond \(t_{\rm E}\): the angular Einstein radius \(\theta_{\rm E}\) and the microlensing parallax \(\pi_{\rm E}\). In practice, \(\theta_{\rm E}\) can be obtained from astrometric microlensing or finite-source effects, while \(\pi_{\rm E}\) is inferred from annual-parallax distortions of the light curve and is defined as
\begin{equation}
\pi_{\rm E}
\equiv
\frac{\pi_{\rm rel}}{\theta_{\rm E}} .
\end{equation}
When both \(\theta_{\rm E}\) and \(\pi_{\rm E}\) are available, the lens mass follows from
\begin{equation}
M_{\rm L}
=
\frac{\theta_{\rm E}}{\kappa\pi_{\rm E}} .
\end{equation}
Long-duration events are especially valuable for compact-object lenses, because they are more likely to show measurable parallax distortions and allow time-resolved astrometric follow-up. However, a long \(t_{\rm E}\) alone is not a unique signature of a dark compact lens, because similar timescales can also arise from a small \(\mu_{\rm rel}\) or a different lens-source geometry.

\paragraph{Implementation in POPKIN}
POPKIN provides a simplified post-processing capability for estimating the microlensing observables of compact-object lenses. It does not impose a fixed sky selection. Instead, users first select the lenses relevant to their survey or science case, and POPKIN then computes the microlensing quantities for the selected catalogue. This design allows different sky regions, foreground-lens distance cuts, and survey footprints to be adopted without modifying the core calculation. For each selected lens, POPKIN computes the relative proper motion \(\mu_{\rm rel}\), angular Einstein radius \(\theta_{\rm E}\), event timescale \(t_{\rm E}\), and expected number of microlensing events.

In the current implementation, source stars are placed at a default bulge distance of \(D_{\rm S}=8~{\rm kpc}\). Their transverse velocities are drawn from a Gaussian distribution with a one-dimensional velocity dispersion of \(80~{\rm km~s^{-1}}\). Following \citet{Wiktorowicz2019}, the observer's transverse velocity is set to \(v_{{\rm obs},y}=230~{\rm km~s^{-1}}\) and \(v_{{\rm obs},z}=15.5~{\rm km~s^{-1}}\). For a lens with transverse velocity \(\boldsymbol{v}_{\rm L}\), source velocity \(\boldsymbol{v}_{\rm S}\), and distance \(D_{\rm L}\), the relative proper motion is
\begin{equation}
\boldsymbol{\mu}_{\rm rel}
=
\frac{1}{4.74}
\left[
\frac{\boldsymbol{v}_{\rm L}-\boldsymbol{v}_{\rm obs}}{D_{\rm L}}
-
\frac{\boldsymbol{v}_{\rm S}-\boldsymbol{v}_{\rm obs}}{D_{\rm S}}
\right],
\end{equation}
where velocities are in \({\rm km~s^{-1}}\), distances are in kpc, and \(\boldsymbol{\mu}_{\rm rel}\) is in \({\rm mas~yr^{-1}}\). The scalar relative proper motion is \(\mu_{\rm rel}=|\boldsymbol{\mu}_{\rm rel}|\). POPKIN then evaluates \(\theta_{\rm E}\) and \(t_{\rm E}\) using the formulae above.

Using a simple geometric estimate \citep[e.g.][]{Wiktorowicz2019}, the expected number of events contributed by the \(i\)-th lens during an observing time \(T_{\rm obs}\) is
\begin{equation}
\label{eq:num_lens}
N_{{\rm ML},i}
=
w_i
\left(
2\theta_{{\rm E},i}\mu_{{\rm rel},i}T_{\rm obs}
+
\pi\theta_{{\rm E},i}^2
\right)
\frac{N_{\star}}{\Omega},
\end{equation}
where \(w_i\) is the population-synthesis weight, \(N_{\star}\) is the number of monitored source stars, and \(\Omega\) is the survey area. The first term represents the area swept out by the Einstein radius during the observing window, and the second term accounts for the instantaneous microlensing cross-section. By default, we adopt \(N_{\star}=1.5\times10^8\), \(\Omega=31~{\rm deg^2}\), and \(T_{\rm obs}=1~{\rm yr}\) for a bulge-field application \citep{Wyrzykowski2015}.

The calculation returns the columns \(\mu_{\rm rel}\), \(\theta_{\rm E}\), \(t_{\rm E}\), and \(N_{\rm ML}\). For comparison with observed BH microlensing samples, POPKIN also includes an empirical BH-lens fraction as a function of \(t_{\rm E}\), digitized from the Mock EWS simulation of \citet{Lam2020}. This relation is used only as an approximate OGLE-EWS-like reference and should not be interpreted as a universal selection function.

\subsection{Running POPKIN}
\label{subsec:use}

\begin{figure*}
    \centering
    \includegraphics[width=\linewidth]{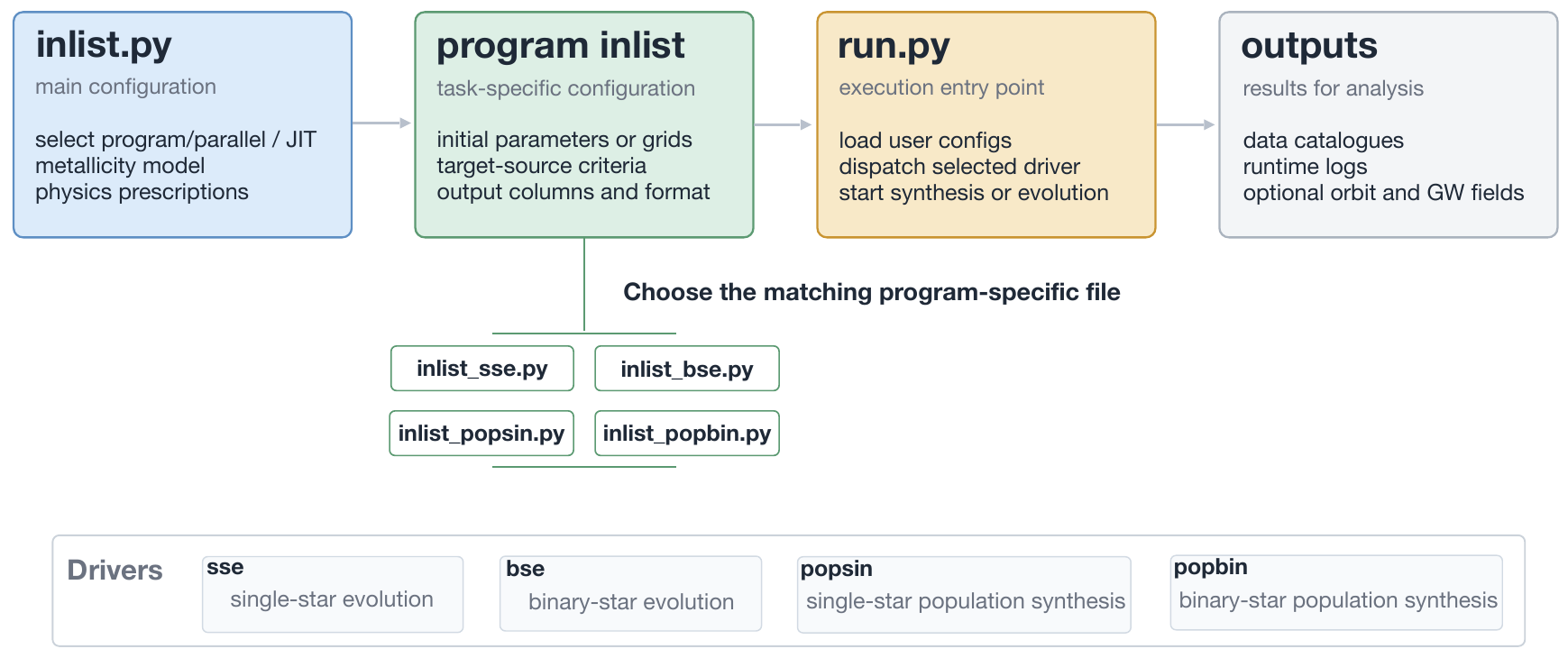}
    \caption{Workflow for running POPKIN. Global options are specified in \texttt{inlist.py}, task-specific options are supplied by the relevant program configuration file, and \texttt{run.py} loads the configuration, executes the selected driver, and writes evolutionary tracks or population-synthesis catalogs with runtime logs.} 
    \label{fig:POPKIN_how_to_use}
\end{figure*}

\subsubsection{Configuration and Execution}
\label{subsec:config}

POPKIN is a configuration-driven code. A typical run is carried out in an independent working directory that contains a main configuration file, \texttt{inlist.py}, and optional program-specific configuration files. The main configuration file selects the calculation mode, such as single-star evolution, binary-star evolution, single-star population synthesis, or binary population synthesis. It also controls global runtime options, including the metallicity prescription, multiprocessing, JIT acceleration, Galactic orbit integration, and GW calculations, as well as the physical prescriptions for stellar and binary evolution, such as stellar winds, mass transfer, SN explosions, natal kicks, and CE evolution. Program-specific files, such as \texttt{inlist\_sse.py}, \texttt{inlist\_bse.py}, \texttt{inlist\_popsin.py}, and \texttt{inlist\_popbin.py}, define the initial parameters, target-source criteria, selected output columns, and output format for each calculation mode. The overall configuration and execution workflow is illustrated in Figure~\ref{fig:POPKIN_how_to_use}.

We recommend installing and running POPKIN in an isolated \texttt{conda} environment to avoid conflicts among dependencies such as \texttt{NumPy}, \texttt{SciPy}, \texttt{Numba}, \texttt{Astropy}, \texttt{galpy}, and \texttt{pandas}. After the environment is created and the required packages are installed, a run is launched by executing \texttt{run.py} in the working directory. This script loads the user configuration, initializes the output and logging directories, and dispatches the selected driver. Runtime information, warnings, errors, and timing diagnostics are written to log files, so that the execution history and configuration of each run can be inspected.

\subsubsection{Evolving Individual Stellar Systems}
\label{subsec:sse}

As in other rapid population-synthesis codes, single-star and binary-star evolution can be run separately for individual systems. A single-star calculation is specified by the initial stellar mass and metallicity and follows the star from the ZAMS to its final evolutionary state. A binary calculation additionally requires the component masses and initial orbital parameters and follows both stars together with their orbital evolution. Both calculation modes use the same physical prescriptions described in Sections~\ref{subsec:single} and~\ref{subsec:binary} and subsequently adopted in the population-synthesis calculations.

Each calculation produces a time-resolved evolutionary track. The single-star output includes the stellar mass, radius, core mass, evolutionary type, remnant properties, and relevant evolutionary events. The binary output additionally records the properties of both components and the orbital evolution, including episodes of mass transfer, CE evolution, SNe, binary disruption, and mergers. Individual tracks are primarily used to inspect particular evolutionary pathways and to verify the adopted physical assumptions before applying them to a population.



\subsubsection{Population Synthesis and Outputs}
\label{subsec:pop}

For statistical studies, POPKIN provides single-star and binary population-synthesis modes. In these modes, the code constructs the initial parameter space, assigns statistical weights to sampled systems, evolves them under the selected metallicity prescription, and connects the resulting evolutionary histories to the Milky Way model. Depending on the configuration, the calculation can also include Galactic orbit integration and GW SNR estimates. Source-selection criteria are specified in the population-synthesis configuration files, allowing targeted populations such as compact remnants, X-ray binaries, compact-binary GW sources, or IBHs to be extracted.

The final data products are written either to the output directory of the working environment or to a user-defined data directory. POPKIN supports multiple output formats, including \texttt{csv}, \texttt{npy}, \texttt{hdf5}, and \texttt{parquet}, and users can select the output columns required for a given scientific application. For large calculations, simplified output can be enabled to retain only key evolutionary transitions, important events, and systems satisfying the target-source criteria, thereby reducing file size. The resulting catalogs can be further analyzed with the available post-processing utilities, such as IBH accretion calculations and the selection of potential GW sources from precomputed SNRs. Additional observable and survey-selection tools will be added in future versions.

\section{Method} \label{sec:method}

We model the Galactic IBH population with POPKIN, which couples single- and binary-star evolution to Galactic birth distributions and orbital integration, as described in Section~\ref{sec:popkin}. Here we summarize the physical assumptions and initial parameter grids adopted for the calculations presented in this work.

\subsection{Physical Assumptions} \label{subsec:physic}

\begin{table*}
\caption{Summary of the ten population-synthesis models considered in this work.}
\label{tab:model_summary}
\centering
\footnotesize
\setlength{\tabcolsep}{5pt}
\begin{tabular}{lllll}
\hline\hline
Model & SN explosion  & Mass transfer &
CE evolution \\
\hline
M25(0.50)/Fiducial & \citet{Maltsev2025}, $f_{\rm fb}=0.50$ &  Rotation-dependent \citep{Shao2014}& $\alpha_{\rm CE}=1.0$ \\
M25(0.25) & \citet{Maltsev2025}, $f_{\rm fb}=0.25$  & -- & -- \\
M25(0.75) & \citet{Maltsev2025}, $f_{\rm fb}=0.75$ & -- & -- \\
F12(R)    & Rapid recipe of \citet{Fryer2012}    & -- & -- \\
F12(D)    & Delayed recipe of \citet{Fryer2012}   & -- & -- \\
MM20      & \citet{Mandel2020}  & -- & -- \\
MT2       & --  & Half-accretion \citep{Shao2014} & -- \\
MT3       & --  & Thermal-equilibrium-limited \citep{Shao2014} & -- \\
CE03      & --  & -- & $\alpha_{\rm CE}=0.3$ \\
CE30      & --  & -- & $\alpha_{\rm CE}=3.0$ \\
\hline\hline
\end{tabular}

\tablecomments{$f_{\rm fb}$ is the fallback fraction and $\alpha_{\rm CE}$ is the CE ejection efficiency. Mass-transfer prescription refers to transfer between non-degenerate stars. A dash indicates that the corresponding parameter or prescription is inherited from the fiducial model.}
\end{table*}

We consider ten population-synthesis models to quantify the effects of uncertain SN prescriptions, fallback fractions, mass-transfer efficiencies, and CE evolution. Unless stated otherwise, all models adopt the default POPKIN prescriptions for stellar winds, tidal interactions, and other stellar-physics processes. The fiducial model is M25(0.50), and each alternative model differs from the fiducial model in only one physical assumption. The principal differences among the models are described below and summarized in Table~\ref{tab:model_summary}.

\textit{M25(0.50)}: This is the fiducial model used throughout this work. Compact-remnant masses are assigned with the prescription of \citet{Maltsev2025}, which depends on metallicity and the history of mass stripping. For BHs formed through fallback, we adopt a fixed fallback fraction of $f_{\rm fb}=0.5$. Natal kicks are drawn from a Maxwellian distribution with one-dimensional dispersion $\sigma_{\rm CCSN}=217~{\rm km~s^{-1}}$ \citep{Disberg2025} and reduced according to the fallback fraction. For mass transfer between non-degenerate stars, we use the rotation-dependent mass-transfer prescription of \citet{Shao2014}, under which accretion ceases as the accretor approaches critical rotation. During CE evolution, we set the ejection efficiency to $\alpha_{\rm CE}=1$ and adopt the binding-energy parameter $\lambda$ from \citet{WangC2016}.

\textit{M25(0.25)}: This model differs from the fiducial model only in the fallback fraction for BHs formed through fallback, for which we adopt $f_{\rm fb}=0.25$. Relative to M25(0.50), these BHs are less massive and receive larger natal kicks.

\textit{M25(0.75)}: This model differs from the fiducial model only in the fallback fraction for BHs formed through fallback, for which we adopt $f_{\rm fb}=0.75$. Relative to M25(0.50), these BHs are more massive and receive smaller natal kicks.

\textit{F12(R)}: This model replaces the \citet{Maltsev2025} prescription with the rapid SN prescription of \citet{Fryer2012}. The rapid prescription predicts a possible mass gap between NSs and BHs. Its fallback fraction depends on the pre-SN carbon-oxygen core mass, and BHs formed with nearly complete fallback receive negligible natal kicks.

\textit{F12(D)}: This model adopts the delayed SN prescription of \citet{Fryer2012}. Unlike the rapid prescription, the delayed prescription fills in the $\sim2-5\,M_{\odot}$ mass gap and allows BHs to form within this range.

\textit{MM20}: This model adopts the stochastic SN prescription of \citet{Mandel2020}, in which BH masses and natal kicks are sampled probabilistically rather than assigned by a deterministic remnant-mass relation. In this framework, a BH can also become isolated through binary disruption caused by mass loss alone \citep{Blaauw1961}, even when no natal kick is imparted.

\textit{MT2}: This model adopts a fixed mass-transfer efficiency of $50\%$ for mass transfer between non-degenerate stars, so that the accretor retains one half of the material lost by the donor \citep{Shao2014}.

\textit{MT3}: This model adopts the thermal-timescale-limited mass-transfer prescription of \citet{Shao2014}. The mass-transfer efficiency is set by the ratio of the mass-transfer timescale to the accretor's thermal timescale and can approach unity, allowing nearly conservative mass transfer.

\textit{CE03}: This model adopts a CE ejection efficiency of $\alpha_{\rm CE}=0.3$, motivated by constraints from Galactic double-WD binaries \citep{Scherbak2023}.

\textit{CE30}: This model adopts a CE ejection efficiency of $\alpha_{\rm CE}=3.0$, a value that can reproduce some post-CE systems, such as IK Peg \citep{Davis2010}.

\subsection{Initial Parameters} \label{subsec:InitialParameter}

In our work, we adopt the metallicity-enrichment model described above, sampled at 20 discrete metallicity values spanning \(Z=0.0002\)--\(0.03\). For each metallicity, we evolve \(10^3\) single stars and \(1.25\times10^5\) primordial binaries. The binary grid is defined by the primary mass \(M_1\), secondary mass \(M_2\), and  orbital period \(P\). To focus on potential BH progenitors, we restrict the primary mass to $5-100\,M_{\odot}$. The secondary-mass grid spans $0.1\,M_{\odot}<M_2<M_1$ and is weighted assuming a uniform mass-ratio distribution \citep{Hurley2002}. We adopt the massive-binary period distribution described in Section~\ref{subsec:initial}. Each of the three initial binary parameters is sampled on a logarithmic grid with $N_{\chi}=50$ points, yielding $50^3=1.25\times10^5$ binaries per metallicity.

The single-star grid contains $10^3$ evolutionary tracks per metallicity, with initial masses in the range $5-100\,M_{\odot}$. Both grids are then weighted according to the adopted IMF, binary fraction, and Galactic star-formation and metallicity-evolution histories described in Section~\ref{sec:popkin}.

\section{IBHs in the Milky Way} \label{sec:IBH}

\begin{table*}
\centering
\caption{Predicted properties of the Galactic IBH population for the ten population-synthesis models considered in this work. For each model, the table lists the total number of IBHs, their mean mass, the fractions formed in each Galactic birth component and formation channel, the fractions receiving a natal kick, and the fractions that remain gravitationally bound to the Galaxy. The M25 models are labeled by their fallback fraction in parentheses; F12(R) and F12(D) denote the rapid and delayed prescriptions of \citet{Fryer2012}, respectively, while MM20 denotes the stochastic prescription of \citet{Mandel2020}. MT2 and MT3 vary the mass-transfer prescription, and CE03 and CE30 vary the CE ejection efficiency.}
\label{tab:IBH}

\begin{tabular}{c|c|ccc|cc|c|cc|cc}
\hline

\multicolumn{2}{c|}{\multirow{3}{*}{Model}} &
\multicolumn{6}{c|}{SN explosion} &
\multicolumn{2}{c|}{Mass transfer} &
\multicolumn{2}{c}{CE evolution} \\

\cline{3-12}

\multicolumn{2}{c|}{} &
\multicolumn{3}{c|}{M25} &
\multicolumn{2}{c|}{F12} &
\multirow{2}{*}{MM20} &
\multirow{2}{*}{MT2} &
\multirow{2}{*}{MT3} &
\multirow{2}{*}{CE03} &
\multirow{2}{*}{CE30} \\

\cline{3-7}

\multicolumn{2}{c|}{} &
($0.25$) &
($0.50$) &
($0.75$) &
(R) &
(D) &
& & & & \\

\hline



\multirow{2}{*}{Population}

& Total number ($\times 10^8$)
& $0.382$ & $0.381$ & $0.378$
& $1.362$ & $2.098$ & $1.752$
& $0.431$ & $0.480$ & $0.379$ & $0.380$ \\

\cline{2-12}

& Mean mass ($M_{\odot}$)
& $10.3$ & $11.1$ & $12.0$
& $12.4$ & $9.4$ & $8.5$
& $10.9$ & $12.1$ & $11.2$ & $11.1$ \\

\hline

\multirow{3}{*}{Origin}
& Thick disk
& 48.2\% & 48.4\% & 48.1\%
& 43.1\% & 42.8\% & 43.1\%
& 47.6\% & 48.8\% & 48.4\% & 48.5\% \\

\cline{2-12}

& Thin disk
& 38.3\% & 38.2\% & 38.4\%
& 41.4\% & 42.0\% & 41.9\%
& 38.8\% & 38.1\% & 38.3\% & 38.2\% \\

\cline{2-12}

& Bulge
& 13.5\% & 13.4\% & 13.5\%
& 15.4\% & 15.2\% & 15.0\%
& 13.6\% & 13.1\% & 13.3\% & 13.3\% \\

\hline

\multirow{5}{*}{Channel}
& SN-induced disruption
& 48.1\% & 47.5\% & 47.3\%
& 42.3\% & 47.5\% & 34.4\%
& 43.4\% & 32.0\% & 47.5\% & 48.7\% \\

\cline{2-12}

& Two MS merger
& 27.6\% & 27.7\% & 27.9\%
& 28.8\% & 22.9\% & 27.3\%
& 27.6\% & 26.9\% & 27.8\% & 27.8\% \\

\cline{2-12}

& CE merger
& 11.1\% & 11.6\% & 11.4\%
& 17.2\% & 17.2\% & 21.9\%
& 17.2\% & 31.2\% & 12.5\% & 8.9\% \\

\cline{2-12}

& Two CO merger
& 2.0\% & 2.1\% & 2.1\%
& 0.8\% & 0.7\% & 1.2\%
& 2.0\% & 1.0\% & 0.9\% & 3.4\% \\

\cline{2-12}

& Single-star evolution
& 11.2\% & 11.2\% & 11.3\%
& 11.0\% & 11.7\% & 15.2\%
& 9.9\% & 8.9\% & 11.3\% & 11.2\% \\

\hline

\multirow{2}{*}{Kick}
& Yes
& 30.1\% & 30.0\% & 29.7\%
& 61.6\% & 88.6\% & 31.6\%
& 30.3\% & 27.1\% & 29.8\% & 30.5\% \\

\cline{2-12}

& No
& 69.9\% & 70.0\% & 70.3\%
& 38.4\% & 11.4\% & 68.4\%
& 69.7\% & 72.9\% & 70.3\% & 69.5\% \\

\hline

\multirow{2}{*}{Escape}
& Bound
& 98.3\% & 99.8\% & 99.8\%
& 99.5\% & 98.1\% & 100.0\%
& 99.6\% & 99.6\% & 99.6\% & 99.6\% \\

\cline{2-12}

& Unbound
& 1.7\% & 0.3\% & 0.2\%
& 0.5\% & 1.9\% & 0.0\%
& 0.4\% & 0.5\% & 0.4\% & 0.4\% \\

\hline

\end{tabular}
\end{table*}

Table~\ref{tab:IBH} summarizes the Galactic IBH populations predicted by the ten models described in Section~\ref{subsec:physic}. For each model, it reports the total number and mean mass of IBHs, the fractions formed in each Galactic birth component and formation channel, the fractions with and without natal kicks, and the fractions that remain gravitationally bound to or escape from the Galaxy. Unless otherwise noted, we adopt M25(0.50) as the fiducial model throughout this section; the effects of varying the physical assumptions are discussed in Section~\ref{sec:discussion}.

\subsection{Origins and Formation Channels}

In the fiducial model, the present-day Milky Way contains approximately $3.8\times10^7$ IBHs, with a mean mass of about $11~{\rm M_\odot}$. This abundance is consistent with the recent population-synthesis estimate of \citet{wagg2026}.

Approximately $6\%$ of these IBHs lie outside the main Galactic disk, which we define as $R\leq20~{\rm kpc}$ and $|z|\leq5~{\rm kpc}$. Within $1~{\rm kpc}$ of the Sun, we predict roughly $8\times10^4$ IBHs, making the Solar neighbourhood an important region for observational searches. The thick disk contributes nearly half of the total population. This large contribution arises from its relatively metal-poor stellar population: massive progenitors retain more massive pre-SN carbon-oxygen cores and can form direct-collapse BHs under the \citet{Maltsev2025} prescription. In the more metal-rich thin disk, stronger winds produce less massive carbon-oxygen cores. For massive progenitors, these cores predominantly occupy the intermediate regime, in which our implementation assigns a $10\%$ probability of fallback-BH formation (Figure~\ref{fig:stellar_mass_relation}).

We classify IBHs into five formation channels:

$\textit{(1) SN-induced disruption}$: BHs become isolated when an SN disrupts the binary, either during BH formation or during a subsequent SN of the companion.

$\textit{(2) Two MS merger}$: BHs are produced by mergers of two main-sequence stars triggered by dynamically unstable mass transfer.

$\textit{(3) CE merger}$: BHs form through mergers during CE evolution, including possible Thorne-{\.Z}ytkow-object-like configurations. Following \citet{Hurley2002}, we assume that the compact object can rapidly eject the envelope.

$\textit{(4) Two CO merger}$: BHs form through mergers of two compact objects.

$\textit{(5) Single-star evolution}$: BHs form from single stars.

In the fiducial model, SN-induced disruption is the most common formation channel, accounting for approximately $48\%$ of all IBHs, whereas the two CO merger channel is the least common, contributing only about $2\%$. Single-star evolution accounts for roughly $10\%$ of our IBH population, substantially below the $\sim45\%$ reported by \citet{Olejak2020}. This difference arises primarily from the adopted initial binary fraction; when we instead assume a constant binary fraction of $50\%$, our result becomes consistent with theirs.

More than two-thirds of IBHs form without natal kicks. As shown below, subsequent evolution in the Galactic potential produces distinct present-day kinematic signatures for the kicked and non-kicked populations. Only a small fraction of IBHs become unbound from the Galaxy. Nevertheless, many bound systems reach large Galactocentric radii or heights above the Galactic plane and return only on long orbital timescales.

\subsection{Mass and Spatial Distributions}

\begin{figure}
	\includegraphics[width=\columnwidth]{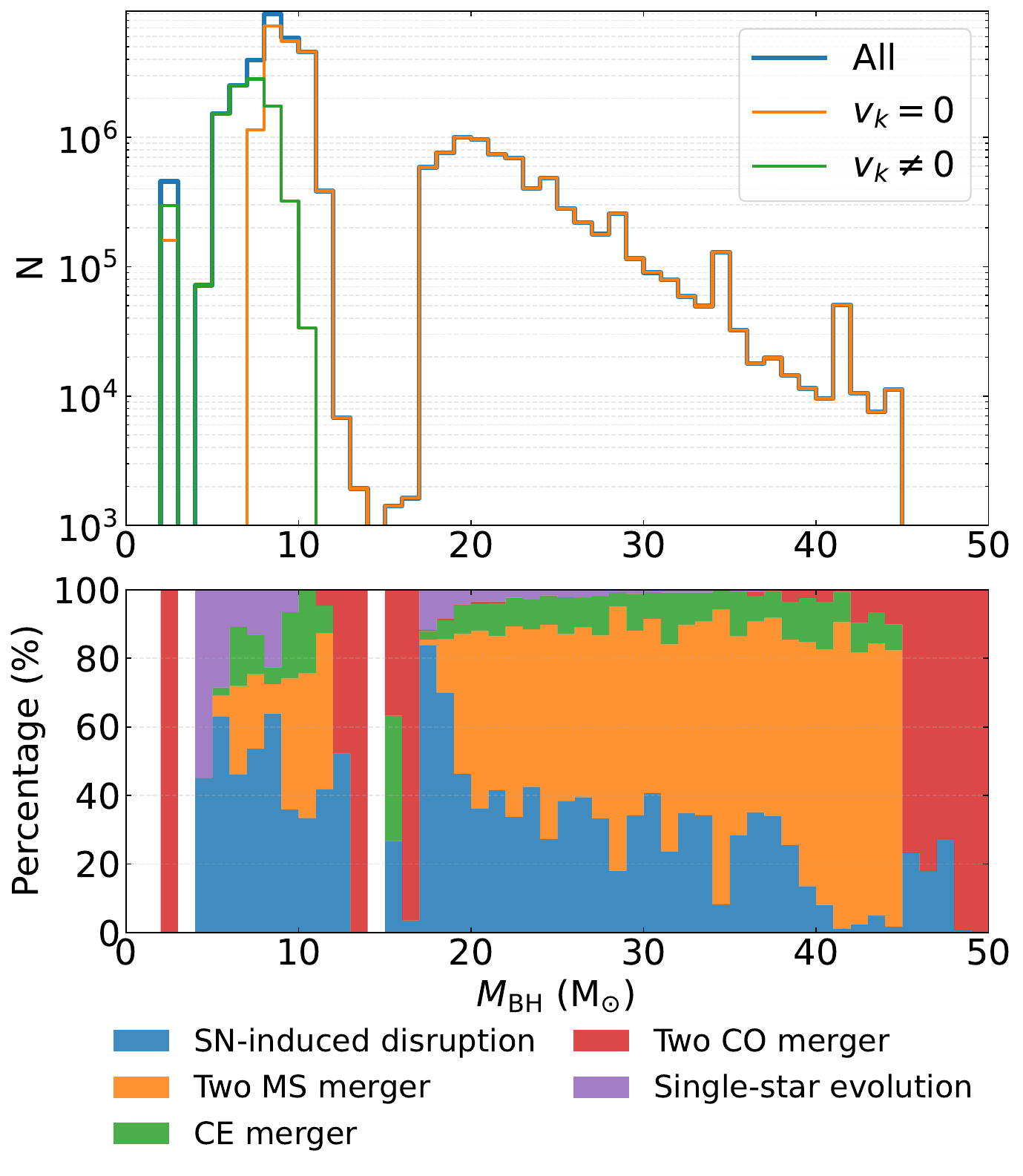}
    \caption{Mass distribution of Galactic IBHs in the fiducial model. In the upper panel, the blue curve shows the full population, while the orange and green curves show IBHs formed without and with natal kicks, respectively. The lower panel shows the fractional contribution of each formation channel as a function of BH mass.} 
    \label{fig:IBH_mass}
\end{figure}

Figure~\ref{fig:IBH_mass} shows the mass distribution of Galactic IBHs. The distribution has two prominent peaks centered near $9\,{\rm M_\odot}$ and $20\,{\rm M_\odot}$. The lower-mass peak contains both BHs formed through fallback and direct-collapse BHs. With the adopted fallback fraction of $f_{\rm fb}=0.5$, the former receive natal kicks and are concentrated primarily at $5-10\,{M_\odot}$. The non-kicked direct-collapse BHs mainly occupy the $7-12\,{ M_\odot}$ range and arise from the first carbon-oxygen core mass regime for direct collapse in the prescription of \citet{Maltsev2025}. The higher-mass peak near $20\,{M_\odot}$ is associated with the second direct-collapse regime.

Between the two peaks, the BH mass distribution is strongly depleted over approximately $13-17\,{ M_\odot}$. This bimodal structure is qualitatively consistent with the result of \citet{Willcox2025} and may contribute to some of the features inferred in the source-frame chirp-mass distribution of GWTC-4 binaries. The comparison is necessarily qualitative, however, because the binary chirp mass depends on both component masses and their mass ratio.

A small population of very low-mass BHs near $2-3\,{M_\odot}$ arises from mergers involving NSs and/or WDs. SN-induced disruption dominates around the lower-mass peak, whereas the two MS merger channel supplies most of the high-mass component over approximately $20-45\,{ M_\odot}$. BHs above $50\,{ M_\odot}$ are rare and are produced predominantly through the two CO merger channel. This scarcity is partly a consequence of the
adopted upper limit of $100\,M_{\odot}$ for the initial primary mass. Moreover,
the progenitors capable of producing BHs in this mass range may enter the regime
affected by pulsational pair-instability and pair-instability SNe, neither
of which is included in our current calculations. The predicted abundance and
formation channels of BHs at the high-mass end should therefore be interpreted
with caution.

\begin{figure}
	\includegraphics[width=\columnwidth]{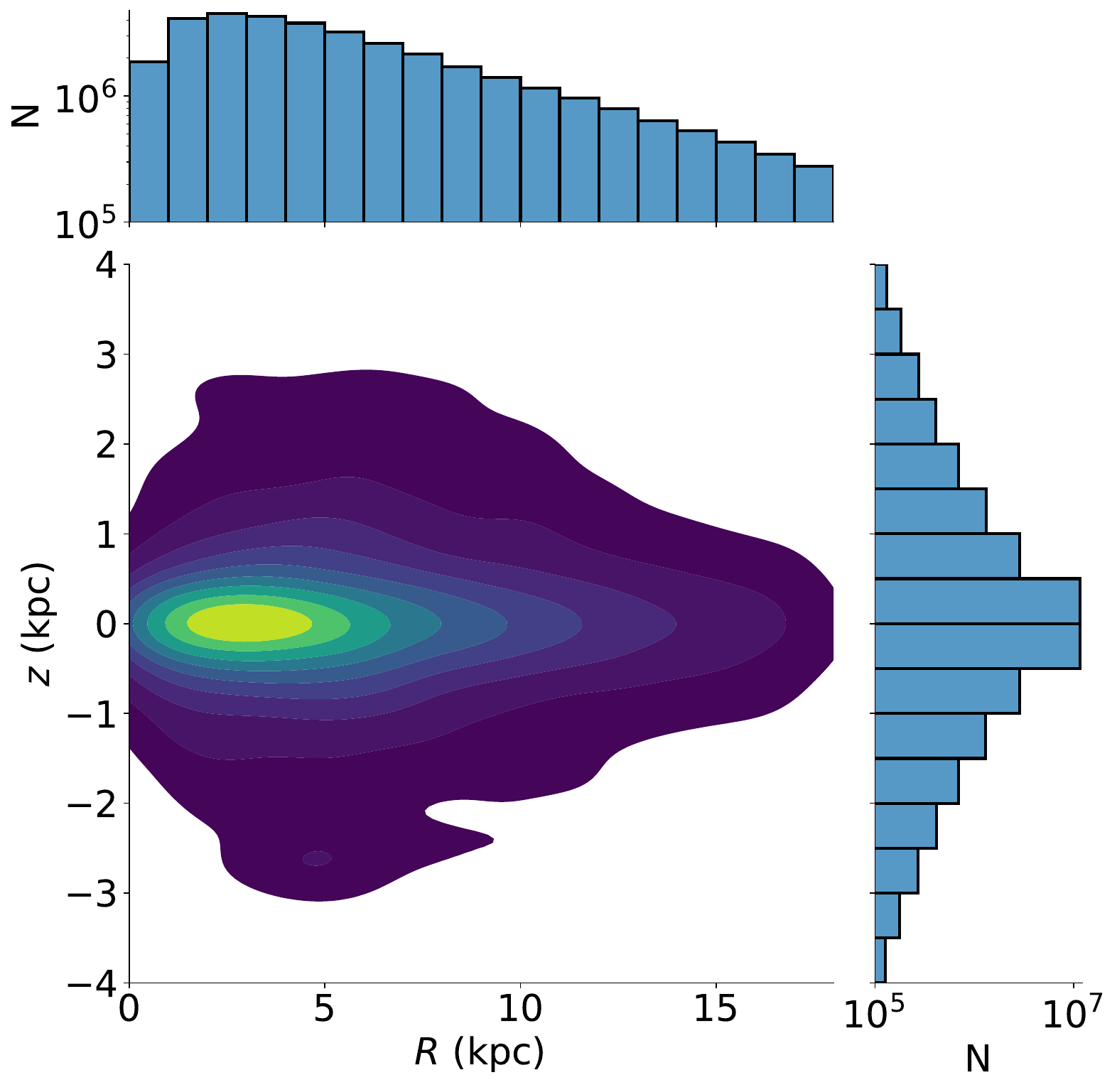}
    \caption{Present-day distribution of Galactic IBHs in Galactocentric radius \(R\) and height \(z\) relative to the Galactic mid-plane. Contours are spaced in 10\% probability intervals, and the marginal panels show the corresponding one-dimensional distributions.} 
    \label{fig:IBH_R_z}
\end{figure}

Figure \ref{fig:IBH_R_z} shows the present-day distribution of IBHs in Galactocentric radius and vertical height. Most systems (\(\sim 62\%\)) lie within $|z|<500~{\rm pc}$, and the radial distribution peaks at $R\simeq 3~{\rm kpc}$. This structure reflects the combined contributions of the thin disk, thick disk, and bulge in the adopted Galactic model.

\begin{figure}
	\includegraphics[width=\columnwidth]{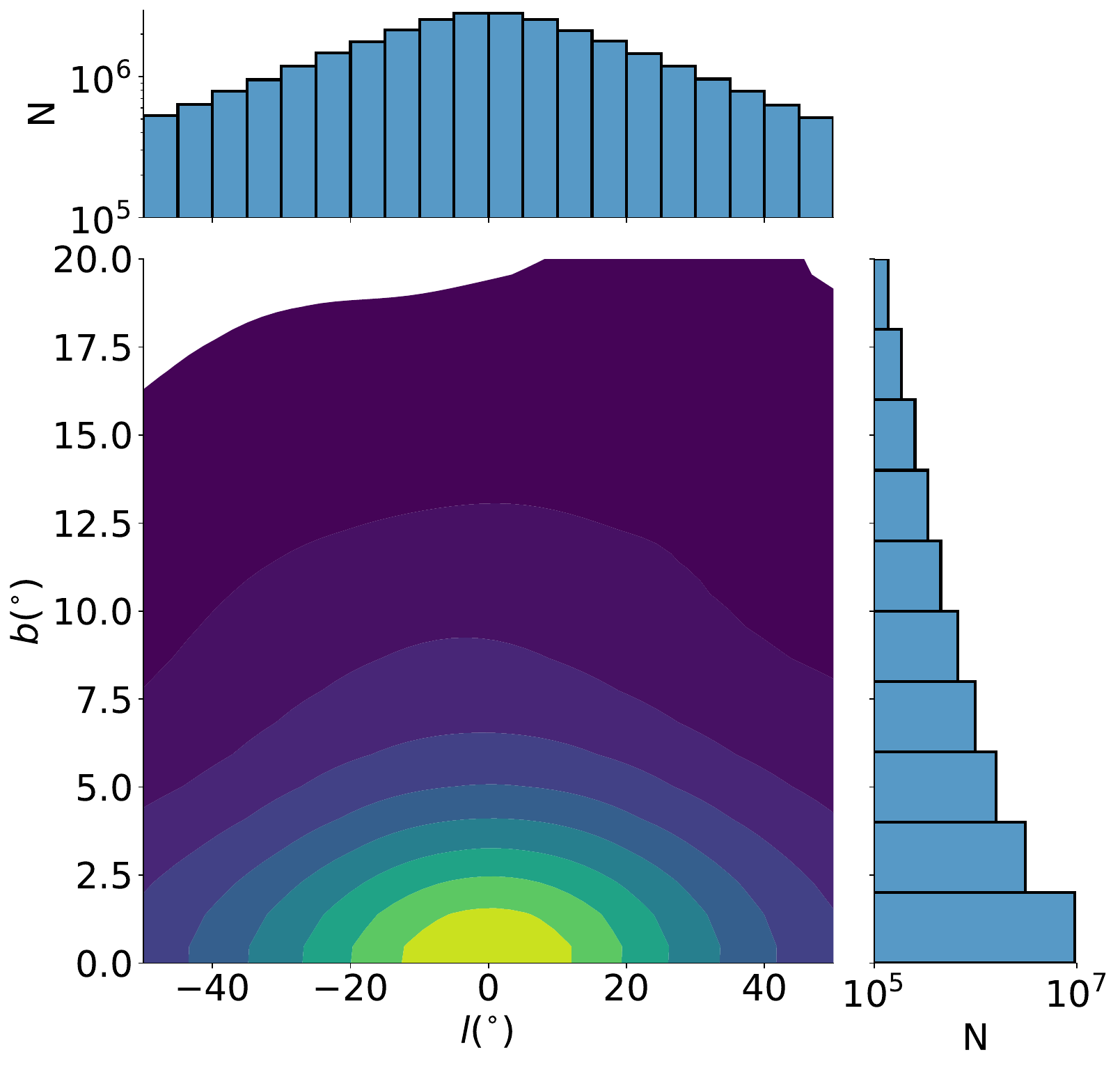}
    \caption{Present-day sky distribution of Galactic IBHs in Galactic longitude \(l\) and latitude \(b\). Contours and marginal histograms are defined as in Figure~\ref{fig:IBH_R_z}.} 
    \label{fig:IBH_ll_bb}
\end{figure}

On the sky, IBHs are concentrated toward low Galactic latitudes and the Galactic center (Figure \ref{fig:IBH_ll_bb}). This distribution differs from that of observable pulsars, which can peak around $|l|\sim 30^{\circ}$ because of the combined effects of spiral-arm structure, birth locations, and age-dependent detectability \citep{Faucher2006}. The imprint of spiral-arm evolution is less direct for IBHs because their observability is governed primarily by their masses, velocities, distances, and local ISM densities rather than by a characteristic electromagnetic lifetime.

\subsection{Orbital Motion}

\begin{figure}
	\includegraphics[width=\columnwidth]{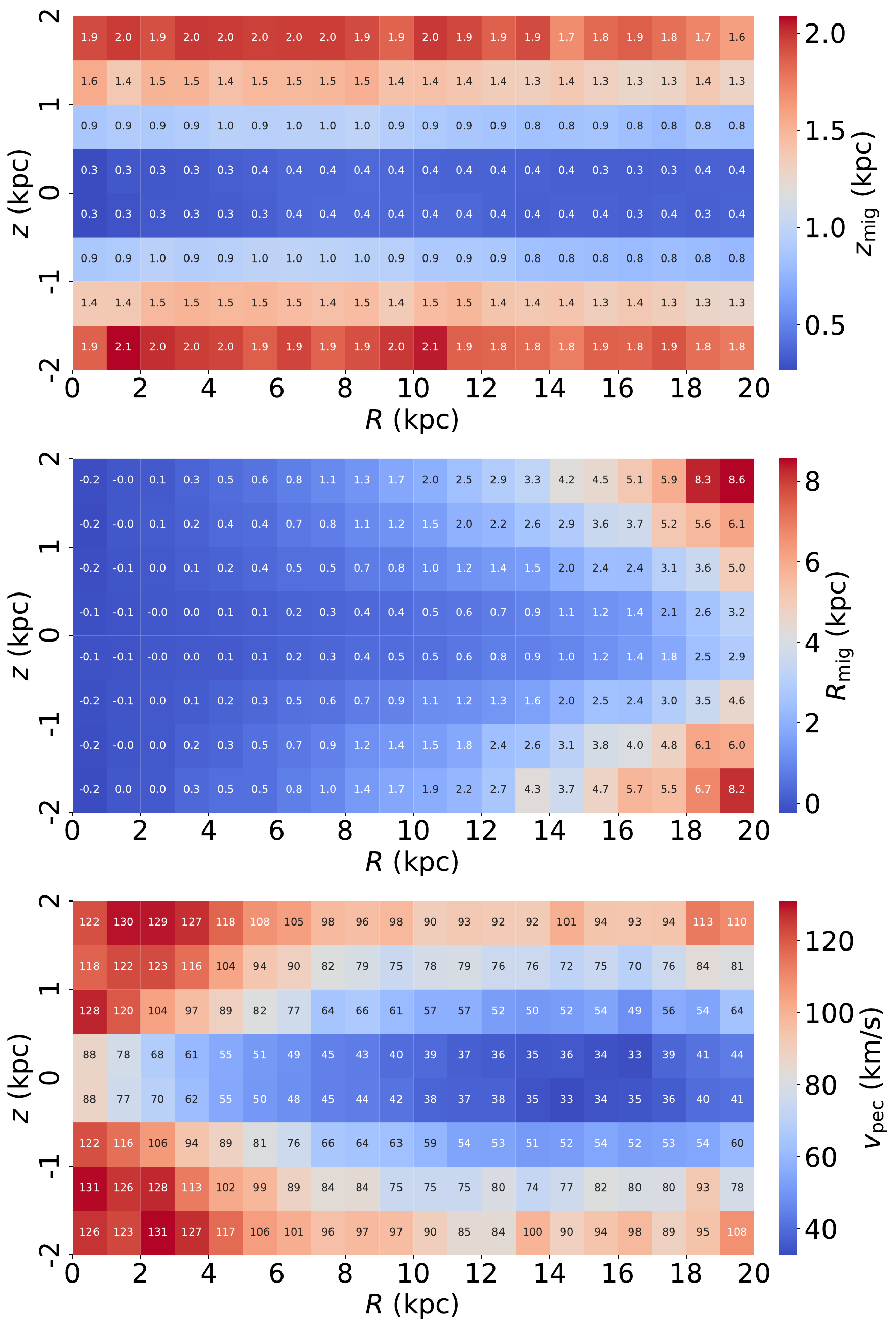}
    \caption{Dynamical properties of Galactic IBHs after orbit integration. From top to bottom, the panels show mean vertical migration, radial migration, and  peculiar velocity in the \(R\)--\(z\) plane. Each cell is labeled with the corresponding value of the quantity shown in that panel.} 
    \label{fig:IBH_dynamical}
\end{figure}

Figure~\ref{fig:IBH_dynamical} summarizes the present-day dynamical properties of Galactic IBHs after orbit integration in the Galactic potential. Here, $(R,z)$ denotes the present-day Galactocentric cylindrical coordinates, while $(R_{\rm ini},z_{\rm ini})$ denotes the corresponding birth coordinates of the progenitor star. The top panel shows the weighted mean vertical migration, $z_{\rm mig}=|z-z_{\rm ini}|$, as a function of $R$ and $z$. The middle panel shows the weighted mean radial migration, $R_{\rm mig}=R-R_{\rm ini}$, where positive and negative values indicate outward and inward migration, respectively. The bottom panel shows the weighted mean peculiar velocity, defined as \citep{Nagarajan2025}
\begin{equation}
v_{\rm pec}=\sqrt{v_R^2+v_z^2+\left[v_T-v_{\rm circ}(R, z=0)\right]^2}.
\end{equation}
We first examine vertical migration and then turn to radial migration and peculiar velocity, which retain more direct signatures of natal kicks.

\subsubsection{Vertical Migration}

\begin{figure}
	\includegraphics[width=\columnwidth]{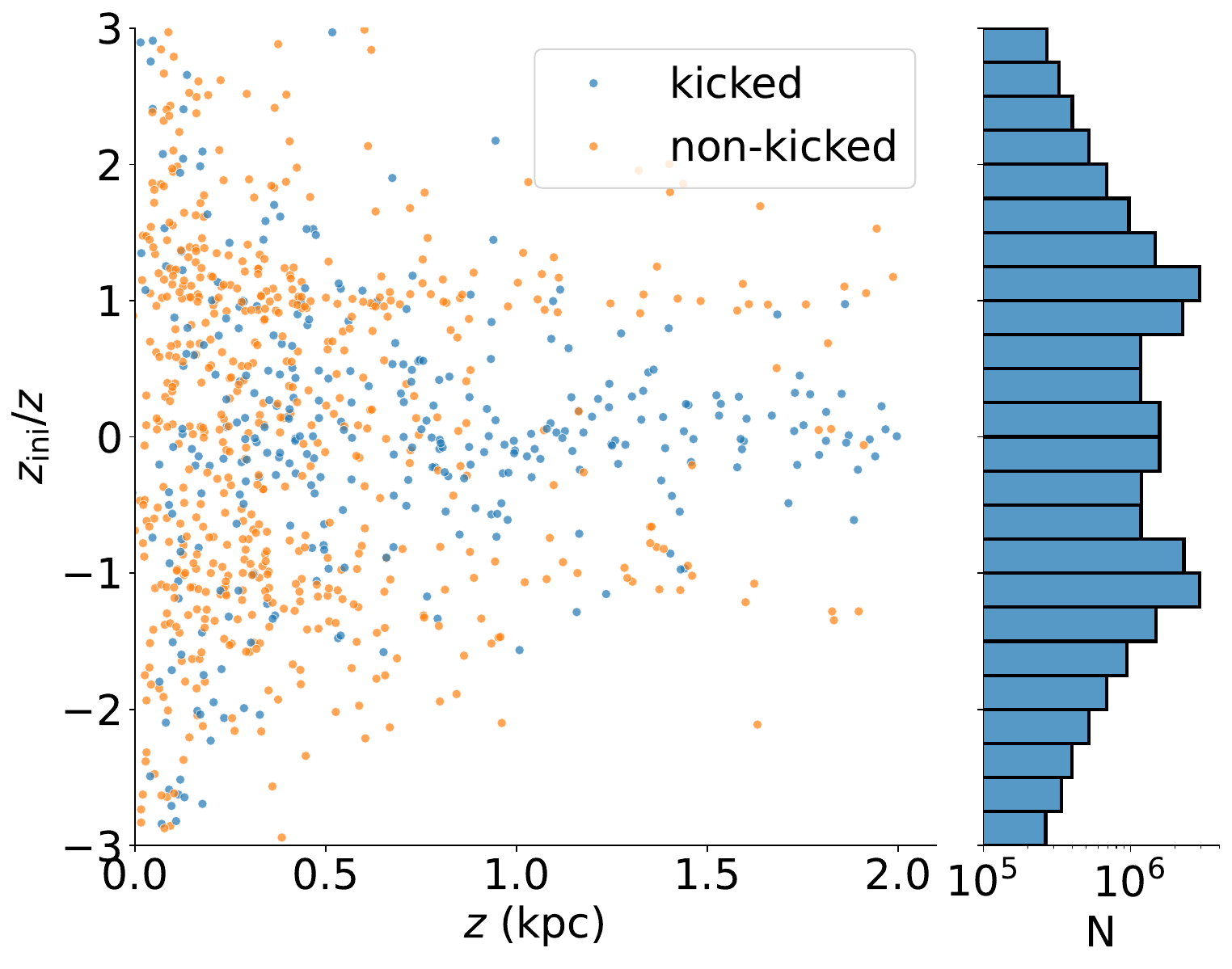}
    \caption{Ratio of initial to present-day height, \(z_{\rm ini}/z\), as a function of present-day height \(z\) for IBHs. Blue and orange points denote systems formed with and without natal kicks, respectively.} 
    \label{fig:IBH_z_ini_z}
\end{figure}

The top panel of Figure~\ref{fig:IBH_dynamical} shows that the mean vertical migration, $z_{\rm mig}$, depends only weakly on Galactocentric radius. Moreover, the ratio of $z_{\rm mig}$ to the present-day absolute height $|z|$ is close to unity. We define this vertical-migration fraction as
\begin{equation}
\label{eq:f_mig_z}
f_{\rm mig, z} = \frac{z_{\rm mig}}{|z|}= \frac{|z-z_{\rm ini}|}{|z|} = \left|1-\frac{z_{\rm ini}}{z}\right|\sim 1.
\end{equation}

Figure \ref{fig:IBH_z_ini_z} shows $z_{\rm ini}/z$ as a function of $z$ for IBHs currently above the mid-plane. Three concentrations appear near $z_{\rm ini}/z=0$ and $\pm 1$. Systems near zero were born close to the mid-plane. Those near unity were born close to their present-day height, whereas those near $-1$ were born at a comparable absolute height on the opposite side of the Galactic plane.

Almost all systems born near the mid-plane and currently located at $z>1~{\rm kpc}$ received natal kicks, because non-kicked systems born near the plane rarely reach such heights. The populations near $z_{\rm ini}/z=\pm1$ are prominent because their vertical velocities near their present-day positions are small, causing them to spend long intervals at similar heights. Together, the three concentrations naturally yield the relation in Equation~(\ref{eq:f_mig_z}). This behavior depends only weakly on whether a BH received a natal kick: kicks primarily change the relative contribution of the mid-plane-born population rather than the geometric relation between $z_{\rm mig}$ and $|z|$.

\subsubsection{Radial Migration and Peculiar Velocity}

Radial migration and peculiar velocity provide more direct diagnostics of natal kicks than vertical migration, because they retain information about both the birth orbit and the velocity impulse imparted during compact-object formation.

In Figure~\ref{fig:IBH_dynamical}, the mean radial migration is negative at $R\lesssim3\,\mathrm{kpc}$ and positive at $R\gtrsim3\,\mathrm{kpc}$, indicating net inward and outward migration, respectively. Its magnitude also increases with height above the mid-plane. Peculiar velocities are high near the Galactic center and at large $|z|$. Values below $50\,\mathrm{km\,s^{-1}}$ occur predominantly near the mid-plane at $R>6\,\mathrm{kpc}$. To identify the origin of these trends, we separate the population according to whether the BH received a natal kick.

\begin{figure}
	\includegraphics[width=\columnwidth]{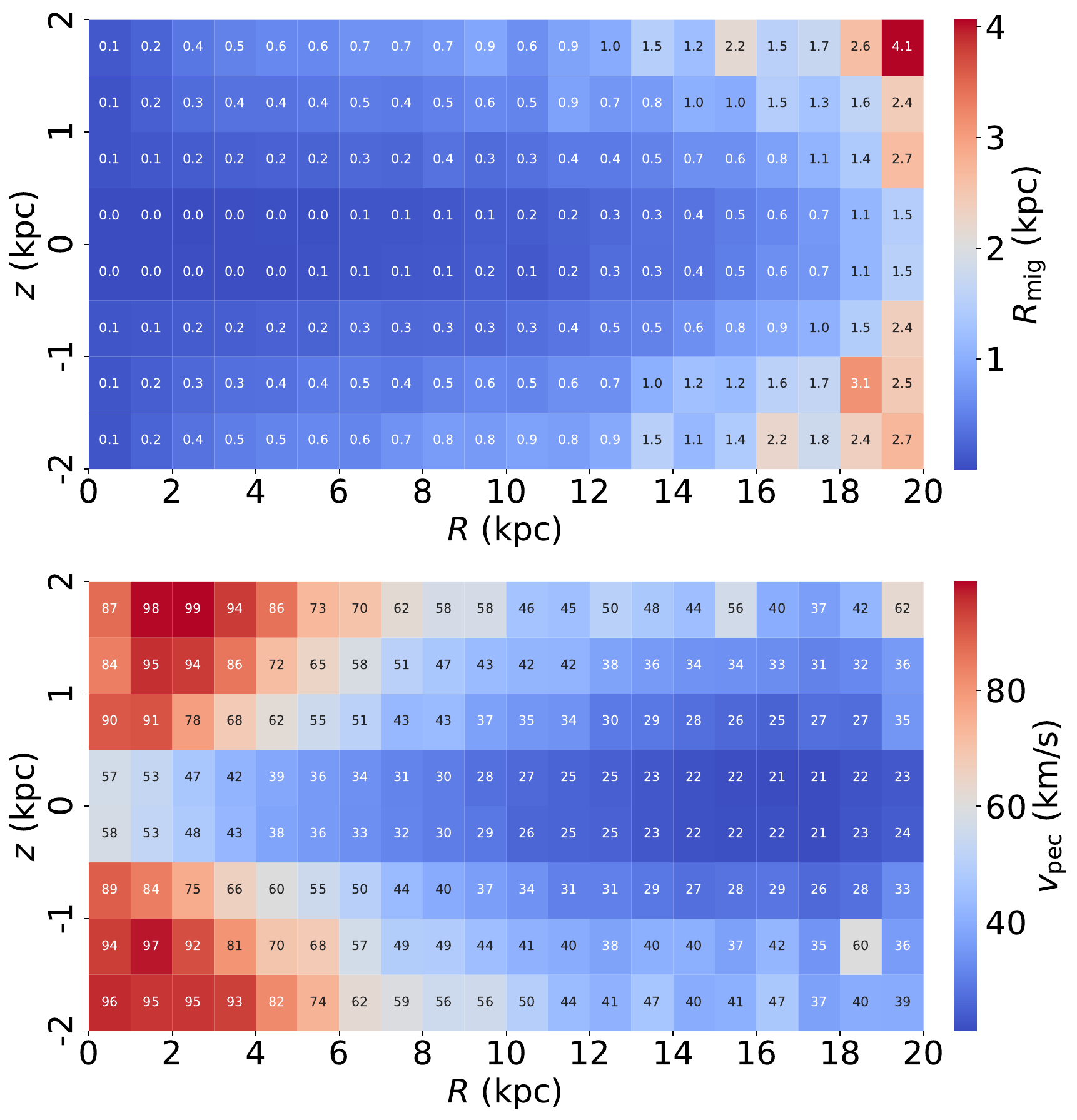}
    \caption{Weighted mean radial migration (top) and peculiar velocity (bottom) in the \(R\)--\(z\) plane for IBHs formed without natal kicks. Each cell is labeled with the corresponding value of the quantity shown in that panel.} 
    \label{fig:IBH_dynamical_no_kick}
\end{figure}

\begin{figure}
	\includegraphics[width=\columnwidth]{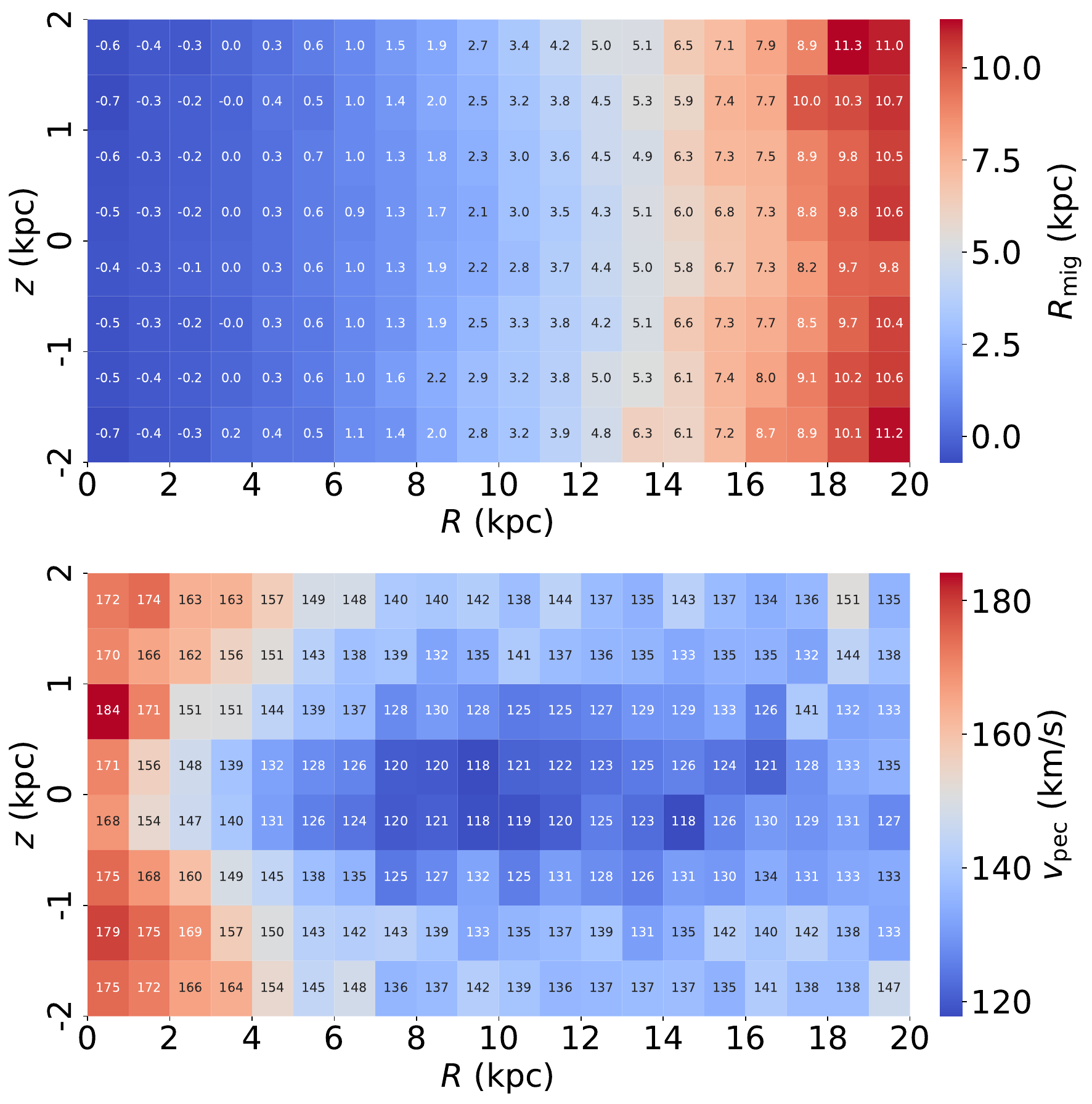}
    \caption{Same as Figure~\ref{fig:IBH_dynamical_no_kick}, but for IBHs formed with natal kicks.}
    \label{fig:IBH_dynamical_with_kick}
\end{figure}

Figure~\ref{fig:IBH_dynamical_no_kick} shows the radial migration and peculiar velocity of IBHs formed without natal kicks. Within $R\lesssim10\,\mathrm{kpc}$, these systems undergo only modest radial migration. Stronger migration occurs mainly near the outer edge of the disk and at large heights, where the population is sparse. Thus, most non-kicked disk IBHs remain on nearly circular orbits. Away from the Galactic center and regions with $|z|>1\,\mathrm{kpc}$, their peculiar velocities are typically $20-30\,\mathrm{km\,s^{-1}}$. Nearby IBHs with low $v_{\rm pec}$ are therefore the most promising candidates for accretion-powered detection.

Figure~\ref{fig:IBH_dynamical_with_kick} shows the corresponding distributions for IBHs formed with natal kicks. These systems undergo substantial radial migration throughout both the inner and outer disk. The typical migration scale is several hundred parsecs in the inner disk and can reach several kiloparsecs in the outer disk. The migration amplitude increases with Galactocentric radius, and many kicked IBHs at $R>10\,\mathrm{kpc}$ originated at smaller Galactocentric radii. Their peculiar velocities generally exceed $120\,\mathrm{km\,s^{-1}}$, suppressing Bondi--Hoyle accretion from the ISM and reducing their detectability as accreting sources. At the same time, these high-velocity systems retain valuable information about natal kicks. We therefore examine the relation between peculiar and natal-kick velocities in more detail below.

\subsubsection{Relations Between \(v_{\rm pec}\) and \(v_{\rm pec,i}\) or \(v_{\rm k}\)}

\begin{figure}
	\includegraphics[width=\columnwidth]{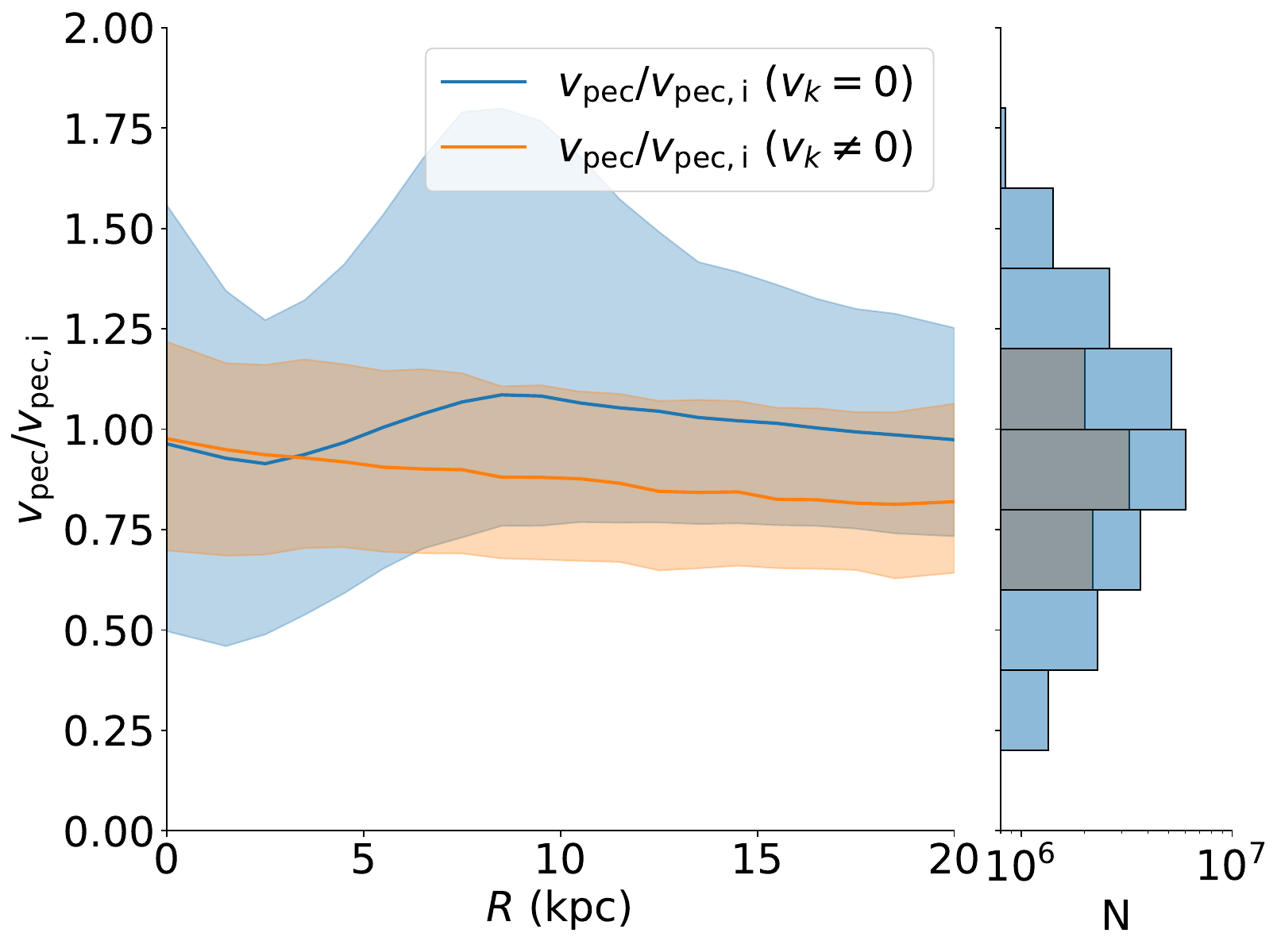}
    \caption{Ratio of present-day to initial peculiar velocity, $v_{\rm pec}/v_{\rm pec,i}$, as a function of Galactocentric radius $R$ for IBHs formed without (blue) and with (orange) natal kicks. Solid lines show the weighted medians, and shaded regions enclose the weighted 16th--84th percentile ranges.}
    \label{fig:IBH_pec_to_pec_i}
\end{figure}

\begin{figure}
	\includegraphics[width=\columnwidth]{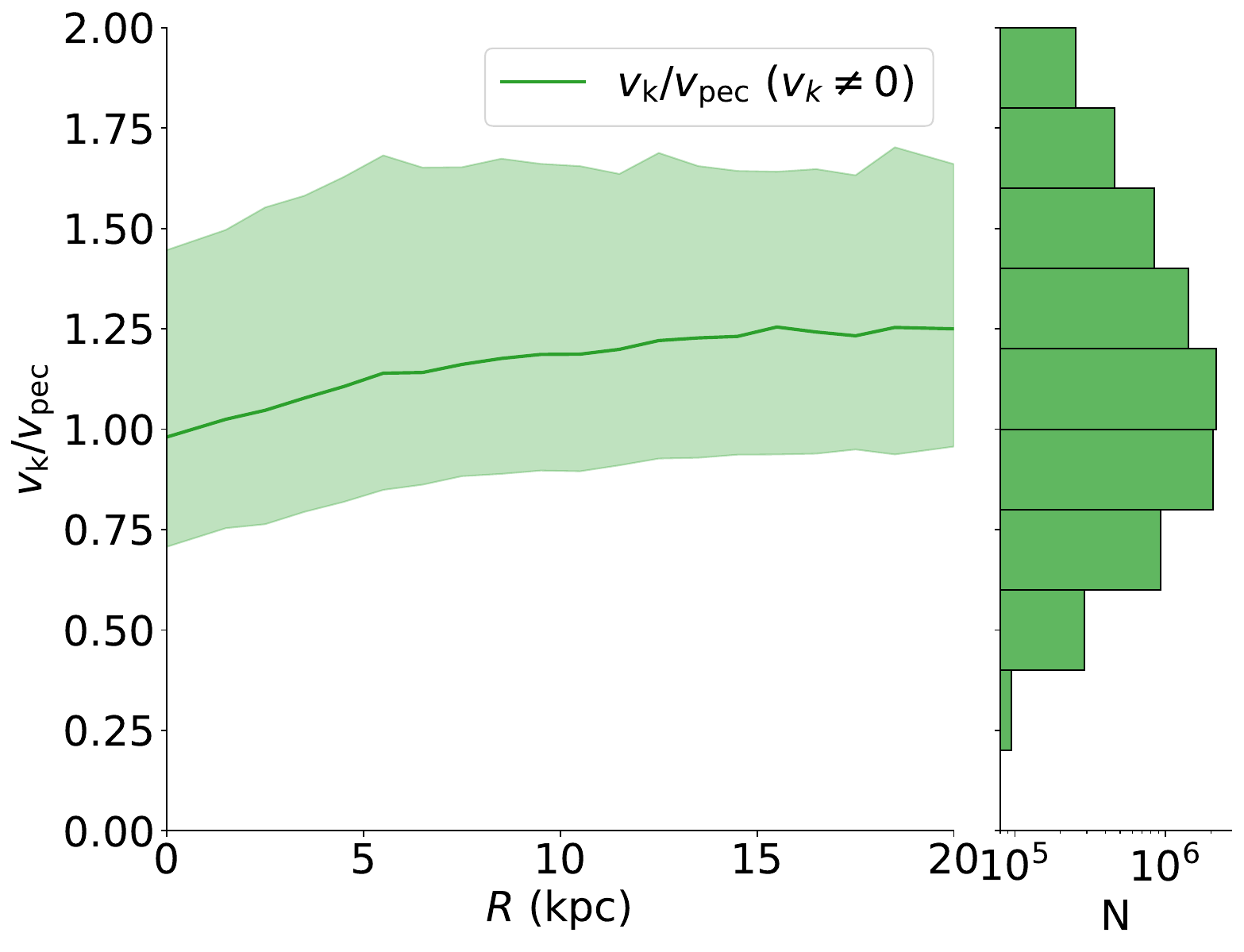}
    \caption{Ratio of natal-kick velocity to present-day peculiar velocity, $v_{\rm k}/v_{\rm pec}$, as a function of Galactocentric radius $R$ for IBHs formed with natal kicks. The solid line shows the weighted median, and the shaded region encloses the weighted 16th--84th percentile range.} 
    \label{fig:IBH_kick_to_pec}
\end{figure}

Figure~\ref{fig:IBH_pec_to_pec_i} shows $v_{\rm pec}/v_{\rm pec,i}$ as a function of Galactocentric radius, where $v_{\rm pec,i}$ denotes the initial peculiar velocity at the time of IBH formation. For IBHs formed without natal kicks, the weighted median remains close to unity across most of the Galaxy. It is slightly below unity in the innermost region, rises modestly above unity near $R\simeq10\,{\rm kpc}$, and approaches unity again at larger radii. Thus, in the absence of natal kicks, Galactic orbital evolution redistributes the velocity components without strongly changing the characteristic peculiar velocity of the population.

The kicked population behaves differently. Its weighted median $v_{\rm pec}/v_{\rm pec,i}$ is generally below unity and decreases mildly with increasing Galactocentric radius. Present-day peculiar velocities are therefore typically smaller than the initial values immediately after BH formation, indicating that subsequent orbital evolution in the Galactic potential partially weakens the initial kinematic signature. This behavior is consistent with the damping of natal-kick signatures discussed by \citet{Disberg2024}.

To connect present-day peculiar velocities to natal kicks, Figure~\ref{fig:IBH_kick_to_pec} shows $v_{\rm k}/v_{\rm pec}$ as a function of Galactocentric radius for kicked IBHs. The weighted median is close to unity in the inner Galaxy and increases gradually with radius, reaching $v_{\rm k}/v_{\rm pec}\simeq1.25$ in the outer disk. Thus, natal-kick velocities are typically comparable to present-day peculiar velocities, with the ratio becoming modestly larger at larger radii. The $v_{\rm k}/v_{\rm pec}$ distribution spans a broad 16th--84th percentile range, reflecting substantial object-to-object scatter arising from variations in natal-kick magnitudes, orbital velocities inherited from the progenitor binaries, and subsequent orbital evolution in the Galactic potential. Present-day peculiar velocities therefore cannot uniquely determine the natal kick of an individual IBH, but they can provide statistical constraints on the underlying kick distribution. We next examine how these relations depend on BH mass.

\subsubsection{Overview of IBHs without Natal Kicks}

\begin{figure*}
	\centering
	\includegraphics[width=18cm]{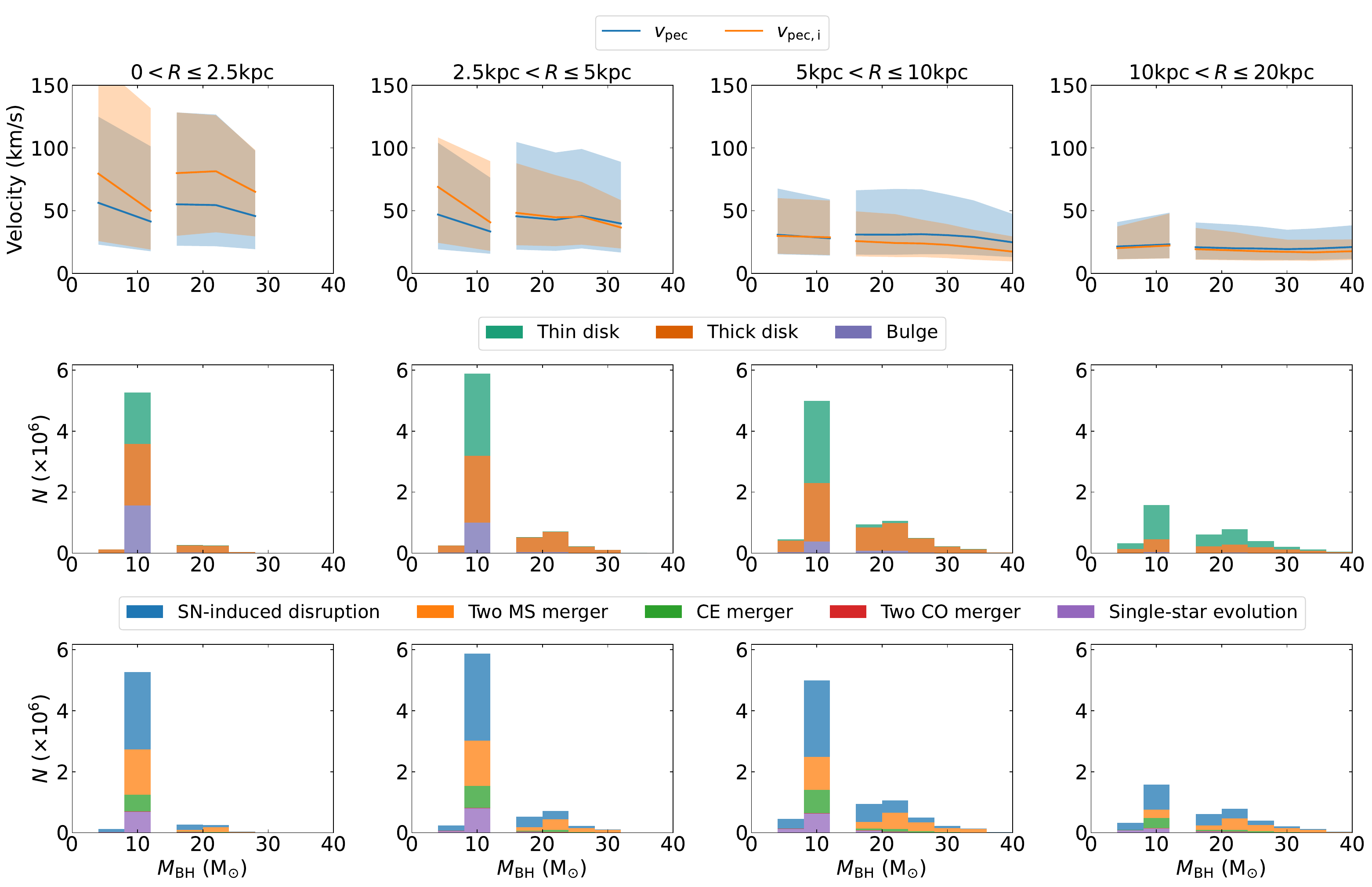}
    \caption{Overview of IBHs formed without natal kicks in four bins of Galactocentric radius, as labeled at the tops of the columns. The top row shows the mass dependence of the present-day and initial peculiar velocities, \(v_{\rm pec}\) and \(v_{\rm pec,i}\), respectively; curves and shaded bands denote the weighted medians and weighted 16th--84th percentile ranges. The middle and bottom rows show the weighted BH mass distributions separated by Galactic birth component and formation channel, respectively.}
    \label{fig:IBH_overview_no_kick}
\end{figure*}

Figure~\ref{fig:IBH_overview_no_kick} presents the properties of IBHs formed without natal kicks in four bins of present-day Galactocentric radius: \(0<R\leq2.5\,\mathrm{kpc}\), \(2.5\,\mathrm{kpc}<R\leq5\,\mathrm{kpc}\), \(5\,\mathrm{kpc}<R\leq10\,\mathrm{kpc}\), and \(10\,\mathrm{kpc}<R\leq20\,\mathrm{kpc}\). The top row shows how the present-day and initial peculiar velocities, \(v_{\rm pec}\) and \(v_{\rm pec,i}\), depend on BH mass. To avoid noisy estimates in sparsely populated mass bins, we omit bins with weighted counts below \(10^4\). The middle and bottom rows show the weighted mass distributions separated by Galactic birth component and formation channel, respectively.

The non-kicked population is dominated by two mass components centered near \(10\,M_{\odot}\) and \(20\,M_{\odot}\). Both consist predominantly of direct-collapse BHs and correspond to the two carbon-oxygen core mass intervals that lead to direct collapse in the prescription of \citet{Maltsev2025}. The fraction of BHs with \(M_{\rm BH}\gtrsim17\,M_{\odot}\) increases toward larger Galactocentric radii, primarily because the mixture of Galactic birth components and their metallicity distributions changes among the radius bins. In the three inner bins (\(R\leq10\,\mathrm{kpc}\)), the high-mass component is supplied almost entirely by thick-disk progenitors, whereas the thin disk dominates in the outermost bin. In the adopted Galactic model, both the old thick disk and the outer thin disk are relatively metal poor (Figure~\ref{fig:Metallicity}).

Lower-metallicity progenitors experience weaker wind mass loss and therefore retain more mass until core collapse, favoring the formation of more massive BHs (Figure~\ref{fig:stellar_mass_relation}). Moreover, in the prescription of \citet{Maltsev2025}, the carbon-oxygen core mass boundary of the second direct-collapse regime shifts to lower values at lower metallicity. Low-metallicity progenitors can therefore enter this regime more readily, both because they develop more massive cores and because the direct-collapse threshold is lower.

Over most of the mass range, both \(v_{\rm pec,i}\) and \(v_{\rm pec}\) generally decrease from the inner to the outer Galactocentric-radius bins, although the detailed trends depend on the mixture of birth components and formation channels.

The initial peculiar velocity reflects both the Galactic environment in which the progenitor forms and its binary-evolution history. For example, in the innermost bin, \(0<R\leq2.5\,\mathrm{kpc}\), BHs near \(20\,M_{\odot}\) have larger \(v_{\rm pec,i}\) than those near \(10\,M_{\odot}\) because the former are dominated by thick-disk progenitors. In the adopted Galactic model, the thick disk is older, kinematically hotter, and has a larger intrinsic scale height, leading to larger initial peculiar velocities. Formation channels further modify this trend. In the \(5\,\mathrm{kpc}<R\leq10\,\mathrm{kpc}\) bin, the gradual decrease in \(v_{\rm pec,i}\) at \(M_{\rm BH}\gtrsim17\,M_{\odot}\) coincides with an increasing contribution from the two MS merger channel. The merger of two MS stars does not itself impart a velocity impulse to the merger product, so the subsequently formed direct-collapse BH retains the systemic motion of its progenitor. By contrast, a BH isolated through SN-induced disruption can inherit a velocity offset from its pre-disruption orbital motion even if the BH itself receives no natal kick.

The present-day peculiar velocity broadly follows the trends in \(v_{\rm pec,i}\), although subsequent orbital evolution in the Galactic potential modifies its amplitude. The kinematics of non-kicked IBHs are therefore determined jointly by their Galactic birth components, binary-evolution pathways, and subsequent motion through the Galaxy.


\subsubsection{Overview of IBHs with Natal Kicks}

\begin{figure*}
	\centering
	\includegraphics[width=18cm]{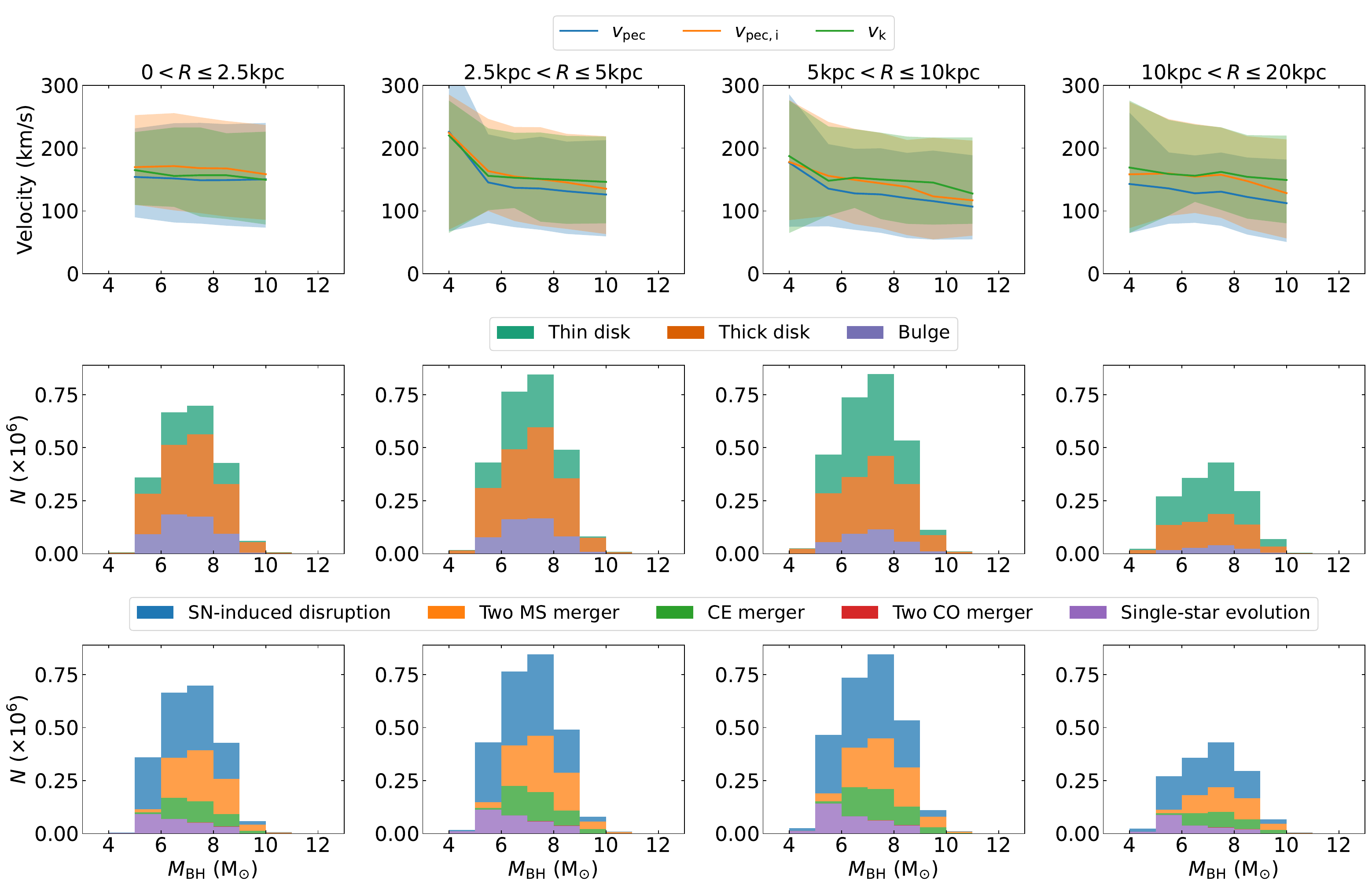}
    \caption{Same as Figure~\ref{fig:IBH_overview_no_kick}, but for IBHs formed with natal kicks. The top row additionally shows the natal-kick velocity \(v_{\rm k}\).}
    \label{fig:IBH_overview_with_kick}
\end{figure*}

Figure~\ref{fig:IBH_overview_with_kick} presents the properties of IBHs formed with natal kicks in the same four bins of Galactocentric radius as Figure~\ref{fig:IBH_overview_no_kick}. In the fiducial model, all BHs with nonzero natal kicks form through fallback with a fixed fallback fraction of \(f_{\rm fb}=0.5\). Their masses are set by their pre-SN helium-core masses and are concentrated primarily at \(\sim 5-10\,M_{\odot}\). Their natal-kick magnitudes are drawn from the same fallback-scaled Maxwellian distribution.

The fractional contributions of the thick disk and bulge decrease with Galactocentric radius, whereas that of the thin disk increases. For the dominant \(5-10\,M_\odot\) component, SN-induced disruption is the primary formation channel. Although the kinematics of kicked IBHs retain imprints of their Galactic birth components and binary-evolution histories, natal kicks provide the principal additional velocity at BH formation. Their present-day peculiar velocities therefore typically reach \(\sim100-200\,\mathrm{km\,s^{-1}}\), substantially exceeding those of non-kicked IBHs.

A distinct low-mass component appears at \(4-5\,M_{\odot}\) in the range \(2.5\,\mathrm{kpc}<R\leq10\,\mathrm{kpc}\) and is dominated by thick-disk progenitors. In the prescription of \citet{Maltsev2025}, the minimum carbon-oxygen core mass required for fallback-BH formation shifts to a lower value at lower metallicity. Metal-poor thick-disk progenitors can therefore form BHs through fallback from lower-mass helium cores. This component receives no contribution from the CE merger or two MS merger channels because, in our calculations, the products of these mergers do not fall within the narrow pre-SN core-mass interval that produces these low-mass fallback BHs. The relatively large \(v_{\rm k}\) and \(v_{\rm pec}\), particularly in the \(2.5\,\mathrm{kpc}<R\leq5\,\mathrm{kpc}\) bin, should be interpreted with caution. Because the population is binned by Galactocentric radius, this bin contains only a subset of this rare component, resulting in a small effective sample size. The large velocities should not be taken as evidence that lower-mass BHs formed through fallback receive intrinsically stronger natal kicks.

For the dominant \(5-10\,M_{\odot}\) component at \(R>2.5\,\mathrm{kpc}\), \(v_{\rm pec,i}\) broadly follows \(v_{\rm k}\), indicating that natal kicks dominate the initial peculiar velocity. The present-day value \(v_{\rm pec}\) is generally smaller than \(v_{\rm pec,i}\) because subsequent evolution in the Galactic potential redistributes orbital energy and modifies the velocity relative to local circular motion. Orbital evolution therefore partially obscures the natal-kick signature in present-day kinematics, as illustrated in Figure~\ref{fig:IBH_kick_to_pec}. Although the relations among \(v_{\rm k}\), \(v_{\rm pec,i}\), and \(v_{\rm pec}\) are statistical rather than deterministic, they can be used to constrain natal kicks from the kinematics of observed IBH populations.

\section{Accretion of IBHs from the ISM} \label{sec:accrete ISM}

Previous estimates of accreting IBH populations have often assumed simplified mass, velocity, or spatial distributions, thereby overlooking the coupled effects of binary evolution and Galactic dynamics. Here we estimate the accretion rates and X-ray fluxes of the dynamically evolved IBH population described above.

\subsection{Accretion Rates} \label{sec:accretion rate}

\begin{figure}
	\includegraphics[width=\columnwidth]{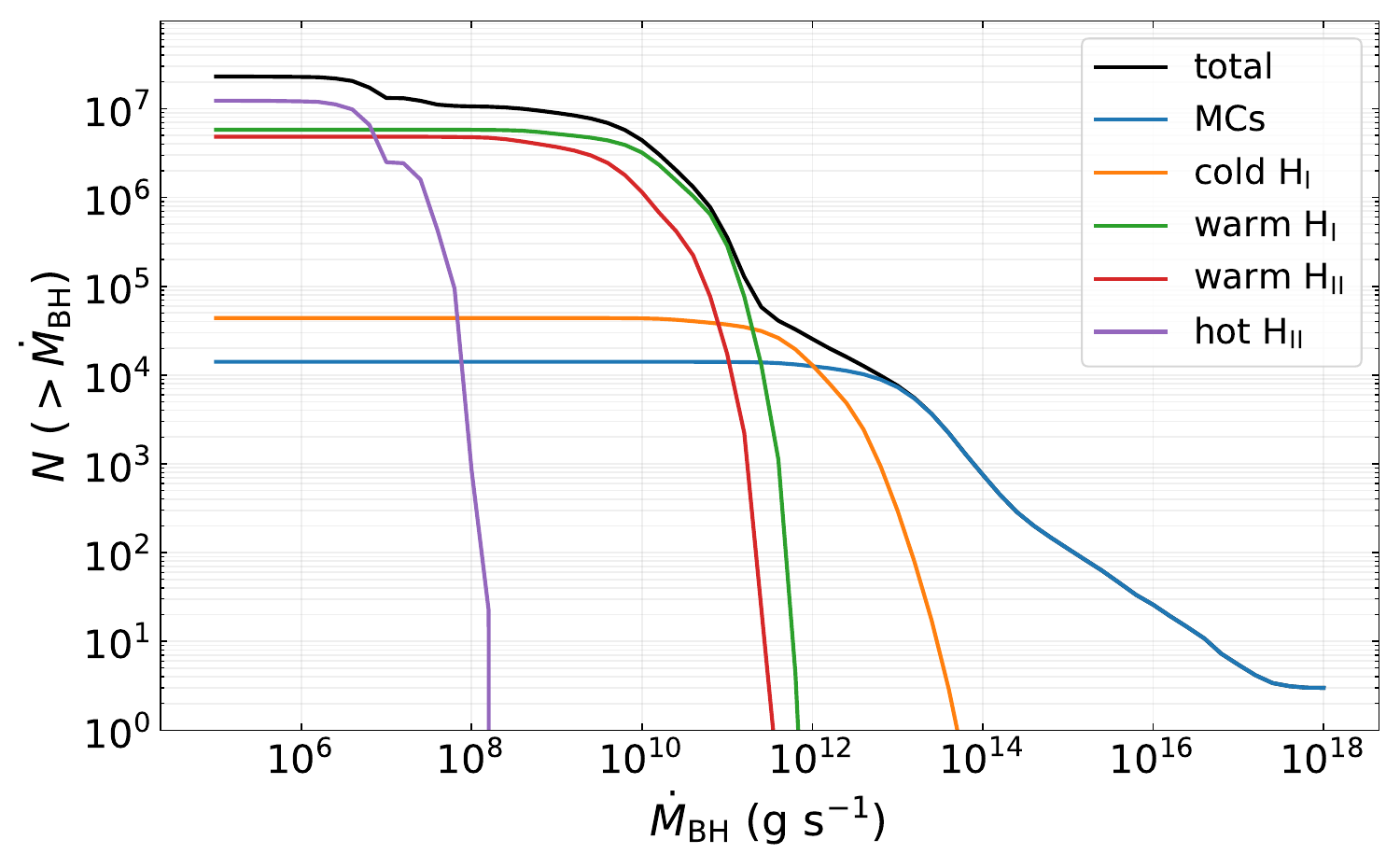}
    \caption{Distribution of accretion rates reaching the BH event horizon for systems embedded in each of the five ISM phases.} 
    \label{fig:IBH_ism_mdot}
\end{figure}

Figure~\ref{fig:IBH_ism_mdot} shows the distribution of the accretion rates reaching the BH event horizon for systems embedded in the five ISM phases. Accretion from hot \(\mathrm{H}_{\text{II}}\) is the weakest in all cases and remains below \(10^{9}\,\mathrm{g\,s^{-1}}\), because its low density and high sound speed strongly suppress Bondi--Hoyle capture. Among the other phases, the highest rates are found in MCs, followed by cold \(\mathrm{H}_{\text{I}}\), warm \(\mathrm{H}_{\text{I}}\), and warm \(\mathrm{H}_{\text{II}}\). Approximately \(10^2\) BHs reach \(\dot{M}_{\rm BH}>10^{15}\,\mathrm{g\,s^{-1}}\), while several systems exceed \(10^{18}\,\mathrm{g\,s^{-1}}\). Nearly all horizon accretion rates lie in the RIAF regime.

\subsection{Detection of IBHs} \label{sec:detection}

For an idealized estimate, we assume that all emitted power lies within the X-ray band \(0.1-100\) keV \citep{Tsuna2018}. The corresponding flux at the Sun's position is
\begin{equation}
F_{\mathrm{X}} \simeq F_{\mathrm{bol}} = \frac{\eta \dot{M}_{\rm BH} c^2}{4 \pi D^2}.
\end{equation}

\begin{figure}
	\includegraphics[width=\columnwidth]{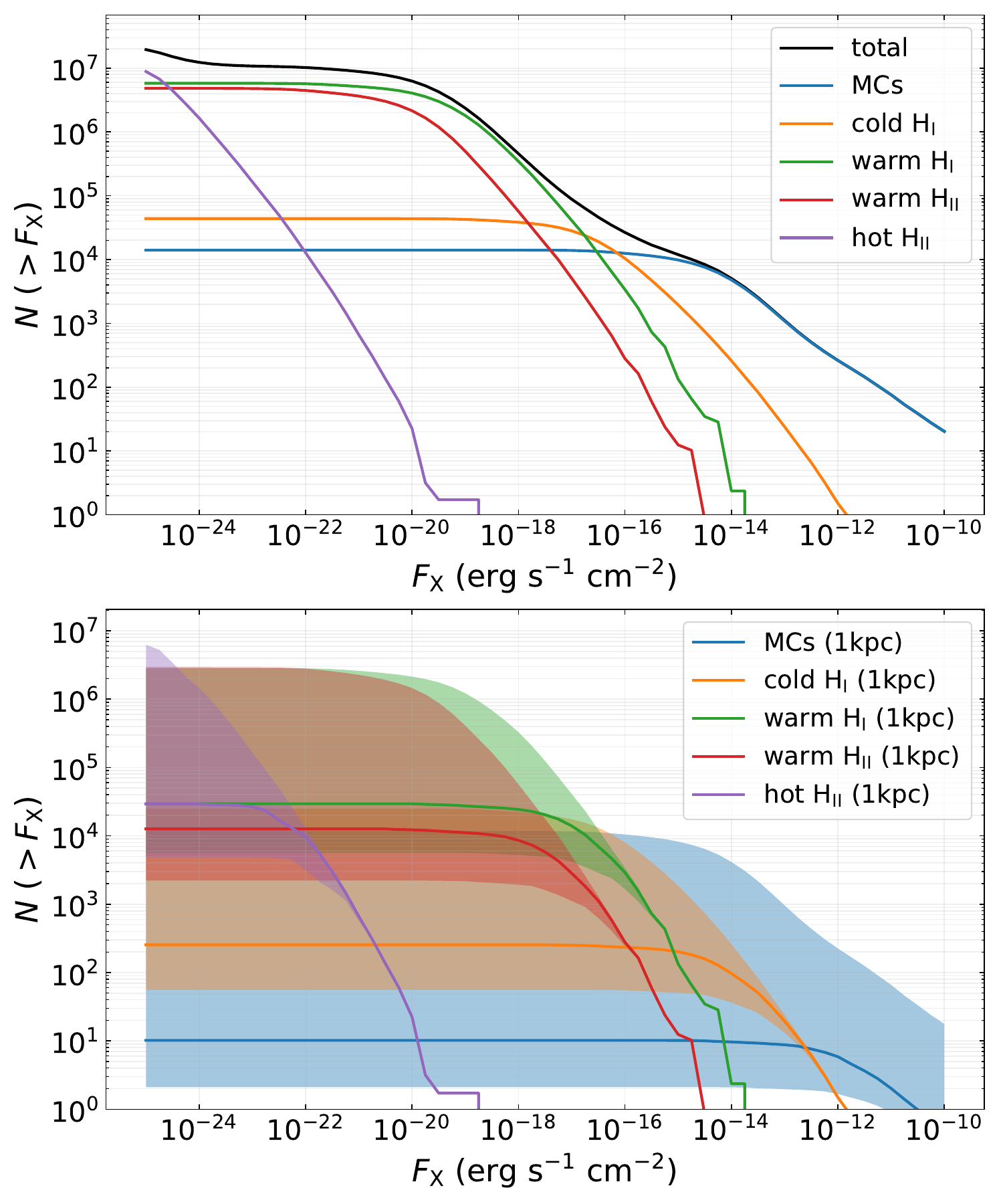}
    \caption{Cumulative number of accreting IBHs as a function of $F_{\rm X}$. The upper panel shows the full Galactic population; the lower panel shows distance-limited samples, with the solid curve for \(D<1\,\mathrm{kpc}\) and the shaded region spanning \(D<500\,\mathrm{pc}\) to \(D<10\,\mathrm{kpc}\).} 
    \label{fig:IBH_ism_flux}
\end{figure}

Figure~\ref{fig:IBH_ism_flux} shows the cumulative number of accreting IBHs as a function of \(F_{\rm X}\). BHs in hot \(\mathrm{H}_{\text{II}}\) are effectively undetectable because their low densities and high effective sound speeds yield extremely faint emission. BHs in MCs, in contrast, dominate the bright end, including sources with \(F_{\rm X}>10^{-12}\,\mathrm{erg\,s^{-1}\,cm^{-2}}\), making this phase the most favorable environment for finding IBH candidates. In total, approximately \(5\times10^3\) BHs exceed \(10^{-14}\,\mathrm{erg\,s^{-1}\,cm^{-2}}\), substantially more than estimated by \citet{Agol2002}. The difference is driven mainly by the radiative-efficiency prescription: \citet{Agol2002} adopted a fixed \(\eta=10^{-5}\), whereas our prescription permits significantly higher efficiencies for part of the low-accretion-rate population.

\begin{figure}
	\includegraphics[width=\columnwidth]{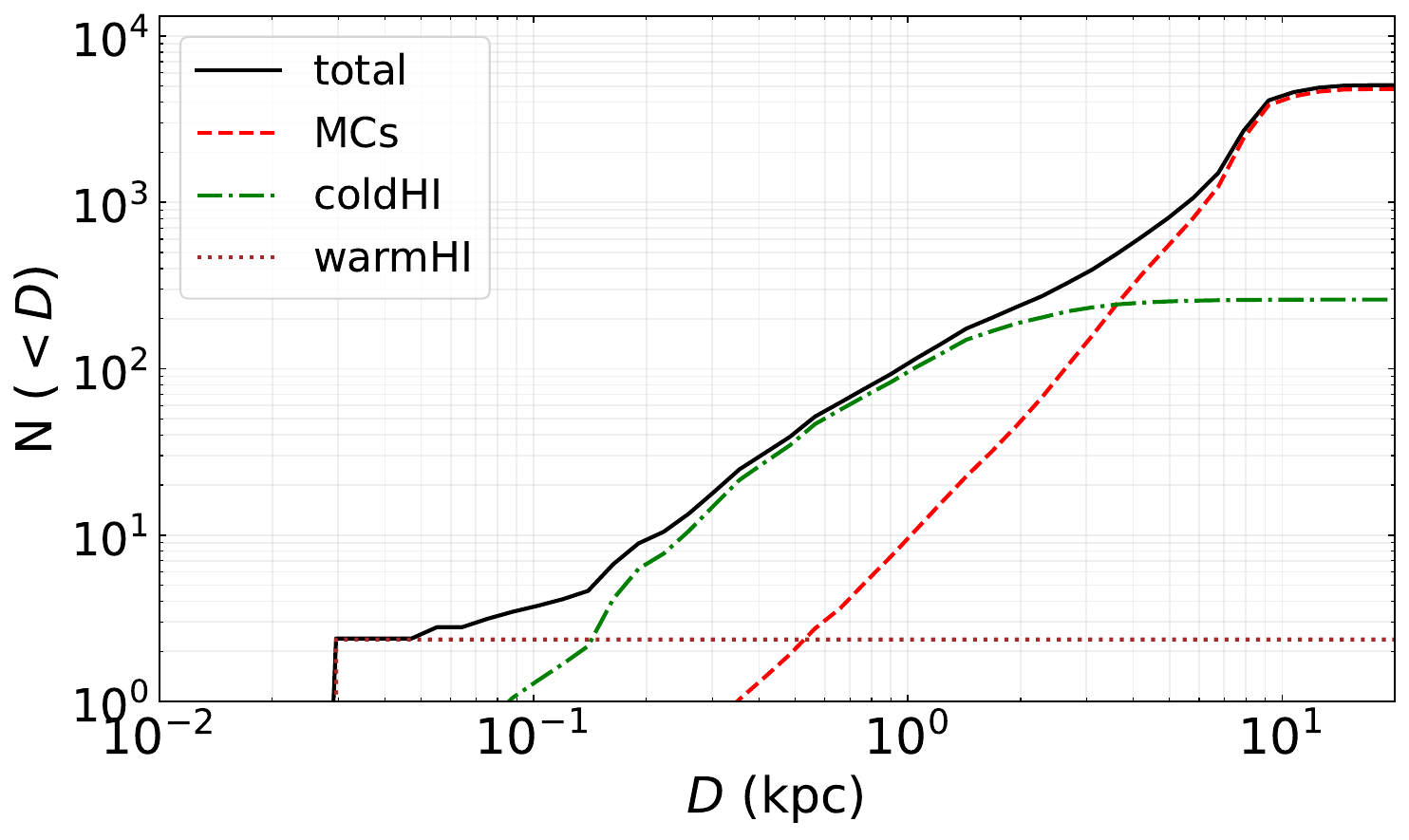}
    \caption{Cumulative number of IBHs with \(F_{\rm X}>10^{-14}\,\mathrm{erg\,s^{-1}\,cm^{-2}}\) as a function of distance from the Sun, shown separately for the three ISM phases that contain sources above this threshold.} 
    \label{fig:IBH_ism_dist}
\end{figure}

The detectability of accreting IBHs also depends strongly on distance. Figure~\ref{fig:IBH_ism_dist} shows the cumulative number of accreting IBHs with \(F_{\rm X}>10^{-14}\,\mathrm{erg\,s^{-1}\,cm^{-2}}\) as a function of their distance from the Sun. The cumulative number is approximately \(40\) within \(500\,\mathrm{pc}\), \(10^2\) within \(1\,\mathrm{kpc}\), and slightly above \(4\times10^3\) within \(10\,\mathrm{kpc}\), compared with nearly \(5\times10^3\) over the full modeled Galactic volume. Thus, most detectable sources lie within \(10\,\mathrm{kpc}\), although a non-negligible fraction are located farther away. Warm and hot \(\mathrm{H}_{\text{II}}\) contribute negligibly at all distances, whereas warm \(\mathrm{H}_{\text{I}}\) contributes only a small number (fewer than 10) of detectable sources. The dominant phase changes with distance. Within \(2\) kpc, cold \(\mathrm{H}_{\text{I}}\) provides the largest contribution, with roughly \(200\) BHs. Beyond \(2\) kpc, its cumulative contribution grows more slowly and is overtaken by that of MCs at approximately \(4\) kpc.

Within 500 pc, MC-hosted BHs are relatively rare because MCs occupy only a small fraction of the local volume, despite their high densities. Their contribution increases by roughly three orders of magnitude from 500 pc to 10 kpc. Consequently, MCs contain more than \(90\%\) of all BHs detectable within 10 kpc. This trend reflects both the increasing line-of-sight abundance of MCs and the larger underlying IBH population toward the Galactic center (Figures~\ref{fig:filling fraction} and \ref{fig:IBH_R_z}). As an illustration, a \(10\,M_{\odot}\) BH at \(10\,\mathrm{kpc}\), accreting at \(\dot{M}_{\rm BH}\sim10^{14}\,\mathrm{g\,s^{-1}}\) with \(\eta\sim10^{-3}\) (Equation~\ref{eq:eta}), would produce a flux of order \(10^{-14}\,\mathrm{erg\,s^{-1}\,cm^{-2}}\). More than \(6\times10^2\) synthetic IBHs have accretion rates above this illustrative value.

\begin{figure}
	\includegraphics[width=\columnwidth]{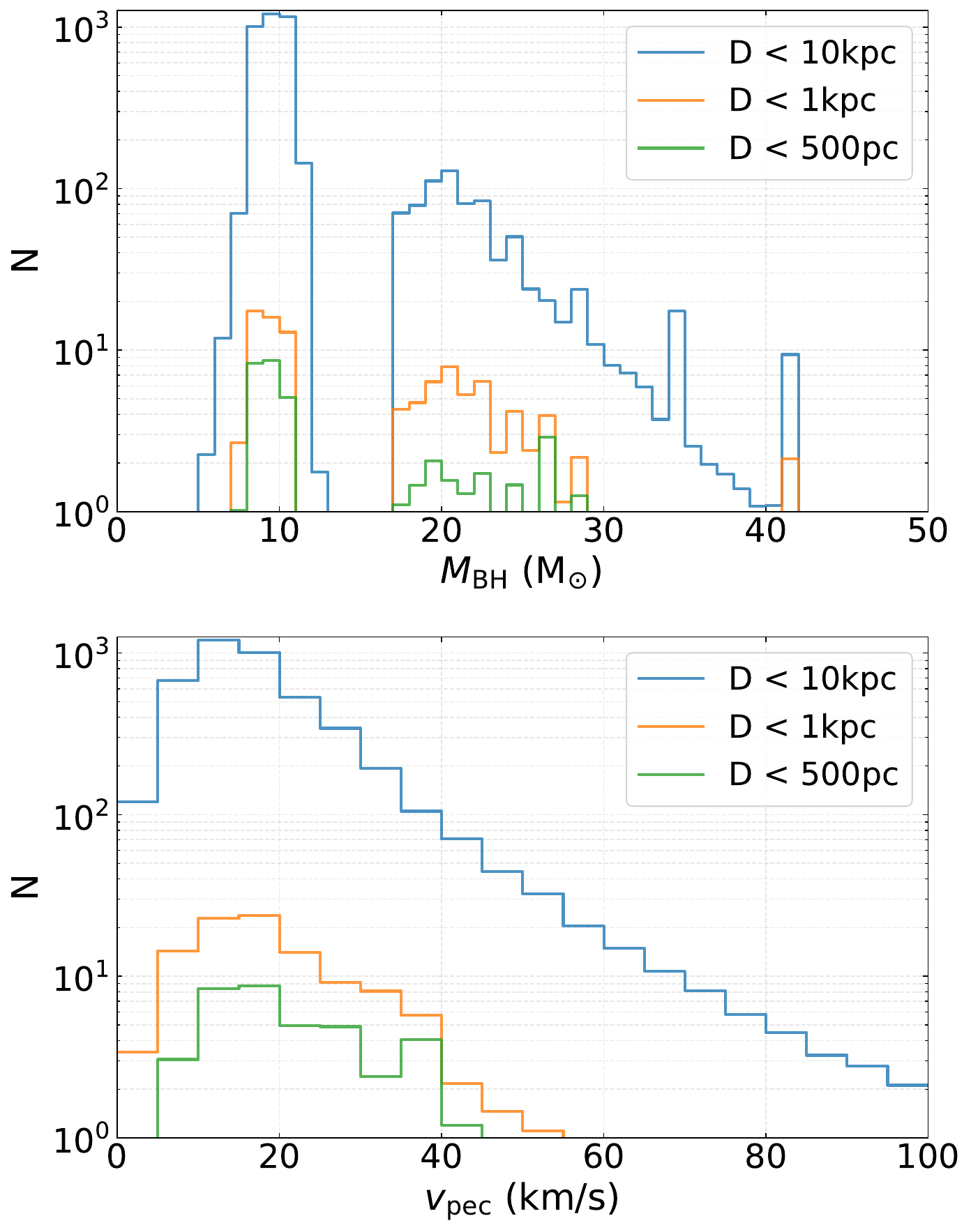}
    \caption{Mass (top) and peculiar-velocity (bottom) distributions of accreting IBHs with \(F_{\rm X}>10^{-14}\,\mathrm{erg\,s^{-1}\,cm^{-2}}\). Curves show samples within 500 pc, 1 kpc, and 10 kpc of the Sun.} 
    \label{fig:IBH_ism_mass_pec}
\end{figure}

Figure~\ref{fig:IBH_ism_mass_pec} shows the masses and peculiar velocities of IBHs with \(F_{\rm X}>10^{-14}\,\mathrm{erg\,s^{-1}\,cm^{-2}}\). The green, orange, and blue curves correspond to samples within 500 pc, 1 kpc, and 10 kpc, respectively. In all three samples, most detectable BHs have peculiar velocities of \(10- 30\,\mathrm{km\,s^{-1}}\), consistent with the non-kicked population shown in Figure~\ref{fig:IBH_dynamical_no_kick}. Their mass distribution likewise retains the characteristic two-peak structure of BHs formed without natal kicks (Figure~\ref{fig:IBH_mass}).

Within \(1\) kpc, the two mass peaks contain comparable numbers of IBHs, indicating a substantial contribution from the higher-mass non-kicked population associated with the outer Galactic disk. Within \(10\) kpc, by contrast, accreting IBHs are dominated by the peak near \(9\,M_{\odot}\), owing to stronger stellar-wind mass loss in the more metal-rich Galactic environment. This radial variation reflects the changing mass distribution of non-kicked IBHs across Galactic regions (Figure~\ref{fig:IBH_overview_no_kick}), which arises from regional differences in the chemical environment and hence in the typical masses of newly formed BHs.

Nearly all detectable IBHs (\(\simeq99\%\)) are predicted to have formed without natal kicks. The nearby sample within 1 kpc is therefore not fully representative of the intrinsic non-kicked population, because the local region is weighted toward a relatively low-metallicity formation history (Figure~\ref{fig:Metallicity}). The sample within 10 kpc more closely traces that intrinsic distribution.

\begin{deluxetable*}{cccccc}
\caption{Predicted numbers of accreting Galactic IBHs detectable within 500 pc, 1 kpc, and 10 kpc in representative X-ray bands of \textit{Chandra}, \textit{NuSTAR}, and \textit{eROSITA}, under the adopted spectral and radiative-efficiency assumptions.}
\label{tab:detector}
\tablehead{
\colhead{Detector} & \colhead{Band} & \colhead{Normalization factor} & \colhead{Sensitive} & \colhead{Exposure time} & \colhead{Number (500pc/1kpc/10kpc)}  
}
\startdata
	Chandra  &   0.2 - 10 keV   & 0.566 &   $4 \times 10^{-15} \mathrm{~erg \, s^{-1} \, cm^{-2}}$  &   $10^5 ~ \rm s$     &  44/129/5241
	\\  \hline
	\multirow{2}{*}{NuSTAR} &   6 - 10 keV  & 0.074  &      $2 \times 10^{-15} \mathrm{~erg \, s^{-1} \, cm^{-2}}$     & $10^6 ~ \rm s$         &     29/65/2487  
	\\ 
        &   10 - 30 keV  & 0.159 &      $1 \times 10^{-14} \mathrm{~erg \, s^{-1} \, cm^{-2}}$    &  $10^6 ~ \rm s$   &      19/38/1379  
    \\ \hline
    \multirow{2}{*}{eROSITA} &   0.5 - 2 keV & 0.201 & $4.4 \times 10^{-14} \mathrm{~erg \, s^{-1} \, cm^{-2}}$   &   $250 ~ \rm s$     &   10/18/564
    \\ 
    	& 2 - 10 keV & 0.233 &  $7.1 \times 10^{-13} \mathrm{~erg \, s^{-1} \, cm^{-2}}$       &     $250 ~ \rm s$      &      2/4/128       
    \\
\enddata
\end{deluxetable*}

To estimate detection numbers for high-energy observatories such as \textit{Chandra}\footnote{\url{https://chandra.harvard.edu/resources/handouts/lithos/chandraA_low.pdf}}, \textit{NuSTAR} \citep{NuSTAR}, \textit{eROSITA} \citep{eROSITA}, and \textit{XRISM} \citep{XRISM}, we also need to assume a spectral shape. Because most systems satisfy \(f_{\dot{M}}<10^{-3}\), corresponding to the hard state of BH binaries at low Eddington ratios, we adopt a nonthermal power-law spectrum, \(F_\nu \propto \nu^{-\alpha_2}\). Observationally, \(\alpha_2\) typically lies between \(0.4\) and \(1.1\) \citep{Remillard2006}. The choice of \(\alpha_2\) changes the predicted number of detectable IBHs by at most a factor of about 2 \citep{Tsuna2018}; following \citet{Agol2002}, we therefore set \(\alpha_2=1\), for which the luminosity per logarithmic photon-energy interval is constant.

Table~\ref{tab:detector} lists the predicted numbers of accreting IBHs detectable in several representative X-ray bands. Within the M25(0.50) model, these values should be regarded as optimistic upper limits. Alternative supernova prescriptions can substantially alter the total Galactic IBH population \citep[see also][]{wagg2026} and, consequently, the number of potentially detectable accreting black holes. The predicted numbers also depend sensitively on the adopted radiative-efficiency prescription. Equation~(\ref{eq:eta}) assumes that electrons receive one half of the energy dissipated viscously in the accretion flow, consistent with detailed modeling of Sgr A* \citep{Yuan2003}. If the electron-heating fraction is smaller, the radiative efficiency at \(f_{\dot{M}}<10^{-4}\) could decrease by one to two orders of magnitude, substantially reducing the predicted numbers. These estimates further assume a quasi-steady accretion flow and do not include temporal variability associated with accretion-disk instabilities. The importance of such variability depends on the structure of the accretion flow and is expected to be limited to a small subset of the most rapidly accreting IBHs. We discuss this issue, together with the prospects for radio detection of low-accretion-rate IBHs, in Section~\ref{sec:discussion_radio}.

The predicted number of sources above a flux threshold does not by itself determine how many IBHs can be identified. The Galactic X-ray sky contains many contaminants, particularly toward the Galactic center, where NSs and other accreting systems are abundant. Separating IBHs from these sources will require detailed spectral modeling and multiwavelength follow-up \citep[e.g.,][]{Murchikova2025,Nosirov2026,Koshimizu2026}; source identification therefore remains a major challenge for future observations.

\section{Gravitational Microlensing of IBHs} \label{sec:lens-BH}
 
\subsection{OGLE-2011-BLG-0462} \label{sec:OGLE-2011}

OGLE-2011-BLG-0462 is the first confirmed IBH identified through microlensing and provides a rare opportunity to probe BH formation and SN physics. We refer to it below as the lens BH. According to \citet{Sahu2025}, it has a mass of \(7.15 \pm 0.83\,M_{\odot}\), lies at a distance of \(1.52 \pm 0.15\,\mathrm{kpc}\), and is located close to the line of sight toward the Galactic center. Its transverse velocity relative to nearby stellar populations is \(51.1 \pm 7.5\,\mathrm{km\,s^{-1}}\). Backward orbit integration by \citet{Andrews2022} indicates that it can reach heights above \(600\,\mathrm{pc}\), consistent with a kinematic thick-disk origin; if it formed in the thin disk, a kick velocity of \(50-100\,\mathrm{km\,s^{-1}}\) would be required.

\begin{figure}
	\includegraphics[width=\columnwidth]{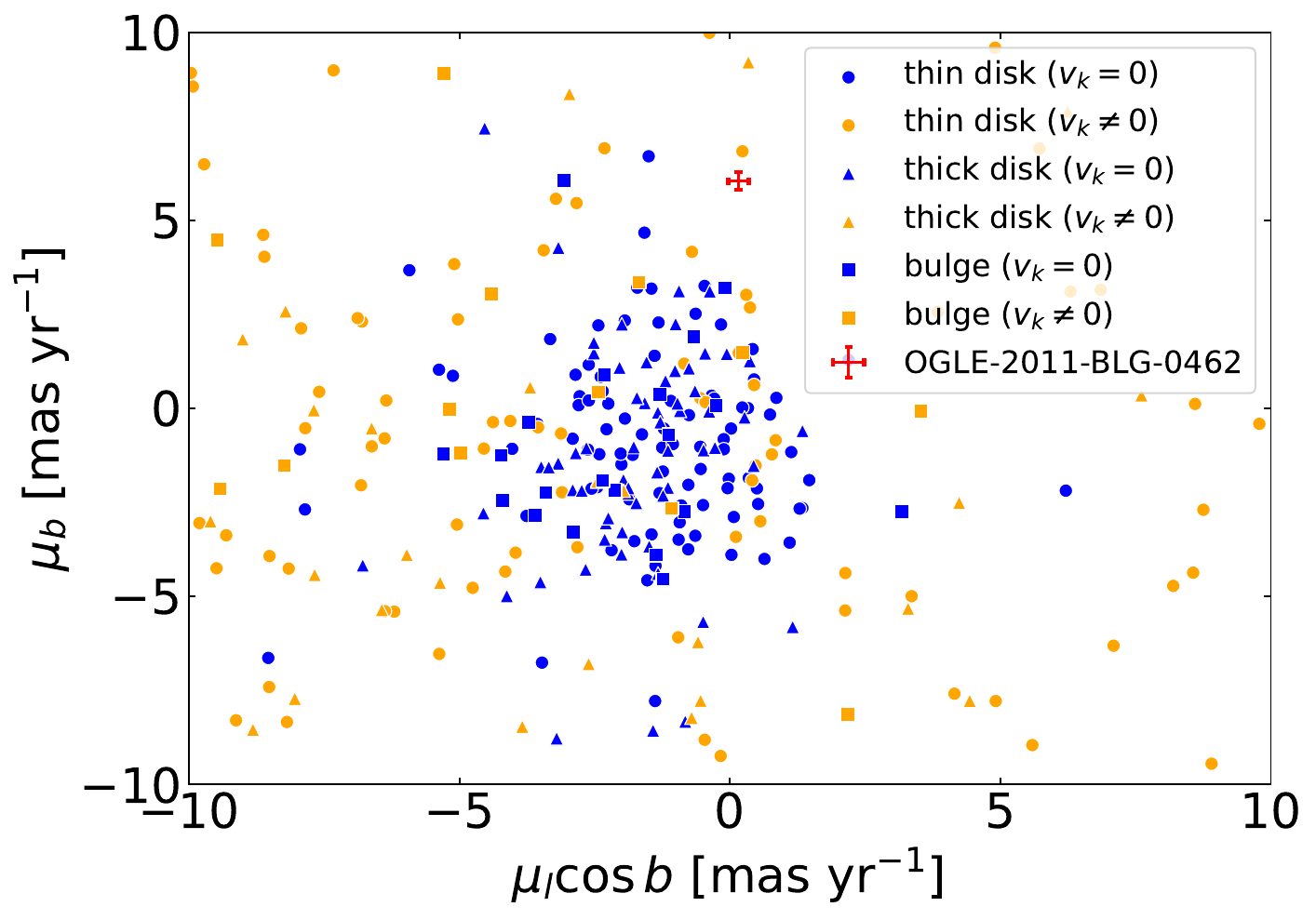}
    \caption{Proper-motion distribution of simulated IBHs selected to resemble OGLE-2011-BLG-0462, with \(6\,M_{\odot}<M_{\rm BH}<8.5\,M_{\odot}\), \(1.2\,\mathrm{kpc}<D<1.8\,\mathrm{kpc}\), \(|l|<20^{\circ}\), and \(|b|<5^{\circ}\). The red error bar marks the measured proper motion of OGLE-2011-BLG-0462 \citep{Sahu2025}.} 
    \label{fig:IBH_lens_ogle_2011_pm}
\end{figure}

To investigate the origin of the lens BH, we selected simulated BHs with comparable distances, sky positions, and masses. The selection requires \(1.2\,\mathrm{kpc}<D<1.8\,\mathrm{kpc}\), \(|l|<20^{\circ}\), \(|b|<5^{\circ}\), and \(6\,M_{\odot}<M_{\rm BH}<8.5\,M_{\odot}\). Figure~\ref{fig:IBH_lens_ogle_2011_pm} shows the proper-motion distributions in Galactic longitude and latitude; blue denotes BHs formed without natal kicks and orange denotes kicked BHs. Non-kicked BHs are concentrated mainly within \(-5\,\mathrm{mas\,yr^{-1}}<\mu_l\cos b<2\,\mathrm{mas\,yr^{-1}}\) and \(|\mu_b|<5\,\mathrm{mas\,yr}^{-1}\), whereas kicked BHs occupy a substantially broader region. Values of \(\mu_l\cos b<-5\,\mathrm{mas\,yr}^{-1}\) indicate counter-rotation, while \(\mu_l\cos b>2\,\mathrm{mas\,yr}^{-1}\) corresponds to motion faster than the local rotation curve. Similarly, \(|\mu_b|>5\,\mathrm{mas\,yr}^{-1}\) indicates substantial vertical motion. The measured proper motion of the lens BH, \((\mu_l\cos b,\mu_b)=(+0.158\pm0.190,+6.042\pm0.233)\,\mathrm{mas\,yr}^{-1}\) \citep{Sahu2025}, is therefore more consistent with the kicked population and a thin-disk origin. A thick-disk origin nevertheless remains possible, because thick-disk BHs can also have large \(|\mu_b|\) while moving toward the midplane.

\begin{figure}
	\includegraphics[width=\columnwidth]{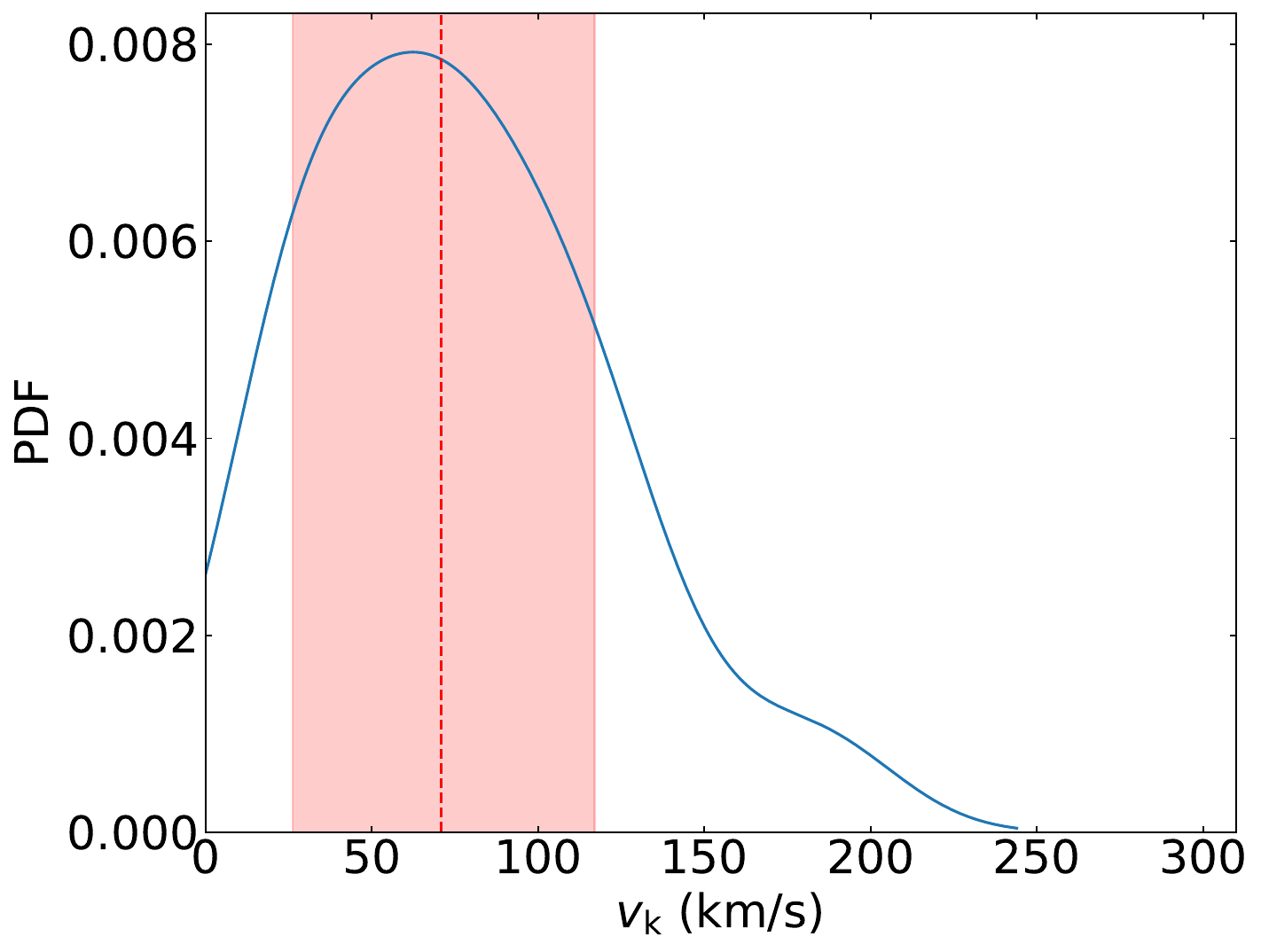}
    \caption{Natal-kick velocity PDF of simulated Galactic IBHs selected as analogs of OGLE-2011-BLG-0462. The shaded region marks the range between the weighted 16th and 84th percentiles, and the vertical dashed line indicates the weighted median.} 
    \label{fig:IBH_lens_ogle_2011_kick}
\end{figure}

To estimate the kick velocity compatible with the lens BH, we selected simulated BHs with similar positions and kinematic properties. The selection requires \(6\,\mathrm{kpc}<R<7\,\mathrm{kpc}\), \(|z|<0.1\,\mathrm{kpc}\), \(6\,M_{\odot}<M_{\rm BH}<8.5\,M_{\odot}\), \(220\,\mathrm{km\,s^{-1}}<v_T<245\,\mathrm{km\,s^{-1}}\), and \(40\,\mathrm{km\,s^{-1}}<|v_z|<60\,\mathrm{km\,s^{-1}}\). We impose no constraint on \(v_R\), because no radial-velocity measurement is available for the lens BH. Figure~\ref{fig:IBH_lens_ogle_2011_kick} shows the kick-velocity distribution of the selected kicked BHs. The inferred kick velocity lies within the 16th--84th percentile interval of the synthetic analogs, \(v_{\rm k}\simeq26-117\,\mathrm{km\,s^{-1}}\), whose weighted median is \(\simeq70.8\,\mathrm{km\,s^{-1}}\).

\subsection{OGLE-III-like Bulge Survey}  \label{sec:OGLE-III}

\begin{figure}
	\includegraphics[width=\columnwidth]{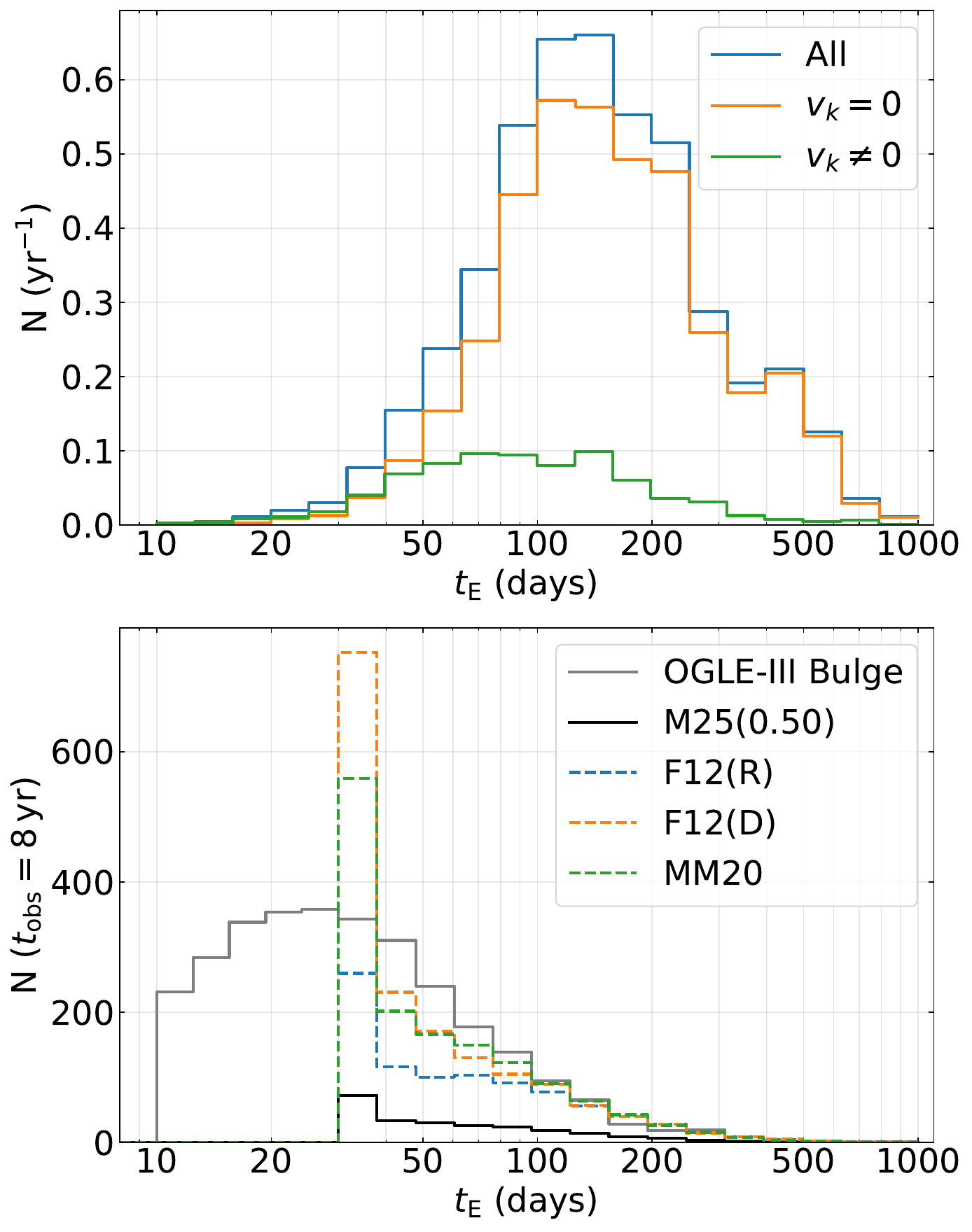}
    \caption{Microlensing timescale distributions in an OGLE-III-like bulge survey. The upper panel shows the predicted BH-lensing event rate in the fiducial M25(0.50) model, separated into IBHs formed without and with natal kicks. The lower panel compares the OGLE-III event distribution with the predicted eight-year total-event distributions for the M25(0.50), F12(R), F12(D), and MM20 SN prescriptions. No OGLE-III detection-efficiency correction is applied.}
    \label{fig:IBH_lens_ogle_iii}
\end{figure}

We compute the microlensing observables of the simulated IBHs with the implementation in \textsc{POPKIN}. Following \citet{Wiktorowicz2019}, we adopt the OGLE-III bulge footprint of \(31\,\mathrm{deg^2}\) and the \(1.5\times10^8\) monitored sources reported by \citet{Wyrzykowski2015}. We approximate the footprint by an elliptical region of the same area centered at \((l,b)=(0^\circ,-2^\circ)\), which preserves the survey area while approximately accounting for its asymmetric latitude coverage. Source stars are placed at \(D_{\rm S}=8\,\mathrm{kpc}\), and lenses are required to lie between the observer and the bulge.

For the fiducial M25(0.50) model, we predict a geometric BH-lensing rate of \(\sim4.7\,\mathrm{yr}^{-1}\) within the OGLE-III-like footprint, corresponding to approximately \(25\) events over an eight-year survey. This is not simply eight times the one-year rate because, in the effective cross section of Equation~(\ref{eq:num_lens}), only the sweep term \(2\theta_{\rm E}\mu_{\rm rel}T_{\rm obs}\) scales linearly with survey duration; the static geometric term \(\pi\theta_{\rm E}^{2}\) is independent of \(T_{\rm obs}\). Because we do not include the OGLE-III detection efficiency, these values represent geometric expectations rather than predicted observed counts. The one-year rate is lower than the \(\sim14\,\mathrm{yr}^{-1}\) estimate of \citet{Wiktorowicz2019}, who adopted the rapid SN prescription. With F12(R), the predicted rate increases to \(\sim20\,\mathrm{yr}^{-1}\), comparable to but somewhat higher than their estimate. The difference may reflect our inclusion of the single-star evolution channel and Galactic orbital evolution, both omitted from their simulation.

The upper panel of Figure~\ref{fig:IBH_lens_ogle_iii} shows that the predicted BH-lensing events are concentrated at long timescales, with non-kicked IBHs providing most of the contribution. Their Einstein crossing times are generally longer than those of kicked IBHs because they tend to be more massive, which increases \(\theta_{\rm E}\), and to move more slowly, which reduces \(\mu_{\rm rel}\). Both effects increase \(t_{\rm E}=\theta_{\rm E}/\mu_{\rm rel}\).

To compare with the observed OGLE-III bulge timescale distribution, we use the class-A standard microlensing events in the catalog of \citet{Wyrzykowski2015}\footnote{\url{https://ogle.astrouw.edu.pl/}}. We then use the BH-lens fraction as a function of \(t_{\rm E}\) from the Mock EWS simulation of \citet{Lam2020} to convert our predicted BH-lens counts into an approximate total-event distribution. Specifically, in each \(t_{\rm E}\) bin, we divide the predicted BH-lens counts by this fraction. This conversion is approximate because the fraction was derived for a different compact-object population and assumes a fixed BH natal kick of \(100\,\mathrm{km\,s^{-1}}\) with random direction. It therefore does not fully capture the mixture of non-kicked and kicked IBHs, the fallback-dependent distribution of natal-kick magnitudes, the dependence on formation channel, or the effects of Galactic orbital evolution included in our calculation

The lower panel of Figure~\ref{fig:IBH_lens_ogle_iii} compares the OGLE-III timescale distribution with predictions from four SN prescriptions\footnote{The comparison is restricted to \(t_{\rm E}\gtrsim30\,\mathrm{days}\), because the BH-lens fraction adopted from \citet{Lam2020} is nearly zero at shorter timescales. The total lens-event rate cannot therefore be reconstructed reliably in this regime; this is a methodological limitation, not evidence for a physical absence of short-timescale events.}. For \(t_{\rm E}\gtrsim30\,\mathrm{days}\), the fiducial M25(0.50) model predicts approximately four times fewer events than observed by OGLE-III. Its predicted counts remain below the observed distribution over most of the reliably reconstructable range, while still producing a tail extending to \(t_{\rm E}\gtrsim150\,\mathrm{days}\).

The predictions depend strongly on the adopted SN prescription. F12(R) gives higher event counts and is closer to the OGLE-III distribution near \(t_{\rm E}\sim30-40\,\mathrm{days}\), whereas F12(D) and MM20 produce an excess in this interval. None of the four prescriptions reproduces the observed distribution over the full available timescale range. The SN prescription is therefore an important source of uncertainty in the predicted BH-lensing rate and timescale distribution. Other sources of uncertainty include the adopted BH-lens fraction, the simulated spatial and kinematic distributions, and the OGLE-III selection function. Because detection efficiency is not included, this comparison should be regarded as a consistency check rather than a direct fit to the OGLE-III data. 

\subsection{Prediction for a Roman-like Survey} \label{sec:Roman}

\begin{deluxetable}{ccc}
\caption{Source-star samples used for the Roman-like GBTDS calculation. The star counts are evaluated in each distance bin for a simulated field of area $0.01\pi~{\rm deg}^{2}$.}
\label{tab:gbtds_region}
\tablehead{
\colhead{GBTDS Region} & \colhead{Distance} & \colhead{Number of source stars} \\
\colhead{} & \colhead{(kpc)} & \colhead{per $0.01\pi~{\rm deg}^{2}$}
}
\startdata
\multirow{6}{*}{\shortstack{North \\ $-0.2^\circ < l < 0.2^\circ$ \\ $-0.5^\circ < b < 0.3^\circ$}}  
 & (2, 4) & $7.30 \times 10^{4}$ \\
 & (4, 6) & $1.39 \times 10^{6}$ \\
 & (6, 8) & $1.07 \times 10^{7}$ \\
 & (8, 10) & $1.65 \times 10^{7}$ \\
 & (10, 12) & $5.20 \times 10^{6}$ \\
 & (12, 14) & $1.19 \times 10^{6}$ \\
\tableline
\multirow{6}{*}{\shortstack{South \\ $-0.6^\circ < l < 1.4^\circ$ \\ $-1.6^\circ < b < -0.8^\circ$}}  
 & (2, 4) & $6.72 \times 10^{4}$ \\
 & (4, 6) & $1.43 \times 10^{6}$ \\
 & (6, 8) & $1.07 \times 10^{7}$ \\
 & (8, 10) & $1.43 \times 10^{7}$ \\
 & (10, 12) & $3.50 \times 10^{6}$ \\
 & (12, 14) & $5.66 \times 10^{5}$ \\
\enddata
\end{deluxetable}

We next estimate the intrinsic microlensing yield of IBHs in a Roman-like Galactic Bulge Time Domain Survey \citep[GBTDS,][]{Penny2019,Lam2023b}. We consider two representative regions near the planned Roman bulge footprint. The northern region lies closer to the Galactic center, with \(-0.2^\circ<l<0.2^\circ\) and \(-0.5^\circ<b<0.3^\circ\), whereas the southern region covers \(-0.6^\circ<l<1.4^\circ\) and \(-1.6^\circ<b<-0.8^\circ\).

For the background source population, we use Galaxia via PopSyCLE \citep{Lam2020} to generate stellar catalogs in the two regions. Following \citet{Kaczmarek2026}, we adopt the optimistic magnitude cut \(F146<23\) to define potential Roman source stars. We divide the sources into six distance bins: \(2-4\), \(4-6\), \(6-8\), \(8-10\), \(10-12\), and \(12-14\) kpc. The corresponding source counts are listed in Table~\ref{tab:gbtds_region}.

For each distance bin, we combine the foreground IBH population with the background source stars and compute the intrinsic microlensing yield over a survey duration of \(5\,\mathrm{yr}\). We normalize the result to an effective Roman-like survey area of approximately \(1.70\,\mathrm{deg}^{2}\). In the fiducial model, we obtain \(356\) intrinsic IBH-lensing events: \(122\) from the northern region and \(234\) from the southern region. Although the southern region covers about five times the area of the northern region, the northern contribution remains substantial because the fields closer to the central bulge contain both more bright source stars and more foreground IBHs. The northern region is therefore expected to contribute a considerable fraction of the lensing events, particularly early in the survey.

\begin{figure}
    \includegraphics[width=\columnwidth]{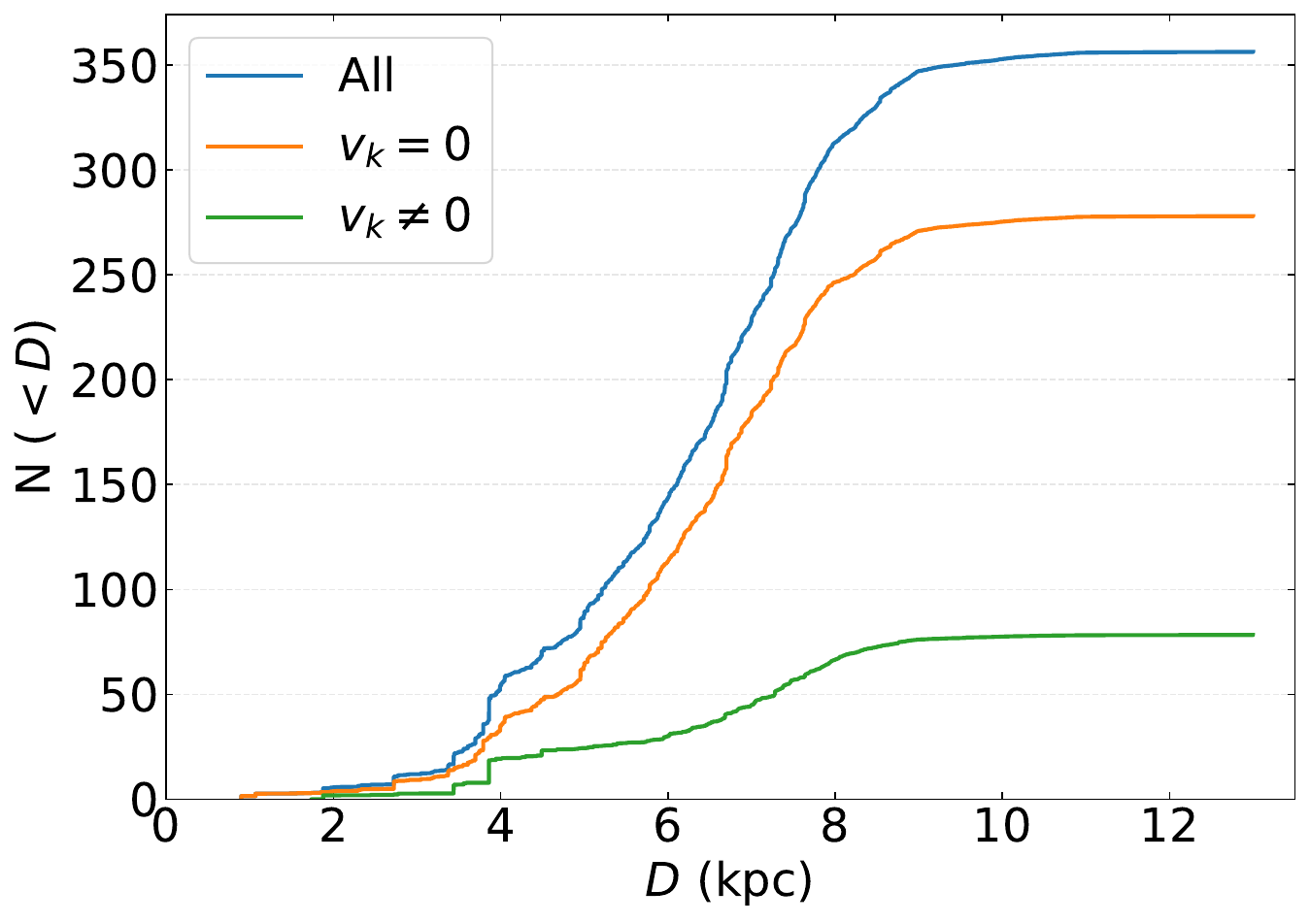}
    \caption{Cumulative distance distribution of intrinsic IBH lensing events in the Roman-like survey region. The blue curve shows all IBH lenses, while the orange and green curves separate IBHs formed without and with natal kicks, respectively.}
    \label{fig:IBH_lens_roman_dist}
\end{figure}

Figure~\ref{fig:IBH_lens_roman_dist} shows that the cumulative number of IBH-lensing events increases most steeply over \(D\simeq5-9\,\mathrm{kpc}\), where the line of sight intersects the densest bulge and inner-disk regions. The non-kicked population dominates the predicted Roman yield at nearly all distances, contributing approximately \(80\%\) of the events. Kicked IBHs contribute the remaining \(\sim20\%\), a non-negligible but clearly subdominant fraction. The relative contributions may vary substantially with the adopted SN prescription.

\begin{figure}
    \includegraphics[width=\columnwidth]{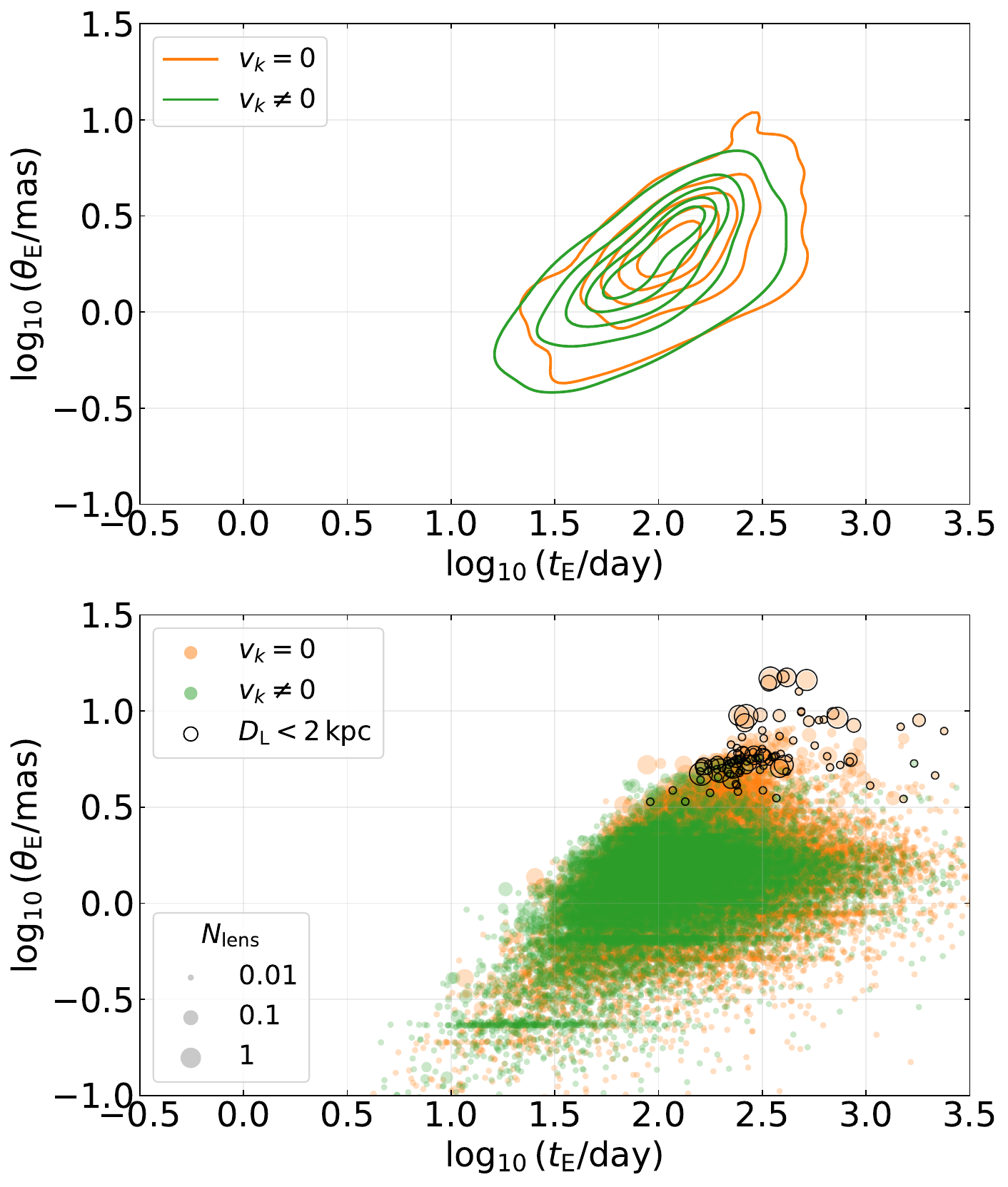}
    \caption{Roman-like IBH lensing events in the $t_{\rm E}-\theta_{\rm E}$ plane. The upper panel shows weighted kernel density estimation contours, and the lower panel shows individual lensing events, with marker sizes proportional to $N_{\rm lens}$. Orange and green denote IBHs formed without and with natal kicks, respectively; black open circles highlight lenses within $2\,{\rm kpc}$.}
    \label{fig:IBH_lens_roman_tE_theta_E}
\end{figure}

Figure~\ref{fig:IBH_lens_roman_tE_theta_E} compares the event properties of the two IBH populations. They overlap strongly in the \(t_{\rm E}-\theta_{\rm E}\) plane, but non-kicked systems are somewhat more concentrated toward longer timescales and larger angular Einstein radii, whereas kicked IBHs extend preferentially toward shorter timescales and smaller \(\theta_{\rm E}\). The lower panel further shows that the most extreme upper-right events are predominantly produced by nearby lenses with \(D_{\rm L}<2\,\mathrm{kpc}\), highlighting the role of lens distance in generating large angular Einstein radii.

The predicted yield of \(356\) events in the fiducial M25(0.50) model is approximately twice the detectable-event yield reported by \citet{Kaczmarek2026}. This difference is expected because our estimate is an intrinsic event yield rather than a detection-level forecast. We do not apply realistic Roman cadence, photometric or astrometric thresholds, blending cuts, finite-source effects, or event-recovery efficiencies \citep[e.g.,][]{Sajadian2023}. Including these selection effects can reduce the number of detectable events by factors of several to tens, as demonstrated by \citet{Kaczmarek2026}.

The predicted yield also depends strongly on the adopted SN prescription. F12(R), F12(D), and MM20 predict \(1571\), \(1793\), and \(1887\) events, respectively. These values should therefore be interpreted as optimistic upper limits for the corresponding prescriptions, rather than as expected numbers of detected events.

\section{Discussion} \label{sec:discussion}

\subsection{Influence of Population-Synthesis Parameters}

\begin{figure*}
	\centering
    \includegraphics[width=0.8\textwidth]{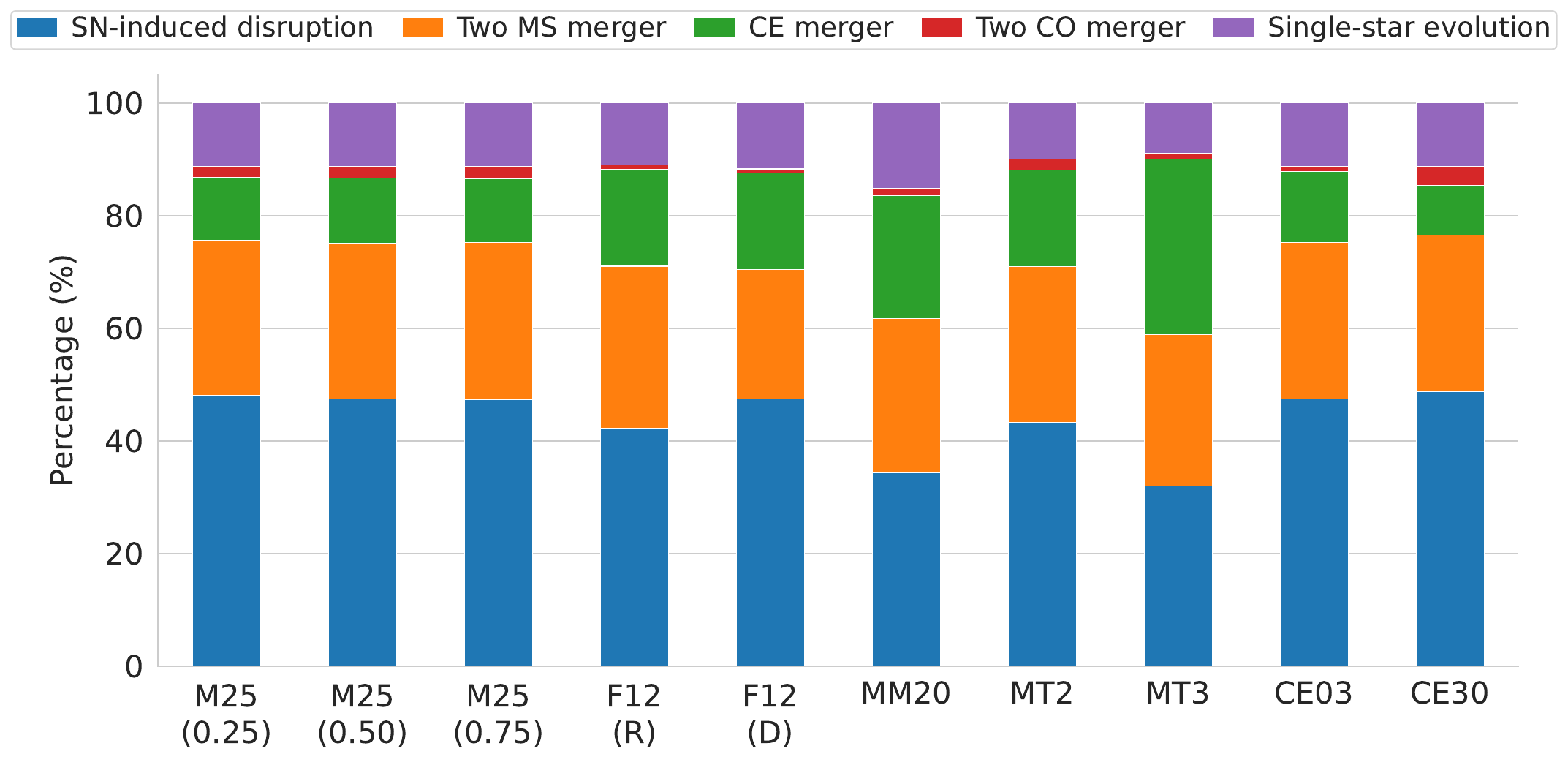}
    \caption{Formation-channel fractions of Galactic IBHs in the ten population-synthesis models considered in this work.} 
    \label{fig:IBH_formation_channel}
\end{figure*}

\begin{figure*}
	\centering
    \includegraphics[width=0.8\textwidth]{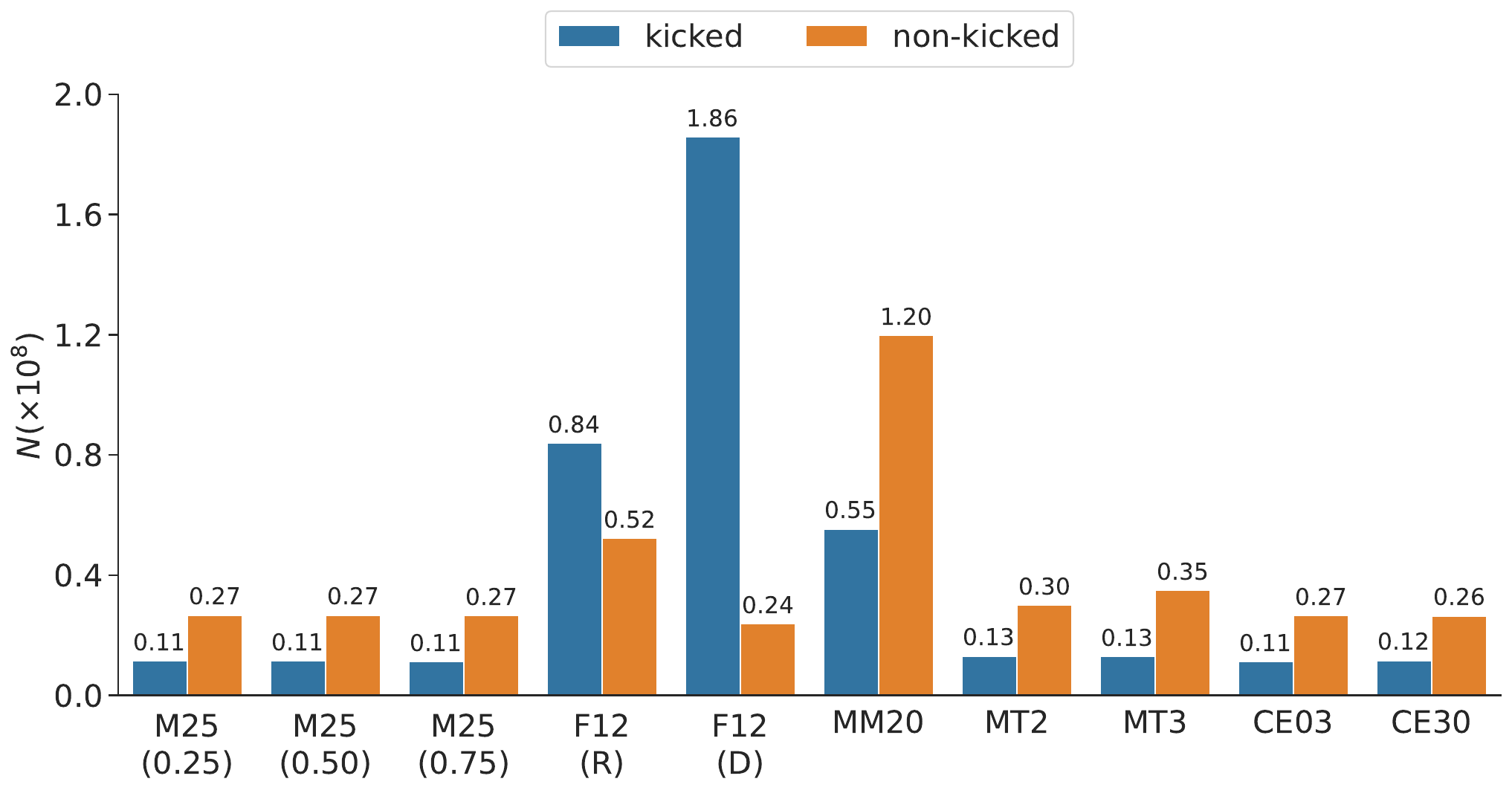}
    \caption{Predicted numbers of Galactic IBHs formed with and without natal kicks for the ten population-synthesis models considered in this work.} 
    \label{fig:IBH_kick_fraction}
\end{figure*}

\begin{figure*}
	\centering
	\includegraphics[width=0.8\textwidth]{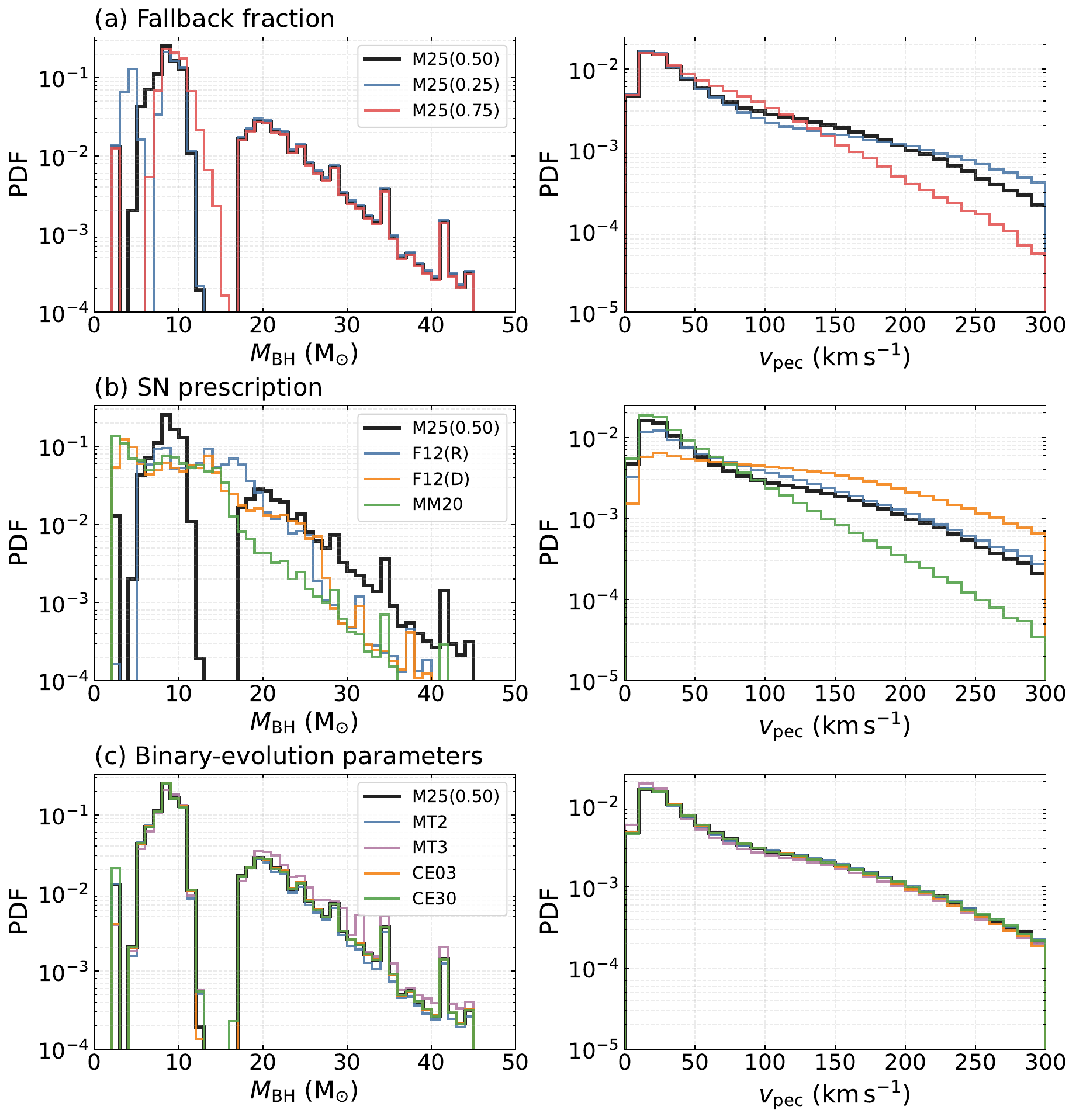}
    \caption{Normalized PDFs of the masses and present-day peculiar velocities of Galactic IBHs for the population-synthesis models considered in this work. The left and right columns show the distributions of \(M_{\rm BH}\) and \(v_{\rm pec}\), respectively. From top to bottom, the rows compare the three M25 models with different fallback fractions, alternative SN prescriptions, and binary-evolution parameters. The fiducial M25(0.50) model is included in each row for reference and is shown by the thicker black curve.}
    \label{fig:IBH_mass_pec_models}
\end{figure*}

To assess the sensitivity of the predicted IBH population to the adopted assumptions, we compare the ten population-synthesis models described in Section~\ref{subsec:physic}; their population-level results are summarized in Table~\ref{tab:IBH}. The comparison spans three types of variation: the fallback fraction \(f_{\rm fb}\) within the M25 model family, the SN prescription (M25, F12(R), F12(D), or MM20), and the binary-evolution parameters. Among these variations, the choice of SN prescription has the largest effect on the total Galactic IBH population. The three M25 models, which differ only in \(f_{\rm fb}\), each yield approximately \(4\times10^7\) IBHs, whereas the F12(R), F12(D), and MM20 models yield \(\sim(1\!-\!2)\times10^8\), consistent with previous population-synthesis studies \citep{Olejak2020,Lam2022}. In the F12(R), F12(D), and MM20 models, the relative contributions of the thin disk, thick disk, and bulge broadly follow the stellar-mass budget of these components. This behavior arises because these SN prescriptions do not couple remnant classification to metallicity as explicitly as \citet{Maltsev2025}.

Figures~\ref{fig:IBH_formation_channel} and~\ref{fig:IBH_kick_fraction} show, respectively, the formation-channel fractions and the numbers of IBHs formed with and without natal kicks in these ten models.

The kicked population dominates only in the F12(R) and F12(D) models; non-kicked IBHs are more numerous in the other models. This contrast is driven primarily by how the different SN prescriptions couple fallback to remnant masses and natal kicks. In our implementation of \citet{Mandel2020}, the BH mass is capped at the pre-SN helium-core mass, and the natal kick is set to zero whenever the remnant mass exceeds the carbon-oxygen core mass; most BHs therefore receive no natal kick. In the prescription of \citet{Maltsev2025}, fallback is assigned probabilistically within the intermediate carbon-oxygen core mass interval, with a fallback probability of \(10\%\) in our implementation. Even when the BH itself receives no natal kick, a binary can be disrupted by symmetric mass loss or by the subsequent SN of its companion; we classify such systems as forming through the SN-induced disruption channel. F12(R) and especially F12(D) produce a larger fraction of fallback BHs, and their fallback-dependent kick prescriptions consequently yield more IBHs with nonzero natal kicks. The kicked population therefore dominates in these two cases, while the delayed prescription also populates the mass gap between NSs and BHs.

Varying the binary-evolution parameters mainly alters the relative contributions of the merger formation channels. MT3 produces the largest contribution from the CE merger channel because thermal-timescale-limited accretion can approach conservative transfer, allowing the accretor to gain substantial mass and making the subsequent CE phase more difficult to survive. Conversely, CE30 enhances the two CO merger channel because its larger \(\alpha_{\rm CE}\) allows more binaries to survive CE evolution and eventually merge as two compact objects.

Figure~\ref{fig:IBH_mass_pec_models} compares the mass and present-day peculiar-velocity distributions of Galactic IBHs across the ten population-synthesis models.

The top row isolates the effect of varying \(f_{\rm fb}\) among the three M25 models. Below \(17\,M_{\odot}\), the mass distribution contains contributions from both fallback and direct-collapse BHs. Changing \(f_{\rm fb}\) alters the masses of fallback BHs and hence their overlap with the direct-collapse component. In the fiducial M25(0.50) model, the combined low-mass population peaks near \(9\,M_{\odot}\) and is concentrated approximately within \(5-12\,M_{\odot}\). For M25(0.75), the larger fallback fraction shifts the fallback contribution to higher masses, extending the combined distribution to approximately \(7-14\,M_{\odot}\) and partially filling the \(13-17\,M_{\odot}\) deficit. In contrast, M25(0.25) produces lighter fallback BHs that extend into the \(2-5\,M_{\odot}\) mass-gap region. The fallback and direct-collapse contributions then become more clearly separated, producing components near \(5\,M_{\odot}\) and \(9\,M_{\odot}\).

Among the M25 models, \(f_{\rm fb}\) has little effect on the low-velocity part of the distribution, which is dominated by direct-collapse BHs and is therefore insensitive to the fallback fraction. Its effect is stronger at high velocities because fallback reduces the natal-kick magnitude. In particular, the M25(0.25) model has a larger fraction of IBHs with \(v_{\rm pec}>200\,\mathrm{km\,s^{-1}}\), whereas this high-velocity tail is substantially weaker in the M25(0.75) model.

The middle row compares the alternative SN prescriptions F12(R), F12(D), and MM20 with the fiducial M25(0.50) model. Relative to M25(0.50), F12(R) retains a pronounced deficit in the \(2-5\,M_{\odot}\) mass-gap region, whereas F12(D) and MM20 populate this interval and shift the mass distribution toward lower BH masses. These latter two prescriptions also produce smoother mass distributions, without the pronounced deficit near \(15\,M_{\odot}\) seen in M25(0.50). The velocity distributions show corresponding differences: F12(D) produces the broadest high-velocity tail, consistent with its larger kicked fraction, whereas MM20 yields more low-velocity IBHs and is consequently more concentrated toward small \(v_{\rm pec}\).

The bottom row shows how variations in the binary-evolution parameters affect these distributions. Changing the mass-transfer efficiency or the CE ejection efficiency produces only modest changes in the BH mass distribution. MT3 yields a slightly larger fraction of higher-mass IBHs, but has little effect on the global \(v_{\rm pec}\) distribution compared with variations in \(f_{\rm fb}\) within M25 or with the SN prescription itself.

Overall, the global mass and peculiar-velocity distributions are most sensitive to the adopted SN prescription, including the choice of \(f_{\rm fb}\) within M25, whereas variations in the binary-evolution parameters produce comparatively minor changes.

\begin{figure}
	\includegraphics[width=\columnwidth]{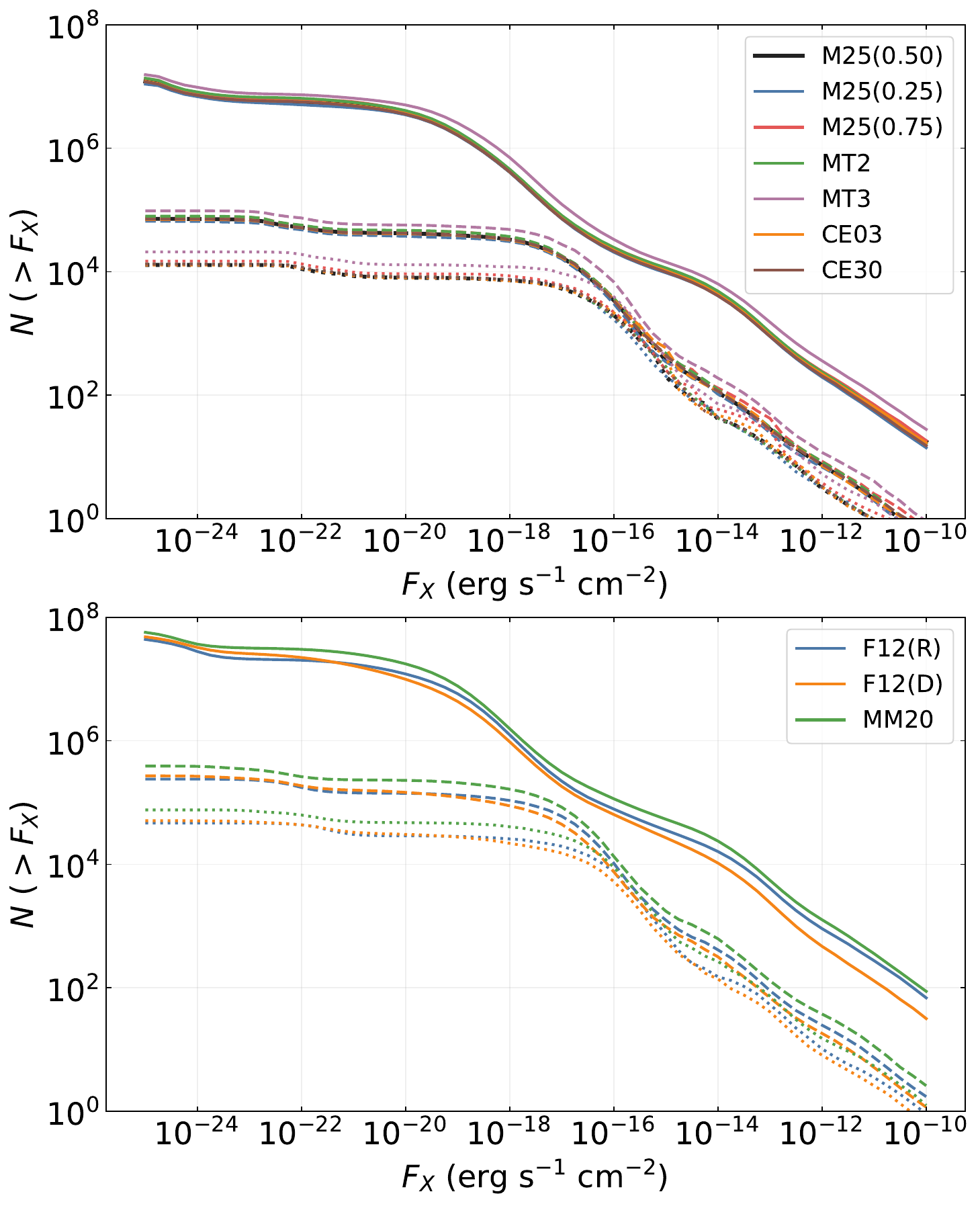}
    \caption{Cumulative numbers of accreting IBHs as a function of \(F_{\rm X}\) for the ten population-synthesis models. The upper panel compares the three M25 models with different fallback fractions and the variations in binary-evolution parameters, while the lower panel compares the F12(R), F12(D), and MM20 SN prescriptions. Line styles denote distance limits of \(500\,\mathrm{pc}\) (dotted), \(1\,\mathrm{kpc}\) (dashed), and \(10\,\mathrm{kpc}\) (solid).}
    \label{fig:IBH_F_models}
\end{figure}

Because these assumptions produce different BH-mass and peculiar-velocity distributions, we next examine their impact on the predicted detectability of Galactic IBHs. Figure~\ref{fig:IBH_F_models} shows the cumulative numbers of accreting IBHs as a function of X-ray flux for distance limits of \(500\,\mathrm{pc}\), \(1\,\mathrm{kpc}\), and \(10\,\mathrm{kpc}\).

The upper panel compares the three M25 models with different fallback fractions and the variations in binary-evolution parameters. At a fixed distance limit, these models produce broadly similar flux distributions. The main exception is MT3, which predicts a slightly larger number of accreting IBHs, consistent with its somewhat larger contribution from massive BHs. The lower panel compares the F12(R), F12(D), and MM20 SN prescriptions. These prescriptions generally predict more accreting IBHs than the \citet{Maltsev2025} prescription over the full flux range, particularly at low \(F_{\rm X}\). This difference primarily reflects their larger total Galactic IBH populations. Among these prescriptions, MM20 predicts the largest number of accreting IBHs. Its more prominent low-velocity population tends to have smaller relative velocities with respect to the ISM, which enhances the accretion rate and improves detectability.

By contrast, uncertainty in the radiative-efficiency prescription changes the predicted counts by orders of magnitude. Confirming an accreting IBH will therefore require improved emission models and multiwavelength follow-up to distinguish these objects from other Galactic X-ray sources \citep{Martinez2025}.

\subsection{Mass Dependence of \(v_{\rm pec}\)}

\begin{figure*}
    \centering
	\includegraphics[width=0.85\textwidth]{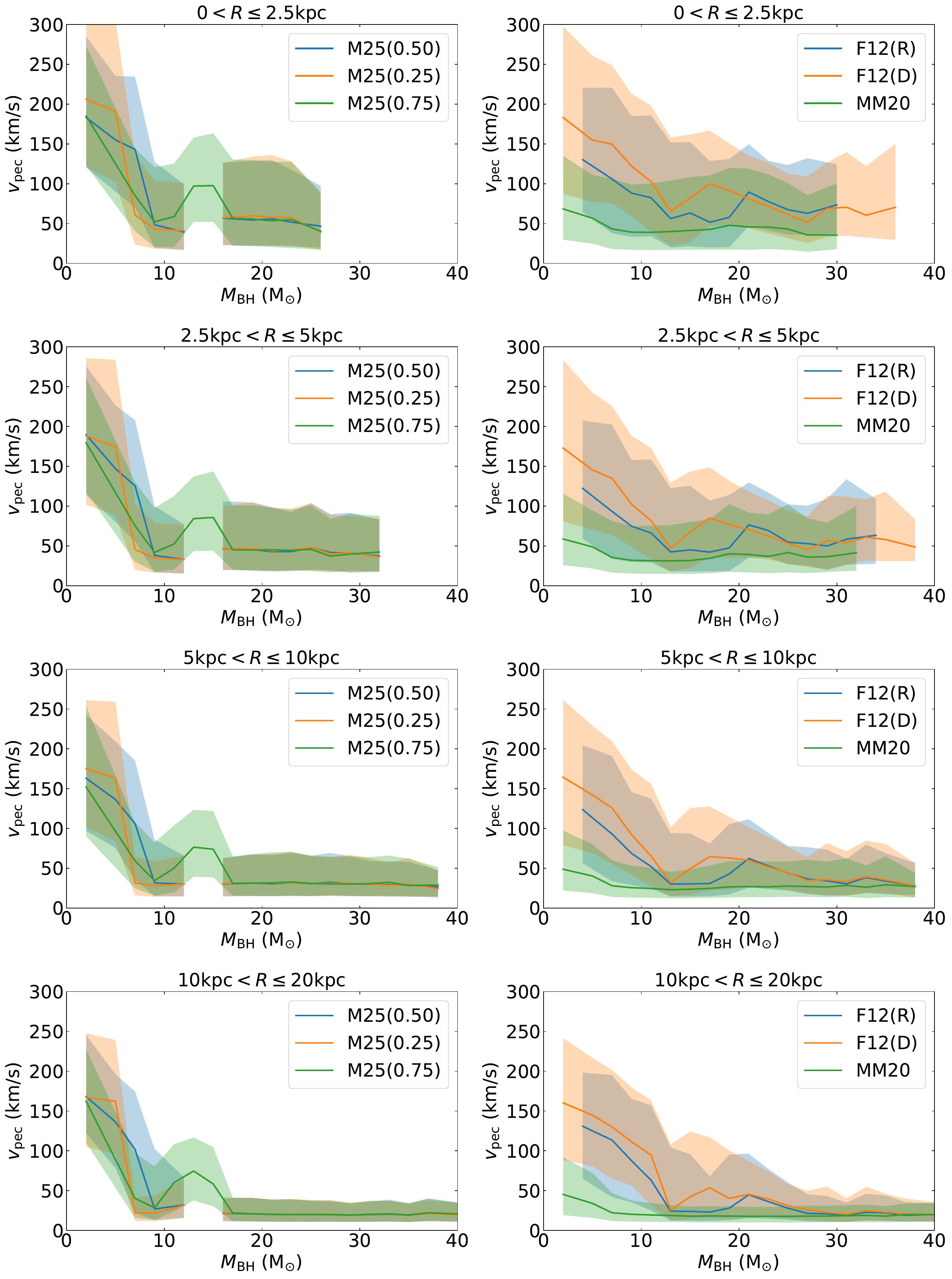}
    \caption{Mass dependence of the present-day peculiar velocity \(v_{\rm pec}\) of Galactic IBHs in four bins in Galactocentric radius. The left column compares the M25 models with fallback fractions of \(0.50\), \(0.25\), and \(0.75\), whereas the right column compares the F12(R), F12(D), and MM20 SN prescriptions. Curves show weighted medians, and shaded regions indicate weighted 16th--84th percentile ranges. Mass bins containing fewer than \(10^4\) weighted IBHs are omitted.}
    \label{fig:IBH_pec_SN}
\end{figure*}

The preceding section considered the mass and peculiar-velocity distributions separately, although both depend on the adopted SN prescription. In F12(R) and F12(D), fallback reduces natal kicks according to the fallback fraction, whereas MM20 assigns remnant masses and natal kicks stochastically. In each M25 model, fallback BHs share a fixed fallback fraction and hence the same fallback scaling of the kick distribution. Their present-day peculiar velocities can nevertheless depend on BH mass because different mass bins contain different mixtures of fallback and direct-collapse BHs, formation channels, and Galactic birth components. Figure~\ref{fig:IBH_pec_SN} shows this mass dependence in four bins in Galactocentric radius.

The left column compares the three M25 models with fallback fractions of \(0.50\), \(0.25\), and \(0.75\). The mass distribution can be divided approximately into three regimes: \(M_{\rm BH}<12\,M_{\odot}\), \(12-16\,M_{\odot}\), and \(M_{\rm BH}>16\,M_{\odot}\). The lowest-mass regime contains overlapping contributions from fallback and direct-collapse BHs. Its decreasing \(v_{\rm pec}\) toward higher BH masses reflects a transition from a population with a larger fallback-BH contribution, and hence nonzero natal kicks, to one increasingly dominated by direct-collapse BHs without kicks. This trend is most pronounced in M25(0.25), whose lower fallback fraction suppresses the natal kicks less efficiently and therefore produces larger peculiar velocities in the low-mass fallback population.

The intermediate \(12\!-\!16\,M_{\odot}\) regime is strongly depleted in M25(0.25) and M25(0.50), but is partially populated in M25(0.75). Increasing the fallback fraction shifts fallback BHs toward larger masses and consequently fills part of this interval. The high-mass component above \(16\,M_{\odot}\) consists predominantly of direct-collapse BHs and is therefore much less sensitive to the assumed fallback fraction. With increasing Galactocentric radius, the overall \(v_{\rm pec}\) decreases, while the high-mass component becomes relatively more prominent. These trends reflect the changing mixture of Galactic birth components and metallicities across the radius bins.

The right column compares the F12(R), F12(D), and MM20 SN prescriptions. MM20 produces the lowest peculiar velocities for BHs below approximately \(10\,M_{\odot}\), with weighted medians generally below \(50\,\mathrm{km\,s^{-1}}\). The F12 prescriptions produce larger velocities in this mass range. In both F12(R) and F12(D), \(v_{\rm pec}\) generally decreases with increasing BH mass at the low-mass end. This trend is consistent with the increasing fallback fraction at larger carbon-oxygen core masses, which suppresses natal kicks more efficiently. F12(D) produces the broadest high-velocity distribution, consistent with its larger fraction of kicked BHs.

At higher masses, the F12(R) and F12(D) curves exhibit several non-monotonic features. These structures result from the combined effects of wind-driven changes in the pre-SN structure and the SN remnant-mass prescription. The Merritt wind can produce a non-monotonic mapping between the ZAMS mass, pre-SN core masses, and final remnant mass. For example, at \(Z=0.01\), the onset of LBV-like mass loss near \(M_{\rm ZAMS}\simeq36\,M_{\odot}\) rapidly strips the hydrogen-rich envelope and alters the subsequent pre-SN structure (see Figure~\ref{fig:stellar_mass_relation}). The F12 prescriptions map these structural changes into variations in the fallback fraction, natal kick, and remnant mass, producing a complex \(M_{\rm BH}\)-\(v_{\rm pec}\) relation. The detailed pattern is further shaped by the mixture of progenitor metallicities, Galactic components, and formation channels.

Overall, the mass dependence of \(v_{\rm pec}\) mainly reflects the relative contributions of fallback and direct-collapse BHs at different masses, as determined by the adopted SN prescription. Mass bins with a larger fallback-BH contribution generally have higher \(v_{\rm pec}\), whereas those dominated by direct-collapse BHs have lower values.


\subsection{Observable Signatures of SN Prescriptions}

\begin{figure*}
    \centering
    \includegraphics[width=0.8\textwidth]{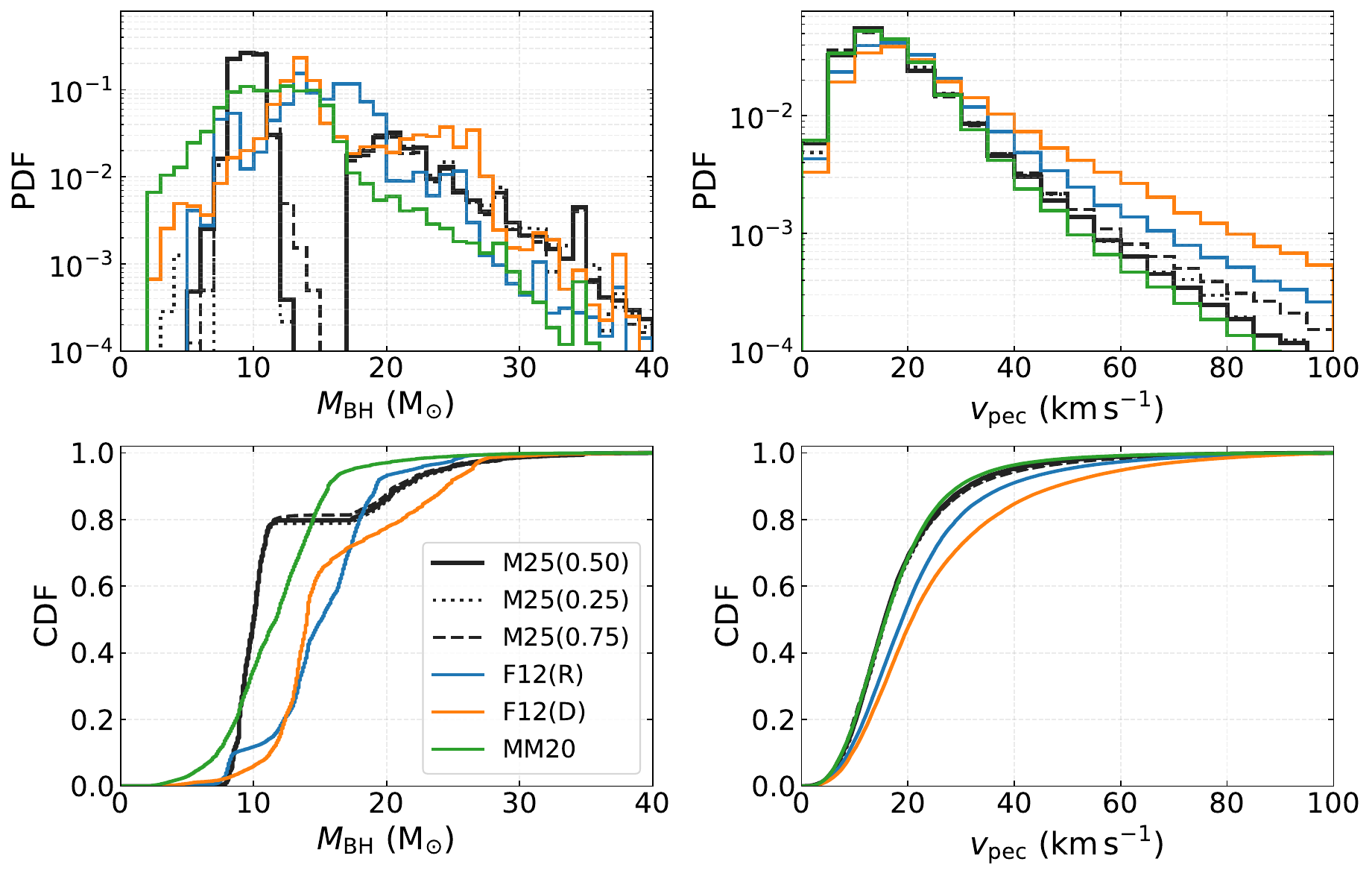}
    \caption{Mass and present-day peculiar-velocity distributions of X-ray-selected Galactic IBHs for six different SN models. The sample is selected with $F_{\rm X}\geq 10^{-14}\,{\rm erg\,s^{-1}\,cm^{-2}}$. The left and right panels show the distributions of $M_{\rm BH}$ and $v_{\rm pec}$, respectively. The upper panels show the normalized PDFs, while the lower panels show the corresponding CDFs.}
    \label{fig:IBH_Xray_mass_pec_SN_models}
\end{figure*}

The results presented above show that the adopted SN prescription can significantly affect the mass and peculiar-velocity distributions of IBHs. We now examine whether these signatures remain visible in observable samples of ISM-accreting and microlensing IBHs.

Figure~\ref{fig:IBH_Xray_mass_pec_SN_models} compares the mass and peculiar-velocity distributions of X-ray-selected IBHs satisfying \(F_{\rm X}\geq10^{-14}\,\mathrm{erg\,s^{-1}\,cm^{-2}}\). Across all six models with different SN-related assumptions, the velocity distributions are strongly concentrated at low \(v_{\rm pec}\), with typical values of \(v_{\rm pec}\sim10-30\,{\rm km\,s^{-1}}\). This concentration reflects the accretion selection, which favors IBHs with small velocities relative to the ISM. The selection therefore largely erases the differences in \(v_{\rm pec}\) produced by the natal-kick prescriptions. MM20 has a slightly weaker high-velocity tail, whereas F12(D) retains a somewhat larger high-velocity contribution. Varying the fallback fraction within the three M25 models has only a minor effect because the selected sample is dominated by low-velocity systems.

The mass distributions retain a clearer imprint of the SN prescription. The \(2\!-\!5\,M_{\odot}\) mass-gap component is largely absent from most X-ray-selected samples, but remains more visible in MM20. This difference may reflect MM20's ability to produce low-mass BHs with relatively low peculiar velocities, whereas mass-gap BHs in the other prescriptions are either less abundant or preferentially excluded by the velocity dependence of accretion. The pronounced \(12\!-\!17\,M_{\odot}\) deficit in the M25 models also remains visible, while F12 and MM20 produce smoother mass distributions. X-ray-selected IBHs alone may nevertheless provide only limited constraints on the SN prescription because the sample is strongly biased toward low velocities, and accreting IBHs are difficult to distinguish reliably from other Galactic X-ray sources and to characterize dynamically.

\begin{figure*}
    \centering
    \includegraphics[width=0.8\textwidth]{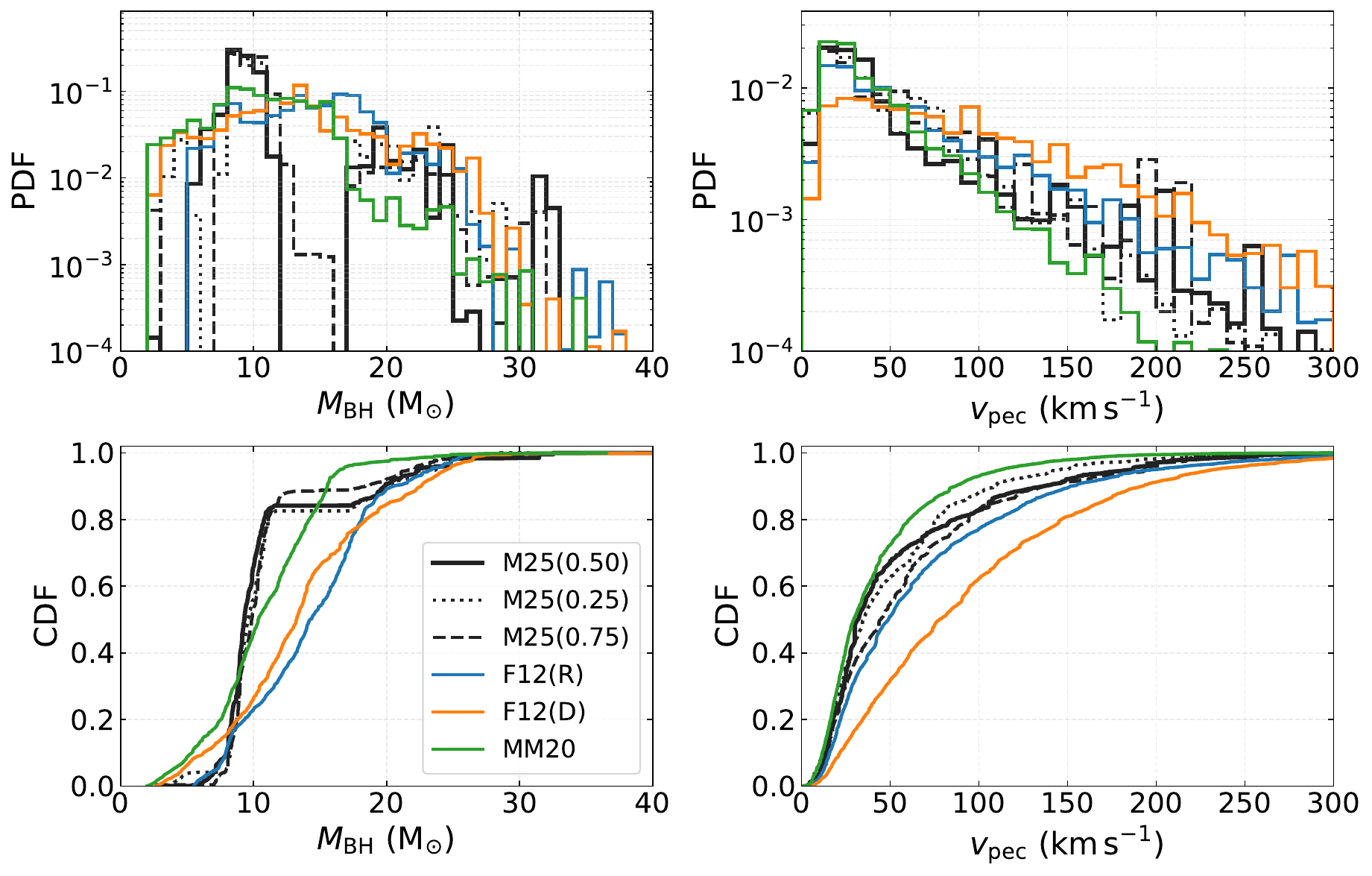}
    \caption{Similar to Figure~\ref{fig:IBH_Xray_mass_pec_SN_models}, but for long-timescale IBH microlensing events selected with \(t_{\rm E}>80\,\mathrm{days}\).}
    \label{fig:IBH_lens_mass_pec_SN_models}
\end{figure*}

Microlensing provides a complementary and less velocity-biased probe. Figure~\ref{fig:IBH_lens_mass_pec_SN_models} shows the mass and peculiar-velocity distributions of long-timescale IBH microlensing events selected with \(t_{\rm E}>80\,\mathrm{days}\). This selection favors events with larger Einstein radii and longer durations, for which parallax and astrometric signals are more likely to be measurable, thereby facilitating constraints on lens masses and kinematics.

The lens-mass distributions retain clear differences between the remnant-mass prescriptions. F12(D) and MM20 each yield a mass-gap BH fraction of order \(10\%\), whereas M25(0.25) yields a smaller fraction. The \(12\!-\!17\,M_{\odot}\) deficit remains particularly pronounced in the three M25 models. The lens velocity distributions also show stronger differences than the X-ray-selected samples: MM20 lenses are more concentrated toward low \(v_{\rm pec}\), whereas F12(D) produces the broadest high-velocity tail. In the cumulative distributions, approximately \(70\%\) of MM20 lenses have \(v_{\rm pec}<50\,\mathrm{km\,s^{-1}}\), compared with roughly \(30\%\) of F12(D) lenses. F12(R) and the M25 prescriptions generally yield intermediate distributions.

\begin{figure*}
    \centering
	\includegraphics[width=0.8\textwidth]{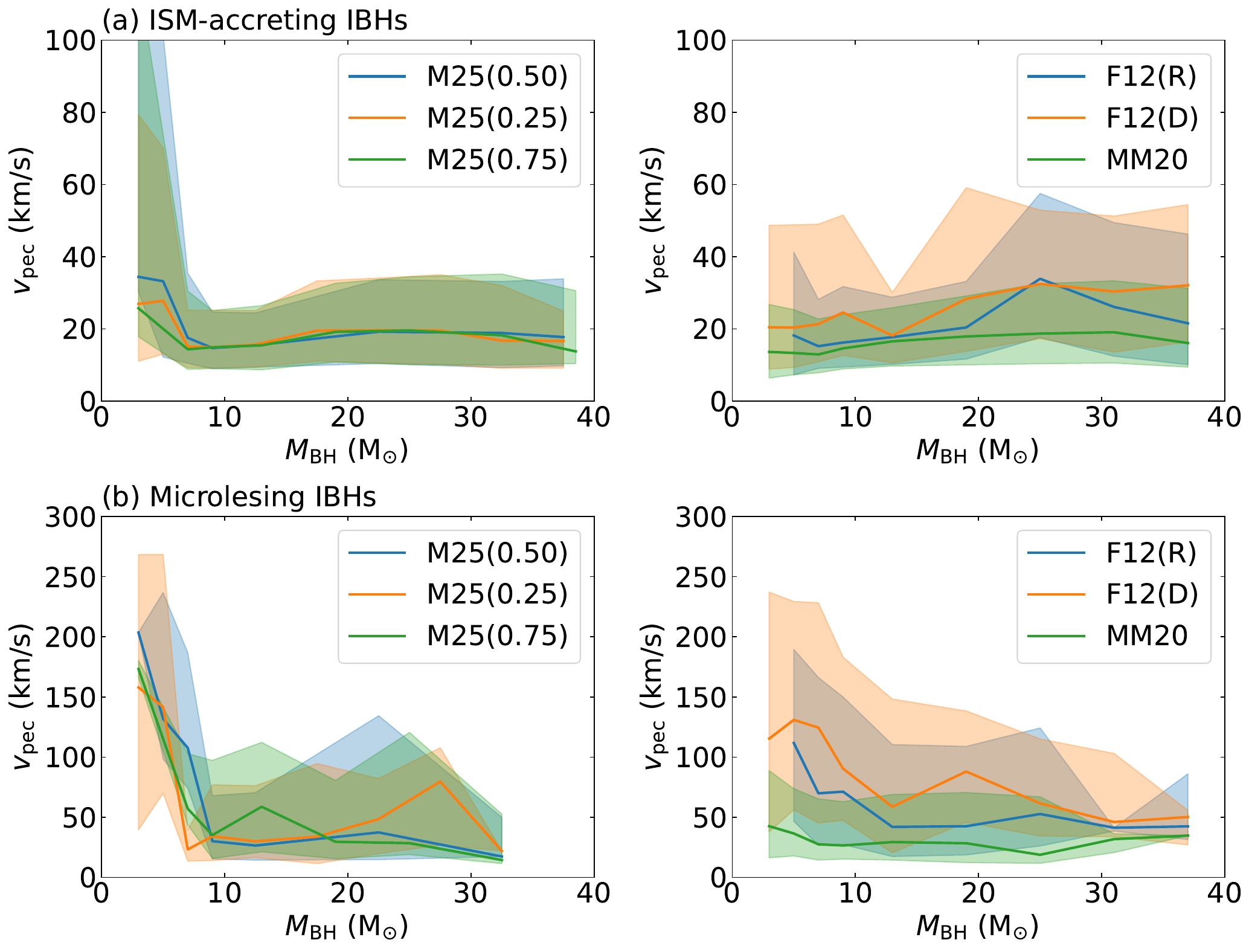}
    \caption{Mass dependence of the present-day peculiar velocity \(v_{\rm pec}\) for observable Galactic IBHs under different SN prescriptions. The upper and lower panels show ISM-accreting IBHs selected with \(F_{\rm X}\geq10^{-14}\,\mathrm{erg\,s^{-1}\,cm^{-2}}\) and long-timescale microlensing events selected with \(t_{\rm E}>80\,\mathrm{days}\), respectively. The left panel compares the three M25 models with fallback fractions of \(0.50\), \(0.25\), and \(0.75\), while the right panel compares the F12(R), F12(D), and MM20 prescriptions. Curves show the weighted median \(v_{\rm pec}\), and shaded regions indicate the weighted 16th--84th percentile ranges.}
    \label{fig:IBH_pec_SN_observation}
\end{figure*}

Figure~\ref{fig:IBH_pec_SN_observation} directly examines the conditional dependence of \(v_{\rm pec}\) on \(M_{\rm BH}\) in the two observable populations. All mass bins are retained in this figure, including bins with small weighted numbers of systems. The corresponding medians and percentile ranges in these bins should therefore be interpreted with caution.

For the ISM-accreting population, the weighted median \(v_{\rm pec}\) remains nearly independent of BH mass over most of the plotted range and is typically around \(15\!-\!25\,\mathrm{km\,s^{-1}}\). The three M25 models are almost indistinguishable, while F12 and MM20 show only modest differences. Thus, accretion-based selection substantially weakens the dependence of
\(v_{\rm pec}\) on BH mass in the observable population.

The long-timescale lensing population retains a clearer mass dependence. At \(M_{\rm BH}\lesssim10\,M_{\odot}\), the M25 and F12 models generally show higher weighted median velocities, often reaching \(50\!-\!300\,\mathrm{km\,s^{-1}}\), whereas MM20 typically yields lower median velocities of \(\lesssim50\,\mathrm{km\,s^{-1}}\). F12(D) has the broadest percentile ranges and a relatively large low-mass velocity contribution, consistent with its larger kicked fraction. Toward higher masses, the curves become more irregular. This behavior reflects both the intrinsic relation between remnant mass and natal kick and the decreasing number of long-timescale lensing events in individual mass bins. The lensing selection in \(t_{\rm E}\), which also depends on lens mass, geometry, and relative proper motion, further complicates this relation.

Taken together, the one-dimensional mass and peculiar-velocity distributions, along with their joint mass--velocity relation, suggest that long-timescale microlensing events retain more information about the adopted SN prescription than ISM-accreting X-ray sources. The most informative diagnostics include the fraction of mass-gap BHs, the high-velocity tail, the presence or absence of the \(12\!-\!17\,M_{\odot}\) deficit, and the peculiar-velocity distribution of low-mass lenses.

\subsection{Robustness of the \(v_{\rm k} / v_{\rm pec}\) Relation}
\label{subsubsec:vk/vpec}

\begin{figure*}
    \centering
	\includegraphics[width=0.80\textwidth]{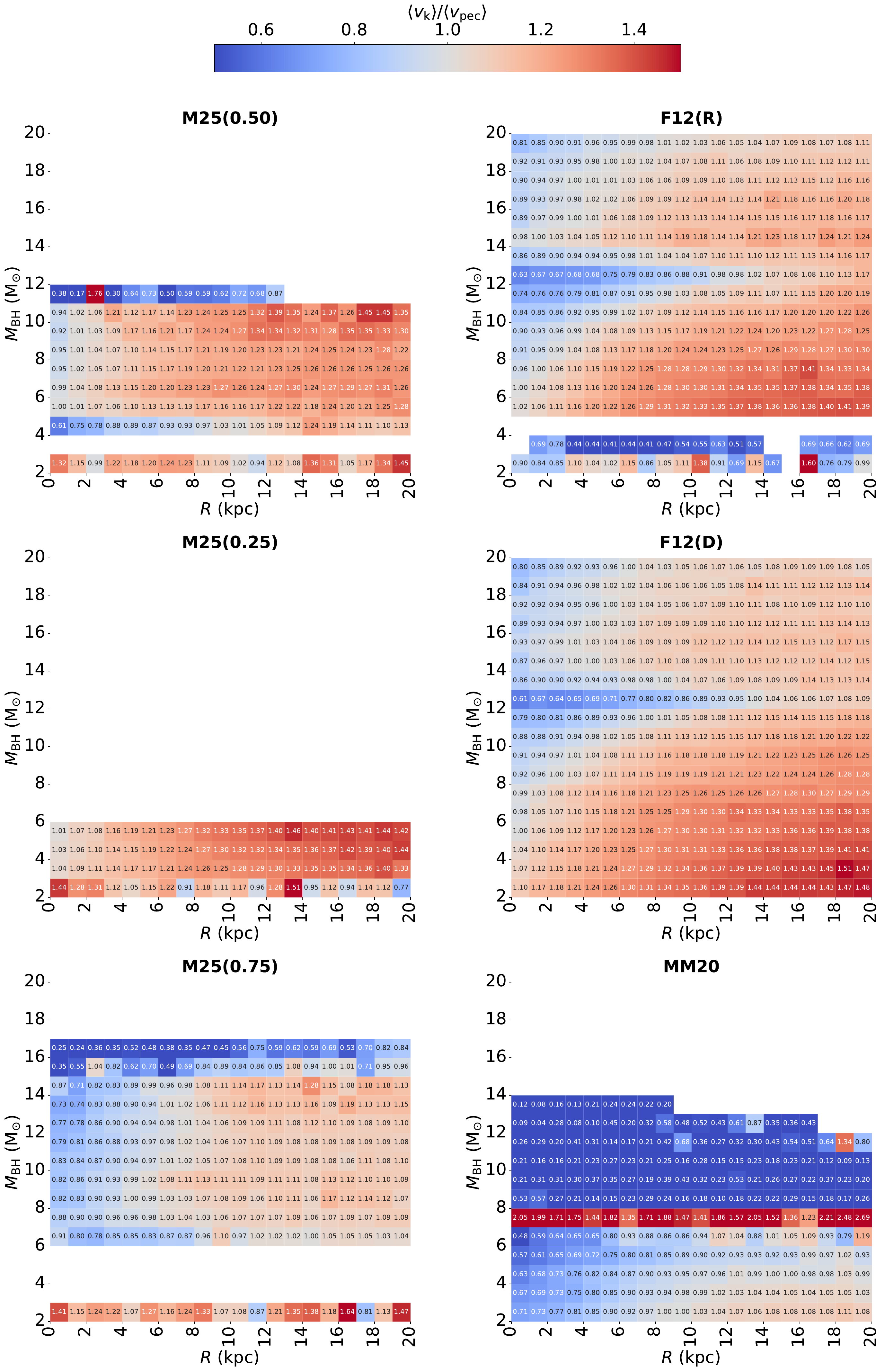}
    \caption{Ratio of the weighted mean natal-kick velocity to the weighted mean present-day peculiar velocity, \(\langle v_{\rm k}\rangle/\langle v_{\rm pec}\rangle\), for kicked IBHs as a function of Galactocentric radius and BH mass. The left column shows the three M25 models with fallback fractions of \(0.50\), \(0.25\), and \(0.75\), from top to bottom, while the right column shows the F12(R), F12(D), and MM20 models. Each cell is labeled with the corresponding value of \(\langle v_{\rm k}\rangle/\langle v_{\rm pec}\rangle\).}
    \label{fig:IBH_kick_pec_models}
\end{figure*}

Constraining the natal kick of an individual BH is difficult, even when its position, mass, and peculiar velocity are accurately measured. The mapping between these observables and \(v_{\rm k}\) depends on the Galactic potential, the birth location, and the formation channel. The natal kick of an individual BH therefore cannot be determined uniquely, as illustrated by the confirmed lensing BH OGLE-2011-BLG-0462 discussed above.

For a sufficiently large sample, however, the underlying kick distribution can be constrained statistically. Figure~\ref{fig:IBH_kick_to_pec} shows the ratio \(v_{\rm k}/v_{\rm pec}\) as a function of Galactocentric radius for kicked IBHs in the fiducial M25(0.50) model. This relation provides a model-dependent statistical calibration between present-day peculiar velocities and natal kicks. It may also depend on BH mass, particularly in models in which the fallback fraction couples the remnant mass to the kick magnitude.

Figure~\ref{fig:IBH_kick_pec_models} shows the ratio of the weighted mean velocities, \(\langle v_{\rm k}\rangle/\langle v_{\rm pec}\rangle\), as a function of Galactocentric radius and BH mass for six SN prescriptions. The relation varies substantially among the models because they assign different remnant masses and kick magnitudes across the BH-mass range. The ratio should therefore be used as a model-specific calibration rather than as a universal conversion between \(v_{\rm pec}\) and \(v_{\rm k}\).

In the fiducial M25(0.50) model, kicked IBHs are concentrated mainly in the \(5-11\,M_{\odot}\) range, where the effective sample size is relatively large and the inferred relation is more reliable. The ratio is generally close to unity in the inner Galaxy and increases to approximately \(1.2-1.3\) toward the outer disk. Its weak mass dependence reflects the changing mixture of formation channels and Galactic birth components across the mass range. The narrow \(2\!-\!3\,M_{\odot}\) component should be excluded from this calibration because these BHs are mainly produced through the channel of two-CO mergers. 

Changing the fallback fraction alters both the masses of fallback BHs and the degree to which their natal kicks are suppressed. M25(0.25) therefore produces lighter fallback BHs with larger kicks, whereas M25(0.75) produces more massive fallback BHs with smaller kicks. Accordingly, the \(v_{\rm k}/v_{\rm pec}\) relation exhibits a stronger radial dependence in M25(0.25) and a weaker dependence in M25(0.75). The relation thus reflects not only the contribution of natal kicks to the peculiar velocity but also the extent of kick-induced radial migration.  

The F12(R) and F12(D) models show broadly similar radial trends, but their mass dependence differs in the \(2\!-\!5\,M_{\odot}\) mass-gap region. In general, \(v_{\rm k}/v_{\rm pec}\) increases toward lower BH masses, reflecting the larger kicks associated with weaker fallback. Local reversals of this trend, such as those around \(12\!-\!13\,M_{\odot}\), can arise from the discontinuous mapping between the progenitor ZAMS mass and the final remnant mass (Figure~\ref{fig:stellar_mass_relation}).

The MM20 model behaves differently. The relation is relatively smooth below \(7\,M_{\odot}\) but becomes less representative of ordinary CCSN-formed BHs at higher masses because the kicked population is dominated by two-CO merger products. In particular, all kicked BHs with \(M_{\rm BH}\geq8\,M_{\odot}\) come from the two-CO merger channel, while this channel contributes approximately \(72\%\) of kicked BHs in the \(7\!-\!8\,M_{\odot}\) range.

Overall, \(v_{\rm k}/v_{\rm pec}\) provides a useful but model-dependent statistical calibration for constraining the natal-kick distribution of BH populations, particularly for sufficiently large samples of ordinary CCSN-formed BHs. The approach is most applicable to candidate kicked-BH populations with high peculiar velocities, for which natal kicks are expected to contribute substantially to the observed kinematics. The calibration should be constructed separately in bins of BH mass and Galactocentric radius; merger-produced BHs and sparsely populated bins should be analyzed separately. A similar strategy may also be useful for NSs, which typically receive larger kicks and are less affected by fallback, although young observable pulsars may not have evolved long enough in the Galactic potential for the same correction to apply directly.

\subsection{Comparison with Previous Work}

Previous studies \citep{Agol2002, Tsuna2018} have investigated the X-ray detectability of Galactic IBHs using simplified prescriptions for the BH population, kinematics, and accretion physics.

In \citet{Agol2002}, the BH masses are assumed to follow a power-law
distribution with an index of \(0.35\) over \(4\!-\!13\,M_{\odot}\).
The relative velocities between the BHs and the ISM are drawn from a Maxwellian distribution with a one-dimensional
velocity dispersion of
\(40\,\mathrm{km\,s^{-1}}\), calibrated using a sample
of 19 BH X-ray binaries. Their model does not include mass loss through
outflows in hot accretion flows. Our fiducial M25(0.50) model predicts approximately an order of magnitude fewer BHs with \(\dot{M}_{\rm BH}>10^{15}\,\mathrm{g\,s^{-1}}\), primarily because it contains only \(\sim4\times10^7\) Galactic IBHs, compared with the \(10^9\) objects assumed by \citet{Agol2002}. Nevertheless, our accretion-rate distribution is flatter at the high-rate end. Whereas \citet{Agol2002} obtain a maximum accretion rate below \(10^{17}\,\mathrm{g\,s^{-1}}\), our model contains several systems with \(\dot{M}_{\rm BH}>10^{18}\,\mathrm{g\,s^{-1}}\). These rare systems have very low peculiar velocities and pass through dense ISM phases, allowing them to sustain high accretion rates even after hot-flow outflows are included. In addition, \citet{Agol2002} adopt a constant radiative efficiency of \(\eta=10^{-5}\), substantially lower than the efficiencies given by the \citet{Xie2012} prescription adopted here. Combining these differences, we predict approximately \(5\times10^3\) BHs with \(F_{\rm X}>10^{-14}\,\mathrm{erg\,s^{-1}\,cm^{-2}}\), several tens of times more than their estimate.

\citet{Tsuna2018} adopt a Gaussian BH-mass distribution with a mean of \(7.8\,M_{\odot}\) and a standard deviation of \(1.2\,M_{\odot}\). They also assume that BH natal kicks are independent of BH mass and draw the kick velocities from a Maxwell--Boltzmann distribution with a three-dimensional mean velocity of \(50\!-\!400\,\mathrm{km\,s^{-1}}\). This treatment omits the fallback dependence expected for massive BHs \citep{Nagarajan2025}. They parameterize the suppression due to hot accretion flows by multiplicative factors of \(10^{-3}\!-\!10^{-1}\) and assume that the radiative efficiency decreases linearly with decreasing accretion rate, which is more pessimistic than the \citet{Xie2012} prescription. They estimate that previous \textit{NuSTAR} surveys could have detected about four IBHs. Our predicted number is larger mainly because the dynamically evolved population contains a substantial non-kicked component with low peculiar velocities and because we adopt a different radiative-efficiency model. The comparison is therefore qualitative, since the two calculations also differ in their BH population models, kinematic prescriptions, and survey assumptions.

\subsection{Accretion Regimes and Radio Detection Prospects} \label{sec:discussion_radio}

The accretion-rate distribution shown in Figure~\ref{fig:IBH_ism_mdot} helps assess whether accretion-disk instabilities are likely to affect the IBH population. For a \(10\,M_{\odot}\) black hole, the Eddington accretion rate is of order \(10^{19}\,\mathrm{g\,s^{-1}}\) for a radiative efficiency of \(\eta=0.1\). Fewer than ten IBHs in our synthetic population have accretion rates above \(10^{17}\,\mathrm{g\,s^{-1}}\). Most systems therefore remain in the low-accretion-rate regime, where accretion is expected to proceed through RIAFs.

The classical hydrogen-ionization disk instability is therefore unlikely to be a generic feature of the IBH population. It could, however, affect a small subset of the most rapidly accreting systems if the captured gas carries sufficient angular momentum to form an extended, relatively cool outer disk (see Section~\ref{subsec:bh_ism}). In a truncated-disk configuration, the outer disk could undergo a thermal-viscous instability while the inner flow remains in an advection-dominated or more general RIAF state \citep{Lasota2001,Meyer2001}. The resulting X-ray emission would be episodic, and the instantaneous number of detectable sources would depend on the outburst duty cycle. Consequently, the steady-state estimates in Table~\ref{tab:detector} may overestimate the time-averaged number of detectable IBHs.

Radio observations may provide a complementary probe of the large population of low-accretion-rate IBHs. Possible radio emission mechanisms include thermal synchrotron radiation from the hot accretion flow, synchrotron emission from a compact jet, and shocks generated when accretion-driven outflows interact with the surrounding interstellar medium \citep{Yuan2014,Tsuna2019}. These components need not be present simultaneously, and their relative importance depends on the magnetic-field configuration, jet or wind power, and local ISM conditions. We do not model the corresponding radio luminosities in this work. Sensitive surveys with facilities such as the Square Kilometre Array (SKA) could nevertheless provide a useful avenue for identifying IBHs whose X-ray emission is strongly suppressed by radiative inefficiency.

\subsection{Uncertainties in CE Mergers}

In our models, BHs originating from the CE-merger channel account for approximately \(10\%-30\%\) of the IBH population. This channel, however, remains physically uncertain.

First, when two non-degenerate stars merge during CE evolution, the final product is uncertain because the stellar-merger process is not fully understood. \citet{Hurley2002} estimate the merger product using a giant mass--radius relation and the binding energy immediately before merger, assuming a constant binding-energy parameter \(\lambda\) for all giants. This treatment is approximate because \(\lambda\) depends on the stellar mass, radius, and evolutionary state \citep{XuXJ2010}. In contrast, \citet{Olejak2020} adopt a fixed fractional mass loss during a CE merger to estimate the mass of the merger product. In this work, we use the method of \citet{Hurley2002} because, despite its limitations, it explicitly links envelope ejection to the released orbital energy.

Second, when an NS or BH enters a CE and does not survive as a binary, the outcome may be a Thorne-{\.Z}ytkow-object-like system. \citet{Hurley2002} argue that such objects are unstable and rapidly eject their envelopes, leaving the NS or BH with little or no mass gain. Results from \citet{MacLeod2015} also suggest that NSs gain only a small fraction of their mass during CE episodes and rarely undergo AIC into BHs. Conversely, \citet{Farmer2023} argue that Thorne-{\.Z}ytkow objects may be more stable than previously believed, increasing the chance of AIC for NSs. If the giant's core were simply added to the NS during the merger, this channel could produce around \(10^6\) IBHs in the putative \(2\!-\!5\,M_{\odot}\) mass gap between NSs and BHs, corresponding to \(\gtrsim1\%\) of all IBHs. To avoid overinterpreting this uncertain channel, we assume that NSs and BHs merging in a CE do not gain mass.


\subsection{Uncertainties in Galactic Structure and Evolution}

Our model for Galactic star formation and chemical evolution \citep{Frankel2018, Wagg2022} provides a useful framework for predicting the IBH population, but several simplifying assumptions should be kept in mind. First, the model assumes that the thick-disk and bulge components evolve entirely in situ, neglecting early merger events such as Gaia-Enceladus \citep{Helmi2018, Belokurov2022}. Such events contributed stars with distinct chemical signatures, typically at lower metallicities, that are not fully captured by our smooth metallicity-evolution model. Because lower metallicity favors the formation of more massive BHs \citep{Belczynski2010b}, our model may underestimate the fraction of high-mass BHs (\(M_{\rm BH}\gtrsim30\,M_\odot\)) in these components.

Second, our orbital-evolution calculation includes natal kicks and motion in the Galactic gravitational potential but does not include radial migration driven by non-axisymmetric Galactic structures. As shown by \citet{Frankel2018}, angular-momentum exchange with spiral arms can produce significant radial migration, which is not captured by our static, axisymmetric potential. These omissions may affect the predicted spatial distribution of BHs. However, the most detectable accreting BHs in our simulations generally receive little or no natal kick and remain close to their birth orbits, so the impact on the dominant detectable X-ray population may be limited.

Incorporating a comprehensive model of thick-disk and bulge formation, merger history, and radial migration is beyond the scope of this work. Future versions of POPKIN will address these limitations by implementing time-dependent Galactic potentials and physically motivated prescriptions for radial migration. These developments will improve predictions for the present-day spatial and kinematic distributions of IBHs in the Milky Way and refine the connection between natal kicks, birth sites, and the observable X-ray and microlensing populations.

\section{Conclusion} \label{sec:conclusion}

We developed POPKIN to investigate the formation, Galactic orbital evolution, and observable signatures of IBHs in the Milky Way. By combining single- and binary-star evolution, the Galactic histories of star formation and metallicity evolution, orbit integration in the Galactic potential, and post-processing for accretion and microlensing observables, we traced how IBH birth properties map onto the present-day population and its observable subsets. Our main results are as follows.

\begin{enumerate}

\item We considered ten population-synthesis models that vary the
SN prescription, fallback fraction, mass-transfer efficiency, and CE ejection efficiency. The total Galactic population of IBHs is controlled primarily by the adopted SN prescription. Models based on the M25 prescription produce approximately \(4\times10^7\) IBHs, whereas the F12(R), F12(D), and MM20 models generally produce \(\sim(1\!-\!2)\times10^8\) IBHs. In the fiducial M25(0.50) model, the Galactic IBH population has a mean mass of about \(11\,M_{\odot}\), a radial distribution peaking near \(R\simeq3\,\mathrm{kpc}\), and roughly \(8\times10^4\) IBHs within \(1\,\mathrm{kpc}\) of the Sun. The population is strongly concentrated toward the Galactic mid-plane, with approximately \(62\%\) of IBHs located within \(|z|<500\,\mathrm{pc}\). SN-induced disruption is the dominant formation channel, while two-MS mergers and CE mergers also contribute substantially. Single-star evolution contributes approximately \(10\%-15\%\), whereas two-CO mergers remain a minor channel overall.

\item The SN prescription strongly affects both the BH mass distribution and the fraction of BHs receiving natal kicks. In the fiducial M25(0.50) model, the mass distribution exhibits two prominent peaks near \(9\,M_{\odot}\) and \(20\,M_{\odot}\), with a pronounced deficit around \(13\!-\!17\,M_{\odot}\). The relative contribution of the high-mass component increases toward larger Galactocentric radii because of the changing mixture of Galactic components and metallicities. In contrast, F12(D) and MM20 populate the \(2-5\,M_{\odot}\) mass-gap region, whereas F12(R) retains a pronounced deficit there. Changing the M25 fallback fraction mainly shifts the mass distribution of fallback BHs and changes their natal-kick magnitudes. Among the models considered here, F12(D) produces the largest fraction of kicked IBHs, approximately \(89\%\), and the largest escape fraction, approximately \(2\%\).

\item Galactic dynamics modifies, but does not erase, the connection between natal kicks and present-day kinematics. Vertical migration increases with current height above the Galactic plane, while radial migration is much more sensitive to natal kicks. IBHs formed without natal kicks usually remain on nearly circular disk orbits and have typical mid-plane peculiar velocities of \(20-30\,{\rm km\,s^{-1}}\), making them favorable candidates for accretion-powered detection. Kicked IBHs instead show stronger radial migration and generally higher peculiar velocities. The relation between \(v_{\rm k}\) and \(v_{\rm pec}\) is therefore statistical rather than deterministic; nevertheless, a mass- and radius-dependent correction can recover the underlying natal-kick distribution more accurately than using the observed \(v_{\rm pec}\) distribution alone, especially for high-\(v_{\rm pec}\) samples.

\item Accretion-powered X-ray searches preferentially select the slowest IBHs. In the fiducial M25(0.50) model, we predict approximately \(5\times10^3\) accreting IBHs with \(F_{\rm X}>10^{-14}\,\mathrm{erg\,s^{-1}\,cm^{-2}}\), including roughly \(40\), \(10^2\), and \(4\times10^3\) sources within 500 pc, 1 kpc, and 10 kpc, respectively. The detectable population is dominated by non-kicked IBHs with \(v_{\rm pec}\sim10\!-\!30\,{\rm km\,s^{-1}}\). Cold \(\mathrm{H}_{\text{I}}\) and warm \(\mathrm{H}_{\text{I}}\) are important locally, while MCs dominate the detectable population at larger distances toward the inner Galaxy. These number estimates should be regarded as optimistic because they depend sensitively on the radiative-efficiency prescription and the treatment of hot accretion flows. Reliable confirmation of accreting IBHs will therefore require multiwavelength follow-up observations and robust discrimination from other Galactic X-ray sources.

\item Microlensing provides a complementary and less velocity-biased route to IBH detection. Our selection of analogs to OGLE-2011-BLG-0462 favors an interpretation in which the confirmed BH lens was born in the thin disk and subsequently received a natal kick, although a thick-disk origin cannot be ruled out. For a Roman-like bulge survey, the fiducial M25(0.50) model predicts approximately \(360\) intrinsic IBH microlensing events over a five-year survey in an effective area of \(1.70\,{\rm deg^2}\). This prediction should be regarded as an optimistic upper limit because it does not include realistic cadence, blending, astrometric-precision, or event-recovery selection effects. Long-timescale microlensing events are particularly useful for distinguishing between SN prescriptions because their one-dimensional and joint mass--velocity distributions retain several diagnostic features. These include the presence and abundance of low-mass BHs in the \(2-5\,M_{\odot}\) range, the \(12\!-\!17\,M_{\odot}\) mass deficit predicted by the M25 prescription, the velocity distribution of low-mass BHs, and the high-\(v_{\rm pec}\) tail associated with natal kicks.
\end{enumerate}


\begin{acknowledgments}

This work was supported by the National Key Research and Development Program of China (Grant No. 2023YFA1607902) and the National Natural Science Foundation of China (Grant Nos. 12041301, 12121003, 12373034, and 12603055). YN thanks Ge-fei Zhao for assistance with the post-supernova orbital calculations. Generative AI (OpenAI Codex) was used under full human supervision to assist with optimization of code used in this work and language editing of the manuscript. The POPKIN code is publicly available at \url{https://github.com/JianguoHe/POPKIN}.

\end{acknowledgments}



\appendix

\section{Calculation of post-SN orbits} \label{appendix:postSNorbit}

This appendix summarizes the prescription adopted for orbital changes induced by
an SN. We denote the exploding star as component 1 and the companion star as
component 2. Quantities before the explosion carry the
subscript ``preSN'', quantities immediately after the impulse carry the
subscript ``postSN'', and the instantaneous velocity perturbations associated
with the explosion carry the subscript ``SN''. Throughout this appendix we work
in the pre-SN center-of-mass frame. The \(x\)-axis points from the companion
toward the exploding star, and the \(z\)-axis is aligned with the pre-SN orbital
angular momentum.

The pre-SN and post-SN masses are denoted by
\(M_{1,\mathrm{preSN}}\), \(M_{2,\mathrm{preSN}}\),
\(M_{1,\mathrm{postSN}}\), and \(M_{2,\mathrm{postSN}}\), respectively. The
post-SN total mass is
\begin{equation}
M_{\mathrm{postSN}}
=
M_{1,\mathrm{postSN}}+M_{2,\mathrm{postSN}} .
\end{equation}
The pre-SN center-of-mass velocities are
\(\boldsymbol{v}_{1,\mathrm{preSN}}\) and
\(\boldsymbol{v}_{2,\mathrm{preSN}}\). The exploding star receives a natal kick
\(\Delta\boldsymbol{v}_{1,\mathrm{SN}}\), and the companion may receive an
impact velocity \(\Delta\boldsymbol{v}_{2,\mathrm{SN}}\). We assume that the
impact velocity is anti-parallel to the instantaneous separation vector:
\begin{equation}
\Delta\boldsymbol{v}_{2,\mathrm{SN}}
=
-v_{\mathrm{imp}}
\frac{\boldsymbol{R}_{\mathrm{SN}}}{R_{\mathrm{SN}}}.
\end{equation}
The velocities immediately after the impulse are therefore
\begin{equation}
\boldsymbol{u}_{1,\mathrm{postSN}}
=
\boldsymbol{v}_{1,\mathrm{preSN}}
+
\Delta\boldsymbol{v}_{1,\mathrm{SN}},
\end{equation}
\begin{equation}
\boldsymbol{u}_{2,\mathrm{postSN}}
=
\boldsymbol{v}_{2,\mathrm{preSN}}
+
\Delta\boldsymbol{v}_{2,\mathrm{SN}} .
\end{equation}

The relative position at the explosion is assumed not to change during the
impulse:
\begin{equation}
\boldsymbol{R}_{\mathrm{SN}}
=
\boldsymbol{r}_{1,\mathrm{preSN}}
-
\boldsymbol{r}_{2,\mathrm{preSN}},
\end{equation}
where \(\boldsymbol{R}_{\mathrm{SN}}\) points from the companion toward the
exploding star. The pre-SN and post-SN relative velocities are
\begin{equation}
\boldsymbol{u}_{\mathrm{rel,preSN}}
=
\boldsymbol{v}_{1,\mathrm{preSN}}
-
\boldsymbol{v}_{2,\mathrm{preSN}},
\end{equation}
\begin{equation}
\boldsymbol{u}_{\mathrm{rel,postSN}}
=
\boldsymbol{u}_{1,\mathrm{postSN}}
-
\boldsymbol{u}_{2,\mathrm{postSN}}
=
\boldsymbol{u}_{\mathrm{rel,preSN}}
+
\Delta\boldsymbol{v}_{1,\mathrm{SN}}
-
\Delta\boldsymbol{v}_{2,\mathrm{SN}} .
\end{equation}
Here \(R_{\mathrm{SN}}=|\boldsymbol{R}_{\mathrm{SN}}|\), and bold symbols denote
vectors while the corresponding non-bold symbols denote their magnitudes.

For eccentric pre-SN binaries, the explosion phase is sampled uniformly in
time. For \(e_{\mathrm{preSN}}<1\), the pre-SN mean anomaly satisfies
\begin{equation}
\mathcal{M}_{\mathrm{preSN}}
=
E_{\mathrm{preSN}}
-
e_{\mathrm{preSN}}\sin E_{\mathrm{preSN}},
\end{equation}
where \(E_{\mathrm{preSN}}\) is the eccentric anomaly. 
Once \(\boldsymbol{R}_{\mathrm{SN}}\) and
\(\boldsymbol{u}_{\mathrm{rel,postSN}}\) are known, the post-SN orbit is fully
determined.

The post-SN center-of-mass velocity relative to the pre-SN center of mass is
\begin{equation}
\boldsymbol{V}_{\mathrm{CM,postSN}}
=
\frac{
M_{1,\mathrm{postSN}}\boldsymbol{u}_{1,\mathrm{postSN}}
+
M_{2,\mathrm{postSN}}\boldsymbol{u}_{2,\mathrm{postSN}}
}{
M_{\mathrm{postSN}}
}.
\label{eq:app_postSN_CM_velocity}
\end{equation}
This is the systemic velocity offset of the post-SN binary.

For \(M_{1,\mathrm{postSN}}>0\), let \(\boldsymbol{u}_i(t)\) denote the
subsequent velocity of component \(i\), with
\(\boldsymbol{u}_i(0)=\boldsymbol{u}_{i,\mathrm{postSN}}\). The motion after
the impulse is governed by the mutual gravity of the two surviving components:
\begin{equation}
M_{1,\mathrm{postSN}}
\frac{d\boldsymbol{u}_1}{dt}
=
-
\frac{
G M_{1,\mathrm{postSN}}M_{2,\mathrm{postSN}}
}{
r_{\mathrm{rel}}^3
}
\boldsymbol{r}_{\mathrm{rel}},
\end{equation}
\begin{equation}
M_{2,\mathrm{postSN}}
\frac{d\boldsymbol{u}_2}{dt}
=
\frac{
G M_{1,\mathrm{postSN}}M_{2,\mathrm{postSN}}
}{
r_{\mathrm{rel}}^3
}
\boldsymbol{r}_{\mathrm{rel}},
\end{equation}
where
\(\boldsymbol{r}_{\mathrm{rel}}=\boldsymbol{r}_1-\boldsymbol{r}_2\). The
relative equation of motion is therefore
\begin{equation}
\frac{d\boldsymbol{u}_{\mathrm{rel}}}{dt}
=
-
\frac{G M_{\mathrm{postSN}}}{r_{\mathrm{rel}}^3}
\boldsymbol{r}_{\mathrm{rel}} .
\label{eq:app_postSN_relative_acceleration}
\end{equation}

The post-SN specific relative angular momentum and specific orbital energy are
\begin{equation}
\boldsymbol{h}_{\mathrm{postSN}}
=
\boldsymbol{R}_{\mathrm{SN}}
\times
\boldsymbol{u}_{\mathrm{rel,postSN}},
\end{equation}
\begin{equation}
\epsilon_{\mathrm{postSN}}
=
\frac{1}{2}u_{\mathrm{rel,postSN}}^2
-
\frac{G M_{\mathrm{postSN}}}{R_{\mathrm{SN}}}.
\label{eq:app_postSN_energy}
\end{equation}
The corresponding eccentricity is
\begin{equation}
e_{\mathrm{postSN}}
=
\left[
1+
\frac{
2\epsilon_{\mathrm{postSN}}h_{\mathrm{postSN}}^2
}{
\left(GM_{\mathrm{postSN}}\right)^2
}
\right]^{1/2}.
\label{eq:app_postSN_eccentricity}
\end{equation}
For completeness, we also define the total orbital angular momentum,
\begin{equation}
J_{\mathrm{orb,postSN}}
=
\mu_{\mathrm{postSN}}h_{\mathrm{postSN}},
\quad
\mu_{\mathrm{postSN}}
=
\frac{
M_{1,\mathrm{postSN}}M_{2,\mathrm{postSN}}
}{
M_{\mathrm{postSN}}
},
\end{equation}
whose scalar magnitude is retained in our calculations.

When the binary remains bound, the post-SN semi-major axis is
\begin{equation}
a_{\mathrm{postSN}}
=
\frac{
h_{\mathrm{postSN}}^2
}{
G M_{\mathrm{postSN}}\left(1-e_{\mathrm{postSN}}^2\right)
},
\label{eq:app_postSN_semimajor}
\end{equation}
and the periastron distance (the closest approach) is
\begin{equation}
R_{\mathrm{closest}}
=
a_{\mathrm{postSN}}\left(1-e_{\mathrm{postSN}}\right) .
\end{equation}
This corresponds to an elliptic post-SN orbit.

For disrupted systems, POPKIN assigns the final runaway velocities at infinity. To determine the asymptotic direction, we adopt a coordinate construction
in which the instantaneous separation vector defines the \(x\)-axis. The
relative orbit can then be written as
\begin{equation}
\frac{1}{r_{\mathrm{rel}}}
=
A_{\mathrm{SN}}\cos\theta_{\mathrm{SN}}
+
B_{\mathrm{SN}}\sin\theta_{\mathrm{SN}}
+
\frac{G M_{\mathrm{postSN}}}{h_{\mathrm{postSN}}^2},
\end{equation}
with
\begin{equation}
A_{\mathrm{SN}}
=
\frac{1}{R_{\mathrm{SN}}}
-
\frac{G M_{\mathrm{postSN}}}{h_{\mathrm{postSN}}^2},
\end{equation}
\begin{equation}
B_{\mathrm{SN}}
=
-
\frac{
\boldsymbol{R}_{\mathrm{SN}}\cdot
\boldsymbol{u}_{\mathrm{rel,postSN}}
}{
h_{\mathrm{postSN}}R_{\mathrm{SN}}
}.
\end{equation}
We then define
\begin{equation}
\Delta_{\mathrm{SN}}
=
A_{\mathrm{SN}}^2
+
B_{\mathrm{SN}}^2
-
\left(
\frac{G M_{\mathrm{postSN}}}{h_{\mathrm{postSN}}^2}
\right)^2 .
\label{eq:app_postSN_Delta}
\end{equation}
Using Equations~(\ref{eq:app_postSN_energy}) and
(\ref{eq:app_postSN_Delta}), one finds
\begin{equation}
\Delta_{\mathrm{SN}}
=
\frac{2\epsilon_{\mathrm{postSN}}}{h_{\mathrm{postSN}}^2}.
\end{equation}
Thus this discriminant is exactly equivalent to the usual specific-energy
criterion: 
\(\Delta_{\mathrm{SN}}<0\) corresponds to a bound elliptic orbit,
\(\Delta_{\mathrm{SN}}>0\) corresponds to a disrupted hyperbolic orbit, and
\(\Delta_{\mathrm{SN}}=0\) is the parabolic limit, which is not treated as a
separate numerical branch in the current implementation. This coordinate
construction assumes \(h_{\mathrm{postSN}}\neq 0\); purely radial post-SN motion is not treated as a
separate branch here.

For \(\Delta_{\mathrm{SN}}>0\), the relative speed at infinity is
\begin{equation}
u_{\mathrm{rel},\infty}
=
\left(
u_{\mathrm{rel,postSN}}^2
-
\frac{2G M_{\mathrm{postSN}}}{R_{\mathrm{SN}}}
\right)^{1/2}.
\label{eq:app_postSN_vrel_inf_mag}
\end{equation}
For the outgoing branch of the hyperbolic orbit, the asymptotic direction is
given by
\begin{equation}
\sin\phi_{\mathrm{SN}}
=
\frac{
-
\frac{G M_{\mathrm{postSN}}}{h_{\mathrm{postSN}}^2}
B_{\mathrm{SN}}
+
\sqrt{\Delta_{\mathrm{SN}}}\,A_{\mathrm{SN}}
}{
A_{\mathrm{SN}}^2+B_{\mathrm{SN}}^2
},
\end{equation}
\begin{equation}
\cos\phi_{\mathrm{SN}}
=
\frac{
-
\frac{G M_{\mathrm{postSN}}}{h_{\mathrm{postSN}}^2}
A_{\mathrm{SN}}
-
\sqrt{\Delta_{\mathrm{SN}}}\,B_{\mathrm{SN}}
}{
A_{\mathrm{SN}}^2+B_{\mathrm{SN}}^2
}.
\end{equation}
It is convenient to write the asymptotic relative velocity in the form (it requires $h_{\rm postSN}\neq 0$)
\begin{equation}
\boldsymbol{u}_{\mathrm{rel},\infty}
=
k_{\mathrm{SN}}\boldsymbol{R}_{\mathrm{SN}}
+
l_{\mathrm{SN}}\boldsymbol{u}_{\mathrm{rel,postSN}},
\label{eq:app_postSN_vrel_inf_vector}
\end{equation}
where
\begin{equation}
k_{\mathrm{SN}}
=
\frac{u_{\mathrm{rel},\infty}}{R_{\mathrm{SN}}}
\left[
\cos\phi_{\mathrm{SN}}
-
\frac{
\boldsymbol{R}_{\mathrm{SN}}\cdot
\boldsymbol{u}_{\mathrm{rel,postSN}}
}{
h_{\mathrm{postSN}}
}
\sin\phi_{\mathrm{SN}}
\right],
\end{equation}
\begin{equation}
l_{\mathrm{SN}}
=
\frac{
R_{\mathrm{SN}}u_{\mathrm{rel},\infty}
}{
h_{\mathrm{postSN}}
}
\sin\phi_{\mathrm{SN}} .
\end{equation}
The final velocities of the two disrupted components in the pre-SN
center-of-mass frame are
\begin{equation}
\boldsymbol{v}_{1,\infty}^{(\mathrm{preSN})}
=
\boldsymbol{V}_{\mathrm{CM,postSN}}
+
\frac{M_{2,\mathrm{postSN}}}{M_{\mathrm{postSN}}}
\boldsymbol{u}_{\mathrm{rel},\infty},
\label{eq:app_postSN_v1_inf}
\end{equation}
\begin{equation}
\boldsymbol{v}_{2,\infty}^{(\mathrm{preSN})}
=
\boldsymbol{V}_{\mathrm{CM,postSN}}
-
\frac{M_{1,\mathrm{postSN}}}{M_{\mathrm{postSN}}}
\boldsymbol{u}_{\mathrm{rel},\infty}.
\label{eq:app_postSN_v2_inf}
\end{equation}
These expressions satisfy
\begin{equation}
\boldsymbol{u}_{\mathrm{rel},\infty}
=
\boldsymbol{v}_{1,\infty}^{(\mathrm{preSN})}
-
\boldsymbol{v}_{2,\infty}^{(\mathrm{preSN})} .
\end{equation}

Type Ia SNe are treated separately. If the exploding WD is completely
destroyed, that is, \(M_{1,\mathrm{postSN}}=0\), no post-SN two-body orbit is
constructed. In that case the surviving companion velocity is simply
\begin{equation}
\boldsymbol{v}_{2,\mathrm{Ia}}^{(\mathrm{preSN})}
=
\boldsymbol{v}_{2,\mathrm{preSN}}
+
\Delta\boldsymbol{v}_{2,\mathrm{SN}},
\end{equation}
and the post-SN orbital elements are not defined.

\section{Reference frame conversion} \label{appendix:frame}

In POPKIN, SN explosions can change the kinematic state of a system by imparting a recoil velocity to a bound post-SN binary or runaway velocities to unbound components. We evaluate these velocity offsets in the pre-SN center-of-mass frame. In this frame, the positive \(z\)-axis is aligned with the pre-SN orbital angular momentum, the positive \(x\)-axis points from the companion star toward the exploding star, and the \(y\)-axis completes a right-handed coordinate system. To use these velocities in the Galactic orbit integration, we transform them into the local Galactocentric cylindrical velocity basis \((v_R, v_T, v_z)\).

We assume that the pre-SN orbital plane is isotropically oriented with respect to the Galactic frame, and that the line of centers within the orbital plane is random. We therefore apply a random three-dimensional rotation to the velocity-offset vector. In the current implementation, the rotation matrix is generated from a random unit quaternion, which provides a Haar-uniform sampling of the rotation group \({\rm SO}(3)\). Specifically, four independent Gaussian random numbers are drawn to form
\begin{equation}
\boldsymbol{q}
=
(q_0,q_1,q_2,q_3),
\end{equation}
which is then normalized to unit length. The corresponding rotation matrix is
\begin{equation}
\mathcal{R}_{\rm rot}
=
\begin{pmatrix}
1-2(q_2^2+q_3^2) & 2(q_1q_2-q_0q_3) & 2(q_1q_3+q_0q_2) \\
2(q_1q_2+q_0q_3) & 1-2(q_1^2+q_3^2) & 2(q_2q_3-q_0q_1) \\
2(q_1q_3-q_0q_2) & 2(q_2q_3+q_0q_1) & 1-2(q_1^2+q_2^2)
\end{pmatrix}.
\end{equation}

The rotated velocity offset in the local Galactocentric cylindrical basis is
then written as
\begin{equation}
\begin{pmatrix}
\Delta v_R \\
\Delta v_T \\
\Delta v_z
\end{pmatrix}
=
\mathcal{R}_{\rm rot}
\begin{pmatrix}
\Delta v_{x,\mathrm{orb}} \\
\Delta v_{y,\mathrm{orb}} \\
\Delta v_{z,\mathrm{orb}}
\end{pmatrix},
\label{eq:app_frame_rotation}
\end{equation}
where the subscript ``orb'' denotes components in the pre-SN orbital frame.
Equation~(\ref{eq:app_frame_rotation}) rotates only the SN-induced velocity
offset; the resulting cylindrical components are then added to the system's
Galactocentric velocity at the corresponding evolutionary time. The subsequent
Galactic orbit is integrated with \textsc{galpy}.


\bibliography{article}{}
\bibliographystyle{aasjournalv7}



\end{document}